\documentclass[12pt]{article}
\RequirePackage{amsthm,amsmath,amsfonts,amssymb}

\usepackage{wrapfig}
\usepackage{float}

\usepackage{diagbox}

\usepackage{graphicx,enumerate}
\usepackage{natbib}

\usepackage{delarray}
\usepackage{hyperref, subcaption}
\usepackage{rotating}
\usepackage{url}
\usepackage{colortbl,color}
\usepackage{multirow,amssymb,amsmath, amsfonts,bm}

\usepackage{mathrsfs}
\usepackage{dsfont}
\usepackage{stmaryrd}
\usepackage{amsthm}

\usepackage{multibib}
\usepackage{bbm}
\usepackage{booktabs}
\usepackage{xr}

\numberwithin{equation}{section}
\usepackage[title]{appendix}

\usepackage[utf8]{inputenc}
\usepackage{chngcntr}
\usepackage{apptools}
\AtAppendix{\counterwithin{lem}{section}}
\AtAppendix{\counterwithin{prop}{section}}

\newtheorem{myPro}{Proof}

\newtheorem{example}{Example}

\newtheorem{assumption}{Assumption}

\newtheorem{defin}{Definition}
\newtheorem{lem}{Lemma}
\newtheorem{thm}{Theorem}
\newtheorem{prop}{Proposition}
\newtheorem{cor}{Corollary}

\theoremstyle{remark}
\newtheorem{myRem}{Remark}

\newcommand{\p}{{\rm I}\kern-0.18em{\rm P}}
\newcommand{\1}{{\rm 1}\kern-0.24em{\rm I}}
\newcommand{\E}{{\rm I}\kern-0.18em{\rm E}}
\newcommand{\uderbar}[1]{\underset{\raise0.3em\hbox{$\smash{\scriptscriptstyle-}$}}{#1}}
\def\boxit#1{\vbox{\hrule\hbox{\vrule\kern6pt\vbox{\kern6pt#1\kern6pt}\kern6pt\vrule}\hrule}}

\newcommand{\beginsupplement}{%
        \setcounter{table}{0}
        \renewcommand{\thetable}{\Alph{section}.\arabic{table}}%
        \setcounter{figure}{0}
        \renewcommand{\thefigure}{\Alph{section}.\arabic{figure}}%
     }

\newcommand{\blind}{1}

\begin{document}

\def\spacingset#1{\renewcommand{\baselinestretch}%
{#1}\small\normalsize} \spacingset{1}

%%%%%%%%%%%%%%%%%%%%%%%%%%%%%%%%%%%%%%%%%%%%%%%%%%%%%%%%%%%%%%%%%%%%%%%%%%%%%%

\if1\blind
{
  \title{\bf Non-splitting Neyman-Pearson Classifiers}
  \author{Jingming Wang \\
    Department of Statistics, Harvard University\\
Lucy Xia \\
  Department of ISOM, Hong Kong University of Science and Technology \\
	Zhigang Bao\\
	Department of Mathematics, Hong Kong University of Science and Technology\\
	and \\
	Xin Tong \thanks{Jingming Wang and Lucy Xia contribute equally to the work. J.M. Wang was partially supported by Hong Kong RGC GRF 16301519; L. Xia was partially supported by Hong Kong RGC ECS 26305120; Z.G. Bao   was partially supported by Hong Kong RGC GRF 16301520 and GRF 16305421;  X. Tong was partially supported by U.S. NSF grant DMS 2113500.  The authors thank Cheng Wang for helpful discussion.} \hspace{.2cm}\\
	Department of Data Sciences and Operations, 
	University of Southern California\\
	}
  \maketitle
} \fi

\if0\blind
{
  \bigskip
  \bigskip
  \bigskip
  \begin{center}
    {\LARGE\bf Non-splitting Neyman-Pearson Classifiers}
\end{center}
  \medskip
} \fi

\bigskip

\begin{abstract}
The Neyman-Pearson (NP) binary classification paradigm constrains the more severe type of error (e.g., the type I error) under a preferred level while minimizing the other (e.g., the type II error). This paradigm is suitable for applications such as severe disease diagnosis, fraud detection, among others. A series of NP classifiers have been developed to guarantee the type I error control with high probability.  However, these existing classifiers involve a sample splitting step: a mixture of class 0 and class 1 observations to construct a scoring function and some left-out class 0 observations to construct a threshold. This splitting enables classifier construction built upon independence, but it amounts to insufficient use of data for training and a potentially higher type II error.  Leveraging a canonical linear discriminant analysis (LDA) model, we derive a quantitative CLT for a certain functional of quadratic forms of the inverse of sample and population covariance matrices,  and based on this result,  develop for the first time NP classifiers without splitting the training sample.  Numerical experiments have confirmed the advantages of our new non-splitting parametric strategy.  \\

\end{abstract}

\noindent%
\vfill  
%
%\begin{keywords}
%classification, asymmetric error, Neyman-Pearson (NP) paradigm, NP oracle inequalities, minimum sample size requirement, non-sample-splittiing
%\end{keywords}

%\end{frontmatter}
\newpage
\spacingset{1.9} % DON'T change the spacing!

\section{Introduction}\label{sec:intro}

%\textcolor{red}{A plan for the intro:  (1) Introduce the NP paradigm and a series of work that delivers type I error control with high probability.  (2) why these classifiers need a split, and the challenges and benefit to do non-split.  Describe in detail the theoretical challenges and the novel solutions we provide; in particular, the technique is very different and more delicate then the techniques employed in all previous NP works, thus it is not an incremental work(3) briefly describe the new procedure, why we choose the LDA model.  Use the simulation example in the NSF proposal to show that it is of values. The main theoretical results of the paper.   (4) Explain the status of this paper in the NP line.  Specifically say that the paper is the first in the non-split area.  Potentially more work can be done here.   (spend some efforts here describing the importance of this work, it's not just another method in NP line, but it starts a new block/chapter in the NP literature.)Also, practically, what are the recommendations?  Can we say at least eLDA strictly dominates pNP-LDA?   (5) The organization of the paper.   }

\vspace{-0.05in}

Classification aims to accurately assign class labels (e.g., fraud vs. non-fraud) to new observations (e.g., new credit card transactions) on the basis of labeled observations (e.g., labeled transactions). The prediction is usually not perfect. In transaction fraud detection, two errors might arise: (1) mislabeling a fraudulent transaction as non-fraudulent and (2) mislabeling a non-fraudulent transaction as fraudulent. The consequences of the two errors are different: while declining a legitimate transaction may cause temporary inconvenience for a consumer, approving a fraudulent transaction can result in a substantial financial loss for a credit card company. In severe disease diagnosis (e.g., cancer vs. normal), the asymmetry of the two errors' importance is even greater: while misidentifying a healthy person as ill may cause anxiety and create additional medical expenses, telling cancer patients that they are healthy may cost their lives. In these applications, it is critical to prioritize the control of the more important error.

Most theoretical work on binary classification concerns risk.  Risk is a weighted sum of type I error (i.e., the conditional probability that the predicted label is $1$ given that the true label is $0$) and type II error (i.e., the conditional probability that the predicted label is $0$ given that the true label is $1$), where the weights are marginal probabilities of the two class labels. In the context of transaction fraud detection, coding the fraud class as $0$, we would like to control type I error
% (i.e., the probability of mislabeling fraud as non-fraud) 
 under some small level.  The common \textit{classical paradigm}, which minimizes the risk, does not guarantee delivery of classifiers that have type I error bounded by the preferred level.  To address this concern, we can employ a general statistical framework for controlling asymmetric errors in binary classification: the \textit{Neyman-Pearson (NP) classification paradigm}, which seeks a classifier that minimizes type II error subject to type I error $\leq$ $\alpha$, where $\alpha$ is a user-specified level, usually a small value (e.g., $5\%$ 
 %or $10\%$
 ). The NP framework can achieve the best type II error given a high priority on the type I error. 
 
The NP approach is fundamental in hypothesis testing (justified by the NP lemma), but its use in classification did not occur until the 21st century \citep{cannon2002learning, scott2005neyman}.  In the past ten years, there is significant progress in the theoretical/methodological investigation of NP classification. An incomplete overview includes (i) a theoretical evaluation criterion for NP classifiers: the NP oracle inequalities \citep{rigollet2011neyman}, (ii) classifiers satisfying this criterion under different settings  \citep{tong2013plug, zhao2016neyman, Tong.Xia.Wang.Feng.2020}, and (iii) practical algorithms for constructing NP classifiers \citep{tong2018neyman, Tong.Xia.Wang.Feng.2020}, (iv) generalizations to domain adaptation \citep{Scott.2019} and to multi-class \citep{Tian.Feng.2021}.

 Unlike the oracle classifier under the classical paradigm, which thresholds the regression function at precisely $1/2$, the threshold of the NP oracle is $\alpha$-dependent and needs to be estimated when we construct sample-based classifiers. Threshold determination is the key in NP classification algorithms, because it is subtle to ensure a high probability control on the type I error under $\alpha$ while achieving satisfactory type II error performance.  
 
 For existing NP classification algorithms \citep{tong2013plug, zhao2016neyman, tong2018neyman, Tong.Xia.Wang.Feng.2020}, a sample splitting step is common practice: a mixture of class 0 and class 1 observations to construct a scoring function $\hat s(\cdot)$ (e.g., fitted sigmoid function in logistic regression) and some left-out class 0 observations $\{x^0_1, \cdots, x^0_m\}$ to construct a threshold. Then under proper sampling assumptions, conditioning on $\hat s(\cdot)$, the set $\{s_1:= \hat s(x^0_1), \cdots, s_m:=\hat s (x^0_m)\}$ consists of independent elements. This independence is important in the subsequent threshold determination and classifier construction. Let us take the NP umbrella algorithm \citep{tong2018neyman} as an example: it constructs an NP classifier $\hat\varphi_{\alpha}(\cdot)=\1(\hat s(\cdot)> s_{(k^*)})$, where $\1(\cdot)$ is the indicator function,  $s_{(k^*)}$ is the $k^*$th order statistic in $\{s_1, \cdots, s_m\}$ and $k^*=\min\left\{k\in\{1, \cdots, m\}: \sum_{i=k}^m {m\choose j} (1-\alpha)^j \alpha^{m-j}\leq \delta\right\}$. The smallest order was chosen to have the best type II error.  
 The type I error violation rate has been shown to satisfy $\p(R_0(\hat \varphi_{\alpha})> \alpha)\leq \sum_{i=k}^m {m\choose j} (1-\alpha)^j \alpha^{m-j}$, where $R_0$ denotes the (population-level) type I error. Hence with probability at least $1-\delta$, we have $R_0(\hat \varphi_{\alpha})\leq  \alpha$. Without the independence of  $\{s_1, \cdots, s_m\}$, the upper bound on the violation rate does not hold.  Therefore, if we used up all class 0 observations in constructing $\hat s(\cdot)$, this umbrella algorithm fails.  In other NP works \citep{tong2013plug, zhao2016neyman, Tong.Xia.Wang.Feng.2020}, the independence is necessary in threshold determination when applying Vapnik-Chervonenkis inequality, Dvoretzky-Kiefer-Wolfowitz inequality, or constructing classic t-statistics, respectively.  
 
 In general, setting aside part of class 0 sample lowers the quality of the scoring function $\hat s(\cdot)$, and therefore makes  the type II error deteriorate.  This becomes a serious concern when the class 0 sample size is small. A more data-efficient alternative is to use all data to construct the scoring function, but this would lose the critical independence property when constructing the threshold.  Innovating a non-splitting strategy has long been in the ``wish list." This is an important but challenging task. For example, the NP umbrella algorithm, which has no assumption on data distribution and adapts all scoring-type classification methods (e.g., logistic regression, neural nets) to the NP paradigm universally via the non-parametric order statistics approach, has little potential to be extended to the non-splitting scenario, simply because there is no way to characterize the general dependence. To address it, we need to start from tractable distributional assumptions.

Among the commonly used models for classification is the linear discriminant analysis (LDA) model \citep{Hastie.Tibshirani.ea.2009, james2014introduction, Fan.Li.Zhang.Zou.2020}, which assumes that the two class-conditional feature distributions are Gaussian with different means but a common covariance matrix: $\mathcal{N}(\bm{\mu}^0, \Sigma)$ and $\mathcal{N}(\bm{\mu}^1, \Sigma)$. Classifiers based on the LDA model have been popular in the literature \citep{Shao.Wang.ea.2011, Fan.Feng.ea.2011, Witten.Tibshirani.2012,mai2012direct, Hao.Dong.Fan.2015,pan2016ultrahigh, Wang.Jiang.2018,  tony2019high, li2018sparse, JMLR:v21:19-428}. Hence, it is natural to start our inquiry with the LDA model. However, even this canonical model demands novel intermediate technical results that were not available in the literature.  For example, we will need delicate expansion results of \textcolor{black}{quadratic forms of the inverse of sample and population covariance matrices}, which we establish for the first time in this manuscript. %These results are also of standalone interests outside the NP classification paradigm.  {\color{red}[Not sure if we shall sell this.]}

As the first effort to investigate a non-splitting strategy under the NP paradigm, this work addresses basic settings.  We only work in the regime that $p/n\rightarrow [0,1)$, where $p$ is the feature dimensionality and $n$ is the sample size.  We take minimum assumptions on $\Sigma$ and $\bm{\mu}_d := \bm{\mu}^1 - \bm{\mu}^0$: $\bm{\mu}_d^{\top} \Sigma^{-1}\bm{\mu}_d$ is bounded from below.  %the eigenvalues of $\Sigma$ are bounded from both above and below, and the $L_2$ norm of $\bm{\mu}_d$ is bounded from below.  
We do not have specific structural assumptions on $\Sigma$ or $\bm{\mu}_d$ such as sparsity.  With these minimal assumptions, we propose our new classifier \verb+eLDA+ (where \verb+e+ stands for \textit{data efficiency}) based on a quantitative CLT for a certain functional of \textcolor{black}{quadratic forms of the inverse of sample and population covariance matrices} and show that \verb+eLDA+ respects the type I error control with high probability.  Moreover, if $p/n\rightarrow 0$, the excess type II error of \verb+eLDA+, that is the difference between the type II error of \verb+eLDA+ and that of the NP oracle, diminishes as the sample size increases; if $p/n\rightarrow r_0\in (0,1)$, the excess type II error of \verb+eLDA+ diminishes if and only if ${\bm{\mu}}_d^{\top} \Sigma^{-1} {\bm{\mu}}_d$ diverges.  We note in particular that this work is the first one to establish lower bound results on excess type II error under the NP  paradigm.  

%\textcolor{red}{double check the rate and argue why this is a reasonable results}.  \textcolor{red}{can be described as whether $\bm{\mu}_d$ diverges?} 

\begin{wraptable}{r}{0.4\linewidth}
\vspace{-50pt}
\caption{\small \texttt{eLDA} vs. \texttt{pNP-LDA}.\label{tb::simu1}}
\hspace{+5pt}
\begin{tabular}{r|rr}
\hline
&\texttt{eLDA}&\texttt{pNP-LDA}\\
\hline
type I error &.0314&.0037\\
type II error &.4478&.7638\\\hline
\end{tabular}
%\end{table}
\end{wraptable}

In addition to enjoying good theoretical properties, \verb+eLDA+ has numerical advantages.  Here we take a toy example:  $\Sigma = I$, $\bm{\mu}_d = (1.2, 1.2, 1.2)^{\top}$ and $\bm{\mu}^0 = (0,0,0)^{\top}$.  The sample sizes $n_0$ and $n_1$ for classes 0 and 1 respectively are both 50. We set the type I error upper bound $\alpha = 0.05$ and the type I error violation rate target $\delta = 0.1$.  In this situation, if we were to use the NP umbrella algorithm, we would have to reserve at least 45 (i.e., $\lceil \log \delta / \log(1-\alpha)\rceil$) class 0 observations for threshold determination, and thus at most 5 class 0 observations can be used for scoring function training. This is obviously undesirable. So we only compare the newly proposed \verb+eLDA+ with \verb+pNP-LDA+, another LDA based classifier proposed in \cite{Tong.Xia.Wang.Feng.2020} with sample splitting, whose threshold determination explicitly relies on the parametric assumption. In Table \ref{tb::simu1}, the type I error  and type II error were averaged over $1{,}000$ repetitions and evaluated on a very large test set ($50{,}000$ observations from each class) that approximates the population. The result shows that our new non-splitting \verb+eLDA+ classifier clearly outperforms the splitting \verb+pNP-LDA+ classifier by having a much smaller type II error. This example is not a coincidence. When the more generic nonparametric NP umbrella algorithm does not apply (or does not work well) due to sample size limitations, \verb+eLDA+ usually dominates \verb+pNP-LDA+.%\textcolor{red}{wrap the table from the proposal.}  

%Add some sentences about the numerical advantages, such as small sample sizes not applicable by NP umbrella algorithm, or \verb+pNP-LDA+ that is too conservative in terms of type I error.....  Add one example, 
 
 %Argue that this is the first in a series of papers non-split strategy.  The structural assumptions on LDA.    

 The rest of the paper is organized as follows.  In Section \ref{sec:models and set up}, we introduce the essential notations and assumptions.  In Section \ref{sec:mainresults}, we derive the efficient non-splitting NP classifier \verb+eLDA+ and its close relative $\verb+feLDA+$, where \verb+f+ stands for \textit{fixed feature dimension}, and show their main theoretical results. Technical preliminaries are presented in Section \ref{sec:Pre}, followed by the proof of the main theorem in Section \ref{sec:proof of main}.    In Section \ref{sec: numerical}, we present simulation and real data studies.  We provide a short discussion in Section \ref{sec:discussion}. In addition, in Appendix \ref{rem:assump}, we give further remark on our assumptions. The proofs of other theoretical results except for the main theorem are postponed to Appendix \ref{sec:appen_main-results}. In Appendix \ref{sec:appen_pf_tech}, we provide the proofs of the technical preliminaries in  Section \ref{sec:Pre}, followed by the proofs of the key lemmas in the proof of the main theorem in  Appendix \ref{sec:pf_LemProp}.  Finally, Appendix \ref{sec:appen_simu}  collects additional numerical results.

\vspace{-0.2in}
\section{Model and Setups}\label{sec:models and set up}
\vspace{-0.1in}

Let $\phi: \mathcal{X}\subset\mathbb{R}^p \to \{0,1\}$ denote a mapping from the feature space to the label space. The level-$\alpha$ NP oracle $\phi_\alpha^*(\cdot)$ is defined as the solution to the program
$
\min_{R_0(\phi)\leq \alpha} R_1(\phi),
$
where  $R_0(\phi)= \p \{ \phi({\bf x})\neq Y \big| Y=0\}$ and $R_1(\phi)= \p \{ \phi({\bf x})\neq Y \big| Y=1\}$ denote the (population-level) type I and type II errors of $\phi(\cdot)$, respectively.
We assume the linear discriminant analysis (LDA) model, i.e., $({\bf x}|Y=0)\sim \mathcal{N} ({\bm{\mu}}^0, \Sigma)$ and $({\bf x}|Y=1)\sim \mathcal{N} ({\bm{\mu}}^1, \Sigma)$,  where ${\bm{\mu}}^0, {\bm{\mu}}^1\in \mathbb{R}^{p}$ and the common positive definite covariance matrix $\Sigma \in \mathbb{R}^{p\times p}$.  Under the LDA model, the level-$\alpha$ NP oracle classifier can be derived explicitly as 
\begin{align}\label{def:class_NP}
\phi_\alpha^*(x)= \1 \Big( (\Sigma^{-1}{\bm{\mu}}_d)^{\top}  x > \sqrt{{\bm{\mu}}_d^{\top} \Sigma^{-1}{\bm{\mu}}_d} \, \Phi^{-1}(1-\alpha) + {\bm{\mu}}_d^{\top} \Sigma^{-1} {\bm{\mu}}^0\Big)\,,
\end{align}
in which ${\bm{\mu}}_d={\bm{\mu}}^1-{\bm{\mu}}^0$, and $\Phi^{-1}(1-\alpha)$ denotes the $(1-\alpha)$-th quantile of standard normal distribution.

%We want to construct sample-based estimator of the above classifier, whose  type II error is quite close to that of the above NP oracle classifier with type I error under the level $\alpha$.

For readers' convenience, we introduce a few notations together. For any $k\in \mathbb{N}$, let  $I_k$ denote  the identity matrix  of size $k$, ${\bf{1}}_{k}$ denote the  all-one column vector of dimension $k$. For arbitrary two column vectors ${\bf u}, {\bf v}$ of dimensions $a, b$, respectively, and any $a\times b$ matrix $M$, we write $(M)_{{\bf u} {\bf v}}$ as the quadratic form ${\bf u}^{\top} M {\bf v}$. Moreover, we write $M_{ij}$ or $(M)_{ij}$ for $i\in\{ 1,\cdots, a\}$ and $j\in \{ 1,\cdots, b\}$ as the $(i,j)$-th entry of $M$. We use $\Vert A \Vert $ to denote the operator norm for a matrix $A$ and use $\Vert \mathbf{v} \Vert $  to denote the $\ell_2$ norm of a  vector $\mathbf{v}$. For two positive sequences $A_n$ and $B_n$, we adopt the notation $A_n\asymp B_n$ to denote $C^{-1}A_n\leq B_n \leq C A_n$ for some constant $C>1$.
 We will use $c$ or $C$ to represent a generic positive constant which may vary from line to line. 
 
 In the methodology and theory development, we assume that we have access to i.i.d. observations from class 0, $\mathcal{S}^0=\{X^0_1, \cdots, X^0_{n_0}\}$, and i.i.d. observations from class 1, $\mathcal{S}^1=\{X^1_1, \cdots, X^1_{n_1}\}$, where the sample sizes $n_0$ and $n_1$ are non-random positive integers. Moreover, the observations in $\mathcal{S}^0$ and $\mathcal{S}^1$ are independent.  We also assume the following assumption unless specified otherwise.
\begin{assumption}\label{asm}
%Throughout the note, we suppose the following assumptions hold.\\
 
(i) (On feature dimensionality and sample sizes): the dimension of features $p$ and the sample sizes of the  two classes $n_0, n_1$ satisfy  $n_0/n> c_0, n_1/n>c_1$ for some positive constants $c_0$ and $c_1$, and 
$$r\equiv r_n:=p/n\to  r_0 \in [0,1) $$ 
as the sample size $n=n_0+n_1\to \infty$. 
\\
%(ii)(On $\Sigma$): For the common population covariance matrix $\Sigma$, we assume that its eigenvalues are bounded from both below and above, i.e.,
%\begin{align*}
%0<c<\lambda_p(\Sigma)\leq \lambda_1( \Sigma)< C<+\infty \,,
%\end{align*} 
%where $\lambda_i(\Sigma)$ is the $i$-th largest eigenvalue of $\Sigma$.
%\\
%(iii) (On $ {\bm{\mu}}_d$): For the mean difference vector, we assume there exists some constant $c>0$, such that 
%\begin{align*}
% \Vert {\bm{\mu}}_d\Vert\geq c\,.
%\end{align*}
(ii) (On Mahalanobis distance): we assume that
\begin{align}\label{def:Mahala_dis}
\varDelta_{d} :={\bm{\mu}}_d^{\top} \Sigma^{-1} {\bm{\mu}}_d \geq c_2\end{align}
for some positive constant $c_2>0$.

%{\color{red}{It seems that we can extend to $\Vert {\bm{\mu}}_d\Vert >0$. However, for  very small $\Vert {\bm{\mu}}_d\Vert$, the type II error of the NP oracle in (\ref{def:class_NP}) is very close to $1$.}}
\end{assumption}

%\textcolor{red}{In a later stage, some explanations should be added to these assumptions.}
 Assumption \ref{asm} is quite natural and almost minimal to the LDA model about $\Sigma$, ${\bm \mu}^0$, and ${\bm \mu}^1$.  First, our theory strongly depends on the analysis of population and  sample covariance matrices. To make the inverse  sample covariance matrix $\widehat{\Sigma}^{-1}$ well-defined, we have to restrict the ratio $p/n$ strictly smaller than $1$. Moreover, the sample size for either class needs to be comparable to the total sample size; otherwise, the class with a negligible sample size would be treated as noises. %Nevertheless, in practice, if we encounter a majority class and a minority class, we can select partial samples from the majority whose size is comparable to the minority class and then apply our theory. 
Second, since the Mahalanobis distance characterizes the difference between  the two classes, we adopt the common regularity condition in the literature that it is bounded from below by some positive constant. 
% the boundedness of the eigenvalues of population covariance matrix from both below and above is a mild condition which ensures that the quadratic form ${\bm \mu}_d^\top \Sigma^{-1} {\bm \mu}_d\asymp \Vert{\bm \mu}_d\Vert^2$. Thereby we can avoid the impact of the spectrum of $\Sigma$ in our analysis.  We might relax this assumption in theory, but in many of the practical scenarios, the common population covariance matrices fit this condition. We henceforth restrict ourselves to this setting for simplicity. 
 % it includes some extreme cases which gives trivial consequences. For instance, if  ${\bm \mu}_d^\top \Sigma^{-1} {\bm \mu}_d \asymp \lambda_p(\Sigma)^{-1} \to \infty$,  the type II error of oracle classifier  will tend to $0$ (see (\ref{21102901})) and our proposed estimator will not suffer from excess type II error in the very high dimension case since its type II error also tends to $0$. Note that this case is too special and is not of practical interest.
%   Third, we need to assume some minimum separation of the distributions of the two classes. This is characterized by a lower bound on the norm of the mean difference vector. %Without additional information (e.g., sparsity) on $\Sigma$ and ${\bm \mu}^i$ for $i=0,1$, the constant order condition has to be posed for the norm. 
  % Overall, Assumption \ref{asm} is mild to accommodate a broad range of  practical model settings. 

To create a sample-based classifier, the most straightforward strategy is to replace the unknown parameters in (\ref{def:class_NP}) with their sample counterparts. However, this strategy is not appropriate for our inquiry for two reasons:  (i) it is well-known that direct substitutions can result in inaccurate estimates when $p/n\to r_0\in (0,1)$; (ii) we aim for a high probability control on the type I error of the constructed classifier, and for that goal, a naive plug-in will not even work for fixed feature dimensionality.  These two concerns demand that delicate refinements and corrections be made to the sample counterparts.

Before diving into the classifier construction in the next section, we introduce the notations for sample covariance matrix $\widehat \Sigma$ and sample mean vectors $\hat {\bm{\mu}}^a$, $a= 1, 2$, and express them in forms that are more amenable in our analysis. Recall that 
\begin{align*}
\widehat \Sigma = \frac{1}{n_0+n_1-2}\sum_{a=0,1}\sum_{i=1}^{n_a}(X^a_i - \hat {\bm{\mu}}^a)(X^a_i - \hat {\bm{\mu}}^a)^{\top} \,,\quad \hat {\bm{\mu}}^a = \frac{1}{n_a}\left(X^a_1 + \ldots +X^a_{n_a}\right)\,, \quad a=0,1\,.
\end{align*}
We set  the $p$ by $n$ data matrix  by ${X}=(x_{ij})_{p,n}:= ({X}^0, {X}^1)$,  where 
\begin{align*}
{{X}}^a :=\frac{1}{(np)^{1/4}}\Sigma^{-\frac 12} (X_1^a- {\bm{\mu}}^a, \cdots, X_{n_a}^a- {\bm{\mu}}^a), \quad a=0,1\,.
\end{align*}
Note that all entries in the $p\times n$ matrix ${X}$ are i.i.d. Gaussian  with mean $0$ and variance $1/\sqrt{np}$. The scaling ${1}/{(np)^{1/4}}$ is to ensure that the spectrum of $XX^{\top} $ has asymptotically a fixed diameter, making it a convenient choice for technical derivations.  
We define two unit column vectors  of dimension $n$:
\begin{align} \label{def:e_0e_1}
{\bf{e}}_0:= \frac{1}{\sqrt{n_0}} ({\bf{1}}_{n_0}^{\top}, 0, \cdots, 0)^{\top}\,, \quad {\bf{e}}_1:= \frac{1}{\sqrt{n_1}} ( 0, \cdots, 0, {\bf{1}}_{n_1}^{\top} )^{\top}\,.
\end{align}

With the above notations, we can rewrite the sample covariance matrix $\widehat \Sigma$ as 
\begin{align} \label{def:hSigma_matrix}
\widehat \Sigma = \frac {\sqrt{np}}{n-2} \Sigma^{\frac 12}{{X}} \Big( I_n - {{E}}{{E}}^{\top}\Big) {{X}}^{\top} \Sigma^{\frac 12}\,,  \text{ where } E:= ({\bf{e}}_0, {\bf{e}}_1)\,. 
\end{align} 
 For the sample means, we can rewrite them as 
\begin{align}
&\hat {\bm{\mu}}^a =\sqrt{ \frac{n}{n_a}}  \, r^{\frac 14}\Sigma^{\frac 12} {X} {\bf e}_a+ {\bm{\mu}}^a\,,  \qquad a=0,1\,.\label{eq:repre_mu_0}
%&
%\hat {\bm{\mu}}^1 =\frac{1}{n_1} \big((np)^{\frac 14}\Sigma^{\frac 12} {X}^1 + {\bm{\mu}}^1 {\bf 1}_{n_1}^*\big)  {\bf 1}_{n_1}=\sqrt{ \frac{n}{n_1}}   \, r^{\frac 14}\Sigma^{\frac 12} {X} {\bf e}_1+ {\bm{\mu}}^1.\label{eq:repre_mu_1}
\end{align}
Furthermore, we write the sample mean difference vector as  
\begin{align}\label{eq:repre_mu_d}
\hat {\bm{\mu}}_d := \hat{\bm{\mu}}^1 -\hat{\bm{\mu}}^0
= r^{\frac 14}\Sigma^{\frac 12}{X}  {\bf v}_1 + {\bm{\mu}}_d\,, \qquad \text{where }
{\bf{v}_1}:= \begin{pmatrix}
-\frac{\sqrt n}{{n_0}} {\bf 1}_{n_0}\\
\frac{\sqrt n}{{n_1}} {\bf 1}_{n_1}
\end{pmatrix}= -\sqrt{\frac{n}{n_0}} {\bf e}_0 + \sqrt{\frac{n}{n_1}} {\bf e}_1\,.
\end{align}
%in which 
%\begin{align} \label{def:v1 X^E}
%{\bf{v}_1}:= \begin{pmatrix}
%-\frac{\sqrt n}{{n_0}} {\bf 1}_{n_0}\\
%\frac{\sqrt n}{{n_1}} {\bf 1}_{n_1}
%\end{pmatrix}= -\sqrt{\frac{n}{n_0}} {\bf e}_0 + \sqrt{\frac{n}{n_1}} {\bf e}_1\,.
%%{X}^E:= \frac{1}{\sqrt n}  \Big( {\bm{\mu}}^0 {\bf 1}_{n_0}^*,   {\bm{\mu}}^1{\bf 1}_{n_1}^*\Big).
%\end{align}

\section{New Classifiers and Main Theoretical Results}\label{sec:mainresults}
In this section, we propose our new NP classifier \verb+eLDA+ and establish its theoretical properties regarding type I and type II errors. We also construct a variant classifier  \verb+feLDA+ for fixed feature dimensions.

To motivate the construction of \verb+eLDA+, we introduce an \textit{intermediate} level-$\alpha$ NP oracle 
\begin{align} \label{def:inter_NP}
\tilde{\phi}_\alpha^*(x)= \1 \Big( \widehat{A} ^{\top} x > \sqrt{\widehat{A} ^{\top} \Sigma \widehat{A} } \, \Phi^{-1}(1-\alpha) +\widehat{A} ^{\top} {\bm{\mu}}^0\Big)\,,
\end{align}
where $\widehat A= \widehat \Sigma^{-1} \hat {\bm{\mu}}_d$ is a shorthand notation we will frequently use in this manuscript. 
One can easily deduce that the type I error of $\tilde{\phi}_\alpha^*(\cdot)$ in \eqref{def:inter_NP} is exactly $\alpha$. Note that $\tilde{\phi}_\alpha^*(\cdot)$ involves unknown parameters $\Sigma$ and ${\bm{\mu}}^0$, so it is not a sample-based classifier. However, it is still of interest to compare the type II error of $\tilde{\phi}_\alpha^*(\cdot)$ to that of the level-$\alpha$ NP oracle in (\ref{def:class_NP}). 

%First recall the type I error $R_0(\phi)$and type II error $R_1(\phi)$ as 
%\begin{align}\label{def:R01}
%R_0(\phi)= \p \{ \phi({\bf x})\neq Y \big| Y=0\}, \quad R_1(\phi)= \p \{ \phi({\bf x})\neq Y \big| Y=1\}.
%\end{align}

%\textcolor{red}{the next theorem should be modified.  First, the type I error equation holds a.s., this is more than with high probability.  We do not want to leave the impression that this is a high probability thing.  }

\begin{lem}\label{thm:1}
Let $\tilde{\phi}_\alpha^*(\cdot)$ be defined in (\ref{def:inter_NP}). Under Assumption \ref{asm}, the type I error of $\tilde{\phi}_\alpha^*(\cdot)$ is exactly $\alpha$, i.e., $R_0(\tilde{\phi}_\alpha^*\, ) = \alpha\,$. Further if $r = p/n \to 0$, then for any $\varepsilon \in(0, 1/2)$ and $D>0$,  when $n>n(\varepsilon, D)$, we have with probability at least $1- n^{-D}$ ,
 the type II error satisfies  $$R_1(\tilde{\phi}_\alpha^*\, )-  R_1({\phi}_\alpha^*\, ) \leq  C  \Big(r + n^{-\frac12+ \varepsilon}\Big)\sqrt{\varDelta_d}\, \exp\Big(- \frac{c\varDelta_d}{2}\Big)\,$$
for some constants $C, c>0$,  where $C$ may depend on $c_{0,1,2}$ and $\alpha$, and $\varDelta_d$ is defined in (\ref{def:Mahala_dis}).
\end{lem}

%\begin{thm}\label{thm:1}
%Let $\tilde{\phi}_\alpha^*(\cdot)$ be defined in (\ref{def:inter_NP}). Under Assumption \ref{asm}, the following results hold simultaneously  with probability at least $1- n^{-D}$ for some large $D>0$ and small $\varepsilon>0$ when $n>n(\varepsilon, D)$,
%
%(i) for the type I error, we have $$R_0(\tilde{\phi}_\alpha^*\, ) = \alpha\,;$$
%
%(ii) in addition, if $r = p/n \to 0$.  then we have for the type II error,  $$R_1(\tilde{\phi}_\alpha^*\, )-  R_1({\phi}_\alpha^*\, ) \leq  C (r + n^{-\frac 12+\varepsilon})\,.$$
%
%
%\end{thm}

%\textcolor{red}{The next reasoning is not rigorous and detailed enough.  Should supply a few more sentences for argument. Why this matters and how it connects to the rest of the section should be clearly argued.}

Lemma \ref{thm:1} indicates that $R_1(\tilde{\phi}_\alpha^*)-  R_1({\phi}_\alpha^*)$ goes to 0 under Assumption \ref{asm} and $p/n \to 0$.  This prompts us to construct a fully sample-based classifier by modifying the unknown parts of $\tilde{\phi}_\alpha^*(\cdot)$. Towards that, we denote the threshold of $\widehat A^{\top} x$ in $\tilde{\phi}_\alpha^*(\cdot)$ by
 \begin{align}\label{def:F}
F(\Sigma, {\bm{\mu}}^0):= \sqrt{\widehat{A} ^{\top} \Sigma \widehat{A} } \, \Phi^{-1}(1-\alpha) +\widehat{A} ^{\top} {\bm{\mu}}^0\,,
 \end{align}
and denote a sample-based estimate of $F(\Sigma, {\bm{\mu}}^0)$ by $\widehat F(\widehat\Sigma, \hat{\bm{\mu}}^0)$, whose exact form will be introduced shortly. By studying the difference between $ F(\Sigma, {\bm{\mu}}^0)$ and $\widehat F(\widehat\Sigma, \hat{\bm{\mu}}^0)$, we will construct a statistic $\widehat C_\alpha^p$  based on $\widehat F(\widehat\Sigma, \hat{\bm{\mu}}^0)$ (where the superscript $p$ stands for parametric) that is slightly larger than  $F(\Sigma, {\bm{\mu}}^0)$  with high probability.  The proposed classifier \verb+eLDA+ will then be defined by replacing  $F(\Sigma, {\bm{\mu}}^0)$ in (\ref{def:inter_NP}) with $\widehat C_\alpha^p$.

%In this section, we will state the expression for $\widehat F(\widehat\Sigma, \hat{\bm{\mu}}^0)$ and $\widehat C_\alpha^p$,   and introduce the main theorem about the proposed classifier. The detailed proof will be stated  in the next few sections. 

%\textcolor{red}{I think some rationale should still be introduced in this section.}

Concretely, suppose we hope that the probability of type I error of \verb+eLDA+ no larger than $\alpha$ is at least around $1-\delta$, for some small given constant $\delta\in(0,1)$. We define
 \begin{align}
&\widehat{F}(\widehat\Sigma, \hat{\bm{\mu}}^0):= \frac{\sqrt{  \widehat A^{\top} \widehat\Sigma \widehat A}}{1-r} \, \Phi^{-1}(1-\alpha) +  \widehat{A} ^{\top} \hat{\bm{\mu}}^0- \sqrt{\frac{n}{n_0}} \frac{r}{1-r} {\bf v}_1^{\top} {\bf e}_0\,,\label{def:whF}\\
&
\widehat C_\alpha^p:=\widehat{F}(\widehat\Sigma, \hat{\bm{\mu}}^0) +\sqrt{ \frac{\big((1-r) \widehat A^{\top} \widehat\Sigma \widehat A - r\Vert {\bf v}_1\Vert^2\big)  \widehat {V}}{ n}} \, \Phi^{-1}(1-\delta)\,, \label{def:whC}
\end{align}
in which $\widehat {V}=\sum_{i=1}^3 \widehat {V}_i$ and 
\begin{align} \label{def:V123}
&\widehat{V}_1:= \big((1-r) \widehat A^{\top} \widehat\Sigma \widehat A - r\Vert {\bf v}_1\Vert^2\big)\mathsf{C}^2 \Phi_\alpha^2  \frac{2(1+r)}{(1-r)^7}\,, \notag\\
&\widehat{V}_2:=\mathsf{C}^2 \Phi_\alpha^2 \Vert{\bf v}_1\Vert^2 \, \frac{4r(1+r)}{(1-r)^7}  + \frac{n}{n_0(1-r)^3}
+2 \mathsf{C} \Phi_\alpha \Vert{\bf v}_1\Vert \, \sqrt{\frac{n_1}{n_0}}\, \frac{2r}{(1-r)^5}\,,  \notag\\
&\widehat{V}_3:=\frac{\Vert {\bf v}_1\Vert^2}{ (1-r)\widehat A^{\top} \widehat\Sigma \widehat A  - r\Vert {\bf v}_1\Vert^2}  \Big(\mathsf{C}^2 \Phi_\alpha^2 \Vert{\bf v}_1\Vert^2 \, \frac{2r^2(1+r)}{(1-r)^7}  + \frac{(n+n_1)r}{n_0(1-r)^3}
+2 \mathsf{C} \Phi_\alpha \Vert{\bf v}_1\Vert \, \sqrt{\frac{n_1}{n_0}}\, \frac{2r^2}{(1-r)^5}  \Big)\,,
\end{align}
where $\mathsf{C}:=(1-r)(\hat {\bm{\mu}}_d^{\top} \widehat\Sigma^{-1} \hat {\bm{\mu}}_d )^{-\frac 12}/2 \,$, $\Phi_\alpha:=\Phi^{-1}(1-\alpha)$ and $ \widehat A := \widehat \Sigma^{-1} \hat{\bm{\mu}}_d\,$.
%\begin{align}\label{def:C_Phi_a}
%\mathsf{C}:=\frac{1-r}{2\sqrt{   \hat {\bm{\mu}}_d^{\top} \widehat\Sigma^{-1} \hat {\bm{\mu}}_d}}\,,
%\quad
%\Phi_\alpha:=\Phi^{-1}(1-\alpha)\,, \text{ and}\quad \widehat A := \widehat \Sigma^{-1} \hat{\bm{\mu}}_d\,.
%\end{align}

%\textcolor{red}{I think the notation $t$ should be explained a bit in here. Otherwise the audience might be confused.  We might say something like we hope the probability is close to $1-t$....}

To construct $\widehat{F}(\widehat\Sigma, \hat{\bm{\mu}}^0)$ and $\widehat C_\alpha^p$, we start with the analysis of the quadratic forms $\widehat{A} ^{\top} \Sigma \widehat{A}$, $\widehat{A} ^{\top} {\bm{\mu}}^0$ as well as their fully plug-in counterparts $ \widehat A^{\top} \widehat\Sigma \widehat A$, $ \widehat{A} ^{\top} \hat{\bm{\mu}}^0$. Once we obtain their expansions (Lemma \ref{lem:ests}) and compare their leading terms, we have the estimator $\widehat{F}(\widehat\Sigma, \hat{\bm{\mu}}^0)$ in \eqref{def:whF}. However, only having $\widehat{F}(\widehat\Sigma, \hat{\bm{\mu}}^0)$ close to $F(\Sigma, {\bm{\mu}}^0)$ in (\ref{def:F}) is not enough for the construction of an NP classifier. Note that the sign of $\widehat{F}(\widehat\Sigma, \hat{\bm{\mu}}^0)- F(\Sigma, {\bm{\mu}}^0)$ is uncertain. If the error is negative, directly using $\widehat{F}(\widehat\Sigma, \hat{\bm{\mu}}^0)$ as the threshold can actually  push the type I error above $\alpha$, which violates our top priority to maintain the type I error below the pre-specified level $\alpha$. To address this issue, we further study the asymptotic distribution of $\widehat F(\widehat\Sigma, \hat{\bm{\mu}}^0)-F(\Sigma, {\bm{\mu}}^0)$ and involve a proper quantile of this asymptotic distribution in the threshold. This gives the expression of $\widehat C_\alpha^p$ in (\ref{def:whC}). By this construction, we see that $\widehat C_\alpha^p$ is larger than $F(\Sigma, {\bm{\mu}}^0)$ with high probability so that the type I error will be maintained below $\alpha$ with high probability. Thanks to the  closeness  of $\widehat C_\alpha^p$ to $F(\Sigma, {\bm{\mu}}^0)$, the excess type II error of our new classifier \verb+eLDA+  shall be close to that of $\tilde{\phi}_\alpha^*(\cdot)$. Further by  Lemma \ref{thm:1},  we shall expect the excess type II error of  \verb+eLDA+ be close to that of ${\phi}_\alpha^*(\cdot)$, at least when $p/n\to 0$ .

%More precisely, we introduce our main theorem on the type I and II errors of our estimator of the oracle classifier.

Now with the above definitions, we formally introduce the new NP classifier \verb+eLDA+:
$$\hat{\phi}_\alpha(x)= \1 \left( \widehat{A} ^{\top} x > \widehat C_\alpha^p\right)\,,
$$
whose theoretical properties are described in the next theorem. 

%We emphasize here that the intermediate level-$\alpha$ oracle $\tilde{\phi}_\alpha^*(\cdot)$ defined in \eqref{def:inter_NP} plays an important role as a 'bridge' connecting \verb+eLDA+ $\hat{\phi}_\alpha(\cdot)$ to the level-$\alpha$ oracle $\phi^*_{\alpha}(\cdot)$ defined in \eqref{def:class_NP}.  Lemma \ref{thm:1} theoretically guarantees the exactly level-$\alpha$ type I error and a dimishing excess type II error in the case $p/n\to 0$ for $\tilde{\phi}_\alpha^*(\cdot)$. This inspires us to further construct an estimator very close to $\tilde{\phi}_\alpha^*(\cdot)$. Notice that the practically  problematic part of $\tilde{\phi}_\alpha^*(\cdot)$ lies in the threshold, i.e., (\ref{def:F}), which contains unknown parameters $\Sigma$ and ${\bm \mu}_0$. We aim to find an appropriate estimator for (\ref{def:F}). 

%\textcolor{red}{fine tune the above two paragraphs.}

\begin{thm} \label{mainthm}
Suppose that Assumption \ref{asm} holds. For any $\alpha, \delta \in(0,1)$, let  $\hat{\phi}_\alpha(x)= \1 \Big( \widehat{A} ^{\top} x > \widehat C_\alpha^p\Big)$, where $\widehat C_\alpha^p$ is defined in (\ref{def:whC}). Recall $\varDelta_d$  in (\ref{def:Mahala_dis}). Then there exist some positive constants $C_1, C_2>0$, such that for any $\varepsilon\in (0, 1/2)$ and $D>0$, when $n>n(\varepsilon, D)$,  it holds with probability at least $1- \delta - C_1 n^{-\frac 12+ \varepsilon} -C_2 n^{-D} $,

\noindent (i) the type I error satisfies:  $\quad R_0(\hat{\phi}_\alpha)\leq \alpha $;
%\begin{align*}
%R_0(\hat{\phi}_\alpha)\leq \alpha \, ;
%\end{align*}

\noindent (ii) for the type II error,  if $r = p/n \to 0$,
\begin{align} \label{IIerror:p/n-0}
R_1(\hat{\phi}_\alpha) - R_1({\phi}_\alpha^*) \leq  C\Big(r + n^{-\frac12+ \varepsilon}\Big)\sqrt{\varDelta_d}\,  \exp\Big(- \frac{c\varDelta_d}{2}\Big)\, ,
\end{align}
for some constants $C, c>0$, where $C$ may depend on $c_{0,1,2}$ and $\alpha$;
 if $r= p/n\to r_0\in (0, 1)$, 
\begin{align*}
\mathcal{L}\leq R_1(\hat{\phi}_\alpha) - R_1({\phi}_\alpha^*) \leq 
 \mathcal{U}\,,
\end{align*}
where 
\begin{align*}
&\mathcal{L}:=\frac{1}{\sqrt {2\pi }} \exp\Big( -\frac 12 \big(\Phi_\alpha - \delta_1 \sqrt{\varDelta_d}\; \big)^2\Big)  (1-\sqrt{1-r} \, -n^{-\frac{1}{2}+ \varepsilon} )\sqrt{\varDelta_d}\,, \notag\\
&\mathcal{U}:=\frac{1}{\sqrt {2\pi }} \exp\Big( -\frac 12 \big(\Phi_\alpha - \delta_2 \sqrt{\varDelta_d}\; \big)^2\Big)  \Big(1-\frac{\sqrt{1-r}}{\sigma} \, + n^{-\frac{1}{2}+ \varepsilon}\Big)\sqrt{\varDelta_d}\,,
\end{align*}
for $\Phi_\alpha=\Phi^{-1}(1-\alpha)$, and some $\sigma>1$, $\delta_1\in (\sqrt{1-r}\, , 1)$, $\delta_2\in(\sqrt{1-r}/\sigma\, , 1)$.

\end{thm}

\begin{myRem}
We comment on the excess type II error in Theorem \ref{mainthm}. When $p/n\to 0$,  the upper bound can be further bounded from above by a simpler form $C\left(r + n^{-\frac12+ \varepsilon}\right)\varDelta_d^{-\beta/2}$ for arbitrary $\beta\geq 1$. This simpler bound clearly implies that if $\varDelta_d=O(1)$, the excess type II error goes to 0, while if $ \varDelta_d$ diverges, the excess type II error would tend to $0$ at a faster rate compared to the bounded $\varDelta_d$ situation.  In contrast, when $p/n\to r_0\in(0,1)$, we provide explicit forms for both upper and lower bounds of the excess type II error. One can read from the lower bound $\mathcal{L}$ that if $\varDelta_d$ is of constant order, the excess type II error will not decay to $0$ since $\mathcal{L}\asymp 1$. Nevertheless, if $\varDelta_d$ diverges, then $\mathcal{U}\to 0$ and \verb+eLDA+ achieves diminishing excess type II error. In addition,  our Assumption \ref{asm} coincides with the previous margin assumption and detection condition \citep{tong2013plug, zhao2016neyman, Tong.Xia.Wang.Feng.2020} for an NP classifier to achieve a diminishing excess type II error. The detailed  discussion can be found in Appendix \ref{rem:assump}.
\end{myRem}

%\begin{thm} \label{mainthm}
%Suppose that Assumption \ref{asm} holds.  Let  $\hat{\phi}_\alpha(x)= \1 \Big( \widehat{A} ^{\top} x > \widehat C_\alpha^p\Big)$, where $\widehat C_\alpha^p$ is defined in (\ref{def:whC}). Let $t\in(0,1)$ be a small constant. Then there exist some constants $c_1, c_2$, such that for any small $\varepsilon>0$ and large $D>0$, when $n>n(\varepsilon, D)$,  it holds with probability at least $1- t - c_1 n^{-\frac 12+ \varepsilon} -c_2 n^{-D} $,
%
%(i) the type I error satisfies
%\begin{align*}
%R_0(\hat{\phi}_\alpha)\leq \alpha \, ;
%\end{align*}
%
%(ii) in addition, if $r = p/n \to 0$, we have for the type II error, 
%\begin{align*}
%R_1(\hat{\phi}_\alpha) - R_1({\phi}_\alpha^*) \leq  C (r + n^{-\frac 12+\varepsilon})\, .
%\end{align*}
%\end{thm}

%\begin{myRem}

%\textcolor{blue}{The audience might wonder at this point what happens if p/n goes to a non-zero number. Might move a statement about that to this point.}

Next we develop \verb+feLDA+, a variant of \verb+eLDA+, for bounded (or fixed) feature dimensionality $p$. In this case, thanks to  $r=O(1/n)$, we can actually simplify \verb+eLDA+.
%\textcolor{red}{The following few paragraphs needs to be checked again.}
%We remark here that  the type I error in both Theorem \ref{thm:1} and \ref{mainthm} hold in the regime that  $p/n \to r_0\in [0,1)$ and the results for type II error only hold for the case $p/n\to 0$. As such, for the special case of fixed $p$, or equivalently, $p=O(1)$,  both Theorem \ref{thm:1} and \ref{mainthm} hold. Nevertheless, thanks to  the fact that $r=O(1/n)$ in this setting, we can actually simplify the estimated classifier $\hat{\phi}_\alpha(x) $ and further get a simplified version of Theorem  \ref{mainthm}.
%For the special case $p$ is fixed, or equivalently $p=O(1)$, Theorem \ref{thm:1} and \ref{mainthm} remain true. {\color{red}[The two regimes shall be $p\gtrsim n^c$ and $p<n^c$ for an arbitrary (but fixed c), in order to use the local laws for the first regime]} Moreover, thanks to $r=O(1/n)$, we can even simplify the estimated classifier in Theorem \ref{mainthm} 
%The simplification is straightforward by absorbing all the $r$-related  terms in error according to (\ref{def:whF}) and (\ref{def:whC}). 
Concretely,
let $\widetilde{V} = {\Phi_\alpha^2}/{2}   +{n}/{n_0}$.
%\begin{align*}
%\widetilde{V} = \frac{\Phi_\alpha^2}{2}   +\frac{n}{n_0}\,.
%\end{align*}
Further define 
 \begin{align}
&\widetilde{F}(\widehat\Sigma, \hat{\bm{\mu}}^0):= \sqrt{  \widehat A^{\top} \widehat\Sigma \widehat A} \, \Phi^{-1}(1-\alpha) +  \widehat{A} ^{\top} \hat{\bm{\mu}}^0 \,, \label{def:wtF}\\
&\widetilde C_\alpha^p:=\widetilde{F}(\widehat\Sigma, \hat{\bm{\mu}}^0) +\sqrt {\widehat A^{\top} \widehat\Sigma \widehat A} \, \sqrt{\frac{\widetilde{V}}{n}} \, \Phi^{-1}(1-\delta)\,. \label{def:wtC}
\end{align}
Then, we can define an NP classifier  \verb+feLDA+: $\hat{\phi}_\alpha^f(x)= \1 \Big( \widehat{A} ^{\top} x > \widetilde C_\alpha^p\Big)$, and  we have the following corollary.

\begin{cor} \label{thm:prop}
Suppose that  Assumption \ref{asm} holds. Further, we assume that $p= O(1)$.  For $\alpha, \delta\in(0, 1)$, let $\hat{\phi}_\alpha^f(x)= \1 \Big( \widehat{A} ^{\top} x > \widetilde C_\alpha^p\Big)$, where $\widetilde C_\alpha^p$ is defined in \eqref{def:wtC}. Then there exist some constants $C_1, C_2$, such that for any $\varepsilon\in (0,1/2)$ and $D>0$, when $n>n(\varepsilon, D)$,  it holds with probability at least $1- \delta - C_1 n^{-\frac 12+ \varepsilon}-C_2 n^{-D} $,
\begin{align*}
R_0(\hat{\phi}_\alpha^f)\leq \alpha\,, \quad \text{ and }\quad  R_1(\hat{\phi}_\alpha^f) - R_1({\phi}_\alpha^*) \leq Cn^{-\frac12+ \varepsilon}
\sqrt{\varDelta_d} \, \exp\Big(- \frac{c\varDelta_d}{2}\Big)\,
\end{align*}
%and 
%\begin{align*}
%R_1(\hat{\phi}_\alpha^f) - R_1({\phi}_\alpha^*) \leq Cn^{-\frac12+ \varepsilon}
%\sqrt{\varDelta_d} \, \exp\Big(- \frac{c\varDelta_d}{2}\Big)\,
%\end{align*}
for some constants $C, c>0$, where $C$ may depend on $c_{0,1,2}$ and $\alpha$,  and $\varDelta_d$ is defined in (\ref{def:Mahala_dis}).
\end{cor}

Note that there is no essential difference between \verb+eLDA+ and \verb+feLDA+. The definitions of $\widetilde{F}(\widehat\Sigma, \hat{\bm{\mu}}^0)$ and $\widetilde C_\alpha^p$ are merely simplified counterparts of  (\ref{def:whF}) and (\ref{def:whC})  by neglecting terms related to $r$; they are negligible due to the approximate $O(1/n)$ size of $r$. The proof of Corollary \ref{thm:prop} is relegated to Appendix \ref{sec:appen_main-results}. 
%We can then simply conclude the results in Proposition  \ref{thm:prop} from Theorem \ref{mainthm} without a formal proof.

%\textcolor{red}{After the simulation and real data sections are done, should come back and give a hint about roughly when to use each classifier.}

%\textcolor{blue}{One solution is to have Part of Section 4 move here.  Lemma 2 should be moved to here, and the stochastic dominance concept should be introduced here.}

%\textcolor{red}{the classifiers eLDA and feLDA should not have the same notation $\hat \phi_{\alpha}^{*}$.  Also, the star in the data-driven classifiers is better be removed.  If that is done, then feLDA classifier can just take the notation $\hat \phi_{\alpha}^{f}$, and the eLDA classifier takes $\hat \phi_{\alpha}$}

%One will see later that the distinct high probability bounds are caused by the different convergence rates of quadratic forms of Green function under different  regimes (i.e, $p>n^{\varepsilon}$ for some $\varepsilon>0$ and $p<n^{\varepsilon}$ for any $\varepsilon>0$ ).

\section{Technical Preliminaries} \label{sec:Pre}

In this section, we collect a few basic notions in random matrix theory and introduce some preliminary results that serve as technical inputs in our classifier construction process. %\textcolor{red}{After we fix the rest of the chapter, come back and rewrite this paragraph.}

%\subsection{Random matrix basics}

%We will use a few notations and basic results from the random matrix literature.  For the readers' convenience and to fix the notations, here we review some random matrix basics. 

 Recall  the $p\times n$ data matrix ${X}$ whose entries are i.i.d. Gaussian with mean $0$, variance $1/\sqrt{np}$. We introduce its sample covariance matrix $H:=  {X}{X}^{\top} $ and  the matrix $\mathcal{H}:= {X}^{\top}{X}$ which has the same non-trivial eigenvalues as $H$. Their Green functions  are defined by 
\begin{align*}
\mathcal{G}_1(z):= (H - z)^{-1}\,, \qquad   \mathcal{G}_2(z):= (\mathcal{H} - z)^{-1}\,, \qquad  z\in \mathbb{C}^+\,.
\end{align*}
Besides, we denote the normalized  traces of $\mathcal{G}_{1}(z)$ and $\mathcal{G}_{2}(z)$ by 
\begin{align*}
&m_{1n}(z):=\frac{1}{p}\text{Tr}\mathcal{G}_1(z)=\int\frac{1}{x-z}\,dF_{1n}(x)\,,\quad m_{2n}(z):=\frac{1}{n}\text{Tr}\mathcal{G}_2(z)= \int \frac{1}{x-z}\,dF_{2n}(x)\,,
\end{align*}
where  $F_{1n}(x)$, $F_{2n}(x)$ are the empirical spectral distributions of $H$ and $\mathcal{H}$ respectively, i.e.,
\begin{align}
F_{1n}(x):=\frac{1}{p}\sum_{i=1}^p \1(\lambda_i(H)\leq x)\,, \quad F_{2n}(x):=\frac{1}{n}\sum_{i=1}^n\1(\lambda_i(\mathcal{H})\leq x)\,. \nonumber
\end{align}
Here we used $\lambda_i(H)$ and $\lambda_i(\mathcal{H})$  to denote the $i$-th largest eigenvalue of $H$ and $\mathcal{H}$, respectively. Observe that $\lambda_i(H)=\lambda_i(\mathcal{H})$ for $i=1,\cdots, p$. 

It is well-known %since \cite{MP67} 
that $F_{1n}(x)$ and $F_{2n}(x)$ converge weakly (a.s.) to the {\it Marchenko-Pastur} laws 
$\nu_{\text{MP},1}$ and $\nu_{\text{MP},2}$ (respectively) given below
\begin{align}
&\nu_{\text{MP},1}({\rm d}x):=\frac{1}{2\pi x \sqrt r}\sqrt{\big((\lambda_+-x)(x-\lambda_-)\big)_+}{\rm d}x+(1-\frac{1}{ r})_+\delta({\rm d}x)\,,\nonumber\\
&\nu_{\text{MP},2}({\rm d}x):=\frac{\sqrt r}{2\pi x}\sqrt{\big((\lambda_+-x)(x-\lambda_-)\big)_+}{\rm d}x+(1-r)_+\delta({\rm d}x)\,, \label{19071801}
\end{align}
where $\lambda_{\pm}:=\sqrt r+ 1/\sqrt r \pm 2$.  Note that here the parameter $r$ may be $n$-dependent. Hence, the weak convergence (a.s.) shall be understood as $\int g(x) {\rm d} F_{an}(x)-\int g(x) \nu_{\text{MP},a}({\rm d}x) \stackrel{a.s.} \longrightarrow 0 $ for any given bounded continuous function $g:\mathbb{R}\to \mathbb{R}$, for $a=1,2$.  %We denote by $F_a$ the cumulative distribution function of $\nu_{\text{MP},a}$ for $a=1,2$. 
Note that $m_{1n}$ and $m_{2n}$ can be regarded as the Stieltjes transforms of $F_{1n}$ and $F_{2n}$, respectively.  We further define their deterministic counterparts, i.e.,  Stieltjes transforms of $\nu_{\text{MP},1},\nu_{\text{MP},2}$,  by $m_1(z),m_2(z)$,  respectively, i.e., $m_a(z):=\int (x-z)^{-1}\nu_{\text{MP},a}({\rm d}x)$, for $a=1,2$.
% \begin{align*}
% m_a(z):=\int \frac{1}{x-z}\nu_{\text{MP},a}({\rm d}x)\,,\qquad a=1,2\,.
% \end{align*}
 From the definition (\ref{19071801}), it is straightforward to derive  
 \begin{align}
&m_1(z)=\frac{r^{-1/2}-r^{1/2} - z+\mathrm{i}\sqrt{(\lambda_+-z)(z-\lambda_-)}}{2r^{1/2}z}\,,  \nonumber\\
&m_2(z)=\frac{r^{1/2}-r^{-1/2} - z+\mathrm{i}\sqrt{(\lambda_+-z)(z-\lambda_-)}}{2r^{-1/2}z}\,, \label{m1m2}
 \end{align}
where the square root is taken with a branch cut on the negative real axis. Equivalently, we can also characterize $m_1(z),m_2(z)$ as the unique solutions from $\mathbb{C}^+$ to $\mathbb{C}^+$ to the equations
\begin{align}
zr^{1/2}m_1^2+[z-r^{-1/2}+r^{1/2}]m_1+1=0\,, \quad  z r^{-1/2}m_2^2+[z-r^{1/2}+r^{-1/2}]m_2+1=0\,. \label{selfconeqt}
\end{align}

%\textcolor{red}{equation 19 does not has hats but equation 20 does. Is it right? }

%\subsection{Local laws}
In later discussions, we need the estimates of  the quadratic forms of Green functions. % $ (\mathcal{G}_{a}^b)$ and $(\mathcal{G}_{a}^b{X})$  for $a=1,2$ and $b\in \mathbb{N}$. \textcolor{red}{The previous notations were used without a prior definition} 
%Hence, in the sequel, we introduce the so-called isotropic  local law with some consequences which provide good estimates of the aforementioned forms.
 Towards that,  we define  the notion {\it stochastic domination} which was initially  introduced in \cite{EKY2013}.  It provides a precise statement of the form ``$\textsf{X}_N$ is bounded by $\textsf{Y}_N$ up to a small power of $N$ with high probability''.
%{\color{red}{Some of the notations like $\rho, \tau$ may need to be replaced by others to avoid the abuse of notations.}}

%\textcolor{red}{Do $\mathbb{N}$ and $\mathbb{N}$ mean the same thing?  If so, can we just use one of them, and which one do you prefer?  Also, $X$ has been used as the $p\times n$ data matrix in previous sections. Can we use some different notations for the definition of "stochastic dominance"?  I can type them into latex, but I need some suggestion here. }

%\textcolor{red}{In the original paper [1], both X and Y are nonnegative, do we actually want to retain that, why just want Y to be nonnegative.  I changed it in the definition according to [1] but this might not be what we wanted.  }

\begin{defin}(Stochastic domination) \label{def.sd} Let
\begin{align*}
\mathsf{X}=\big(\mathsf{X}_N(u): N\in \mathbb{N}, u\in U_N\big) \quad \text{ and }\quad \mathsf{Y}=\big(\mathsf{Y}_N(u): N\in \mathbb{N}, u\in U_N\big)
\end{align*}
 be two families of random variables, $\mathsf{Y}$ is nonnegative, 
 and $U_N$ is a possibly $N$-dependent parameter set.  We say that $\mathsf{X}$ is stochastically dominated  by $\mathsf{Y}$, uniformly in $u$, if for all small $\varrho>0$ and large $\phi>0$, we have
 \begin{align*}
 \sup_{u\in U_N}\p\big(|\mathsf{X}_N(u)|>N^{\varrho}\mathsf{Y}_N(u)\big)\leq N^{-\phi}
 \end{align*}
 for large $N\geq N_0(\varrho, \phi)$. Throughout the paper, we use the notation $\mathsf{X}=O_\prec(\mathsf{Y})$ or $\mathsf{X}\prec \mathsf{Y}$ when $\mathsf{X}$ is stochastically dominated by $\mathsf{Y}$ uniformly in $u$.   Note that in the special case when $\mathsf{X}$ and $\mathsf{Y}$ are deterministic, $\mathsf{X}\prec \mathsf{Y}$
	means for any given $\varrho>0$,  $|\mathsf{X}_{N}(u)|\leq N^{\varrho}\mathsf{Y}_{N}(u)$ uniformly in $u$, for all sufficiently large $N\geq N_0(\varrho)$.

%In addition,  we also say that an $N$-dependent event $\mathcal{E}\equiv \mathcal{E}(N)$ holds with overwhelming probability if, for any large $\varphi>0$,  
%\begin{align*}
%\p(\mathcal{E}) \geq 1- N^{-\varphi}\,,
%\end{align*}
%for sufficiently large $N \geq N_0( \varphi)$.
\end{defin}

%\textcolor{red}{Jingming, we switched to the ``overwhelming probability" following Prof. Bao's suggestion.  But I don't actually see it is used somewhere in the paper.  Is it so? }

%\textcolor{red}{Note that this ``high-probability" is a different from what the NP papers call ``high probability". One way is to call here ``overwhelming probability" or something else,  and make a note that says `` usually in the random matrix literature, this is call high probability, but to avoid the language ambiguity in this context..... " }

%\textcolor{red}{might need Jingming to clean the $\mathsf{X}$ notations and the overwhelming probability in other places }

\begin{defin} \label{defn_asymptotic}
Two sequences of random vectors, $\mathsf X_N \in \mathbb{R}^k$ and $\mathsf Y_N \in \mathbb{R}^k$, $N\geq 1$,  are \emph{asymptotically equal in distribution}, denoted as $\mathsf X_N \simeq \mathsf Y_N,$ if they are tight and satisfy
$
\lim_{N \rightarrow \infty} \big( \mathbb{E}f(\mathsf X_N)-\mathbb{E}f(\mathsf Y_N) \big)=0 
$, 
for any bounded continuous function $f:\mathbb{R}^k\to \mathbb{R}$. 
\end{defin}

Further, we introduce a basic lemma based on Definition \ref{def.sd}.
\begin{lem} \label{prop_prec} Let $  \mathsf{X}_i=(\mathsf{X}_{N,i}(u):  N \in \mathbb{N}, \ u \in {U}_{N})\,, \   \mathsf{Y}_i=(\mathsf{Y}_{N,i}(u):  N \in \mathbb{N}, \ u \in {U}_{N})$,  $i=1,2$,
%	\begin{equation*}
%   \mathsf{X}_i=(\mathsf{X}_{N,i}(u):  N \in \mathbb{N}, \ u \in {U}_{N})\,, \   \mathsf{Y}_i=(\mathsf{Y}_{N,i}(u):  N \in \mathbb{N}, \ u \in {U}_{N})\,,\quad i=1,2\,,
%	\end{equation*}
	be families of  random variables, where $\mathsf{Y}_i, i=1,2,$ are nonnegative, and ${U}_{N}$ is a possibly $N$-dependent parameter set.	Let 
$
	\Phi=(\Phi_{N}(u): N \in \mathbb{N}, \ u \in {U}_{N})
$
	be a family of deterministic nonnegative quantities. We have the following results:
	
(i)	If $\mathsf{X}_1 \prec \mathsf{Y}_1$ and $\mathsf{X}_2 \prec \mathsf{Y}_2$ then $\mathsf{X}_1+\mathsf{X}_2 \prec \mathsf{Y}_1+\mathsf{Y}_2$ and  $\mathsf{X}_1\mathsf{X}_2 \prec \mathsf{Y}_1 \mathsf{Y}_2$.

 (ii) Suppose $\mathsf{X}_1 \prec \Phi$, and there exists a constant $C>0$ such that  $|\mathsf{X}_{N,1}(u)| \leq N^C$ a.s.\ and $\Phi_{N}(u)\geq N^{-C}$ uniformly in $u$ for all sufficiently large $N$. Then $\E \mathsf{X}_1 \prec \Phi$.
\end{lem}

We introduce the following domain. For a small fixed $\tau$, we define 
\begin{align}\label{19071810*}
\mathcal{D}^0\equiv\mathcal{D}(\tau)^0:= \{ z\in \mathbb{C}^+:-\tau< \Re{z} <  \tau, 0< \Im{z}\leq \tau^{-1}\}\,.
\end{align}
Conventionally, for $a=1,2$, we use $\mathcal{G}_a^{\ell}$ and $\mathcal{G}_a^{(\ell)}$ to represent $\ell$-th power of $\mathcal{G}_a$ and the $\ell$-th derivative of $\mathcal{G}_a$ with respect to $z$, respectively.  
 With these notations, we introduce the following proposition which is known as local laws, which shall be regarded as slight adaptation of the results in \cite{bloemendal2014isotropic}, in the Gaussian case.

 \begin{prop} \label{thm:locallaw}
Let  $\tau>0$ be a small but fixed constant.
% Let  ${\bf u},{\bf v}$ be complex deterministic unit vectors of proper dimensions. 
Under Assumption \ref{asm},  for any given $l\in\mathbb{N}$, we have  
\begin{align}
&\Big|\big( \mathcal{G}_1^{(l)}(z)\big)_{ij} - m_1^{(l)}(z) \delta_{ij} \Big|\prec  {n^{-\frac 12}}r^{\frac{1+l}{2}}\, , \quad \Big|\big(z\mathcal{G}_2(z)\big)^{(l)}_{i'j'} - \big(zm_2(z)\big)^{(l)} \delta_{i'j'} \Big| \prec  n^{-\frac 12}r^{\frac{1+l}{2}}\, , \label{est:locallaw1}\\
&\Big|\big( {X}^{\top}\mathcal{G}_1^{(l)}(z)\big)_{i'i} \Big|\prec n^{-\frac 12}r^{\frac 14+ \frac l 2}\, , \quad 
\Big|\big({X}\big(z\mathcal{G}_2(z) \big)^{(l)}\big)_{ii'} \Big|\prec  n^{-\frac 12} r^{-\frac 14+ \frac l 2}\, , \label{est:locallaw2}\\
&\big|m_{1 n}^{(l)}(z)-m^{(l)}_1(z) \big|\prec n^{-1} r^{\frac l 2}\, , \quad  \big|\big(zm_{2 n}(z)\big)^{(l)}-\big(zm_2(z)\big)^{(l)} \big|\prec  n^{-1}r^{\frac {1+l}{ 2}}\label{est_m12N}\, ,
\end{align}
uniformly in $z\in{\mathcal{D}^0} $ and for any $i, j \in \{1,\cdots, p\}$ and $i' , j' \in \{1,\cdots, n\}$. For $l=0$, the second estimates in  (\ref{est:locallaw2}) and (\ref{est_m12N}) can be improved to 
\begin{align}\label{2021051301}
\Big|\big({X}\big(z\mathcal{G}_2(z) \big)\big)_{ii'} \Big|\prec  n^{-\frac 12} r^{\frac 14}\,,\quad 
 \big|\big(zm_{2 n}(z)\big)-\big(zm_2(z)\big) \big|\prec  n^{-1}r\,.
\end{align}
%where $\kappa=E-\lambda_+$. 
%Further, when $\kappa>K$ for some sufficiently large constant $K>0$ and $|y-1|\geq \tau_0$ for any small but fixed $\tau_0>0$,  (\ref{est_m12N}) can be improved to 
%\begin{align}
%|m_{\alpha N}^{(l)}(z)-m^{(l)}_\alpha(z)|=O_\prec\Big(\frac{1}{N(\kappa+\eta)^{l+2}}\Big).
%\label{est_m12N for large z}
%\end{align}
%\textcolor{red}{Another question is that the $\mathcal{D}^0$ is used above, but not $\mathcal{D}$.  Do we actually need the definition of $\mathcal{D}$?}
\end{prop}

\begin{myRem} \label{rmk:locallaw}
By the orthogonal invariance of Gaussian random matrix,  we get from Proposition \ref{thm:locallaw} that  for ${\bf u},{\bf v}$, any complex deterministic unit vectors of proper dimensions, 
\begin{align}
&|\langle {\bf u},\mathcal{G}_1^{(l)}(z){\bf v}\rangle-m_1^{(l)}(z)\langle {\bf u}, {\bf v}\rangle|\prec  n^{-\frac 12}r^{\frac{1+l}{2}}, \quad   |\langle {\bf u},\big(z\mathcal{G}_2(z)\big)^{(l)}{\bf v}\rangle-\big(zm_2(z)\big)^{(l)}\langle {\bf u}, {\bf v}\rangle|\prec n^{-\frac 12}r^{\frac{1+l}{2}}, \label{est.DG}
\\
&|\langle {\bf u},{X}^{\top}\mathcal{G}_1^{(l)}(z){\bf v}\rangle|\prec  n^{-\frac12}r^{\frac 14+ \frac l 2}, \quad  |\langle  {\bf u},{X}\big(z\mathcal{G}_2(z) \big)^{(l)}{\bf v}\rangle|\prec n^{-\frac12}r^{\frac 14+ \frac l 2}, \label{081501}
%&|m_{1 n}^{(l)}(z)-m^{(l)}_1(z)|\prec p^{-\frac 12}n^{-\frac 12}, \quad  |\big(zm_{2 n}(z)\big)^{(l)}-\big(zm_2(z)\big)^{(l)}|\prec n^{-1}
\end{align}
uniformly for $z\in \mathcal{D}^0$.
\textcolor{black}{We further remark that the estimates above and the ones in Proposition \ref{thm:locallaw} also hold at $z=0$ with error bounds unchanged by the Lipschitz continuity  of $\mathcal{G}_{1}, z\mathcal{G}_2(z)$, $m_1(z)$, and $zm_2(z)$.
And we will use (\ref{est_m12N}),  (\ref{est.DG}), and (\ref{081501}) frequently in technical proofs not only for $z\in \mathcal{D}^0$ but also at $z=0$.}
 %\textcolor{red}{Can we say things more specifically than ``later discussion"?   }

\end{myRem}

\section{Proof of Theorem \ref{mainthm}}\label{sec:proof of main}

In this section, we prove our main theorem, i.e.,  Theorem \ref{mainthm}.  To streamline the proof, we first present two technical results and their proof sketches.

\begin{lem}\label{lem:ests}
Suppose that Assumption \ref{asm} holds. Recall the definition of $\varDelta_d$ in (\ref{def:Mahala_dis}). Let $\widehat A = \widehat \Sigma^{-1}\hat {\bm{\mu}}_d$, then we have 
\begin{align}
&\widehat{A} ^{\top} \Sigma \widehat{A} =   \frac{r}{(1-r)^3}\Vert {\bf v}_1\Vert^2 + \frac{1}{(1-r)^3} \,\varDelta_d + O_\prec \big(n^{-\frac 12}\varDelta_d \big)\,,\label{est:hAShA}\\
&\widehat{A} ^{\top} \widehat\Sigma \widehat{A} =   \frac{r}{1-r} \Vert {\bf v}_1\Vert^2 +  \frac{1}{1-r} \, \varDelta_d  +O_\prec \big(n^{-\frac 12}\varDelta_d )  \big)\,, \label{est:hAhShA} \\
%\end{align}
%and 
%\begin{align}
& \widehat{A}^{\top}  {\bm{\mu}}_d  = \frac{1}{1- r}  \varDelta_d  + O_{\prec} \big(n^{-\frac 12} \varDelta_d\big)\, ,  \label{est:hAmu0} \\
& \widehat{A} ^{\top} \hat{\bm{\mu}}^0 - \widehat{A} ^{\top}  {\bm{\mu}}^0 = \sqrt{\frac{n}{n_0}} \frac{r}{1-r} {\bf v}_1^{\top} {\bf e}_0 + O_\prec \big( n_0^{-\frac 12} \varDelta_d^{\frac 12}\big)\, .   \label{est:hAhmu0}
\end{align}
%
%\begin{align}
%&\widehat{A} ^{\top}  {\bm{\mu}}^0=  \frac{1}{1-r}\,  {\bm{\mu}}_d^{\top} \Sigma^{-1}{\bm{\mu}}^0+ O_\prec(n^{-\frac 12} \Vert{\bm{\mu}}_d\Vert \Vert {\bm{\mu}}^0\Vert )\,,
%\label{est:hAmu0} \\
%&\widehat{A} ^{\top} \hat{\bm{\mu}}^0=  \frac{1}{1-r} {\bm{\mu}}_d^{\top} \Sigma^{-1}{\bm{\mu}}^0 + \sqrt{\frac{n}{n_0}} \frac{r}{1-r} {\bf v}_1^{\top} {\bf e}_0+ O_\prec\big(n^{-\frac 12} \Vert{\bm{\mu}}_d\Vert (\Vert {\bm{\mu}}^0\Vert+1) \big)\,. \label{est:hAhmu0}
%\end{align}
%in which ${\bf v}_1 $ is defined in (\ref{def:v1 X^E}). 
Moreover, counterparts of  (\ref{est:hAhmu0}) also hold  if the triple $({\bm{\mu}}^0, \hat{{\bm{\mu}}}^0, \sqrt{n/n_0} \, {\bf e}_0)$ is replaced by $({\bm{\mu}}^1, \hat{{\bm{\mu}}}^1, \sqrt{n/n_1} \, {\bf e}_1)$ or $({\bm{\mu}}_d, \hat{{\bm{\mu}}}_d , {\bf v}_1)$.
\end{lem}

%\textcolor{red}{Jingming, for the main theorem (Theorem 1) to hold, is there any place that we need $\|\bm{\mu}^a\|/n^{1/4}$ goes to 0 for $a=0,1$? If so, this means $\|\bm{\mu}_d\|$ can not grow too fast.  But this is a little counterintuitive, as a larger $\|\bm{\mu}_d\|$ means that the problem is easier. }

\begin{myRem} \label{rmk:ests}
Lemma \ref{lem:ests} hints that we can use $\widehat{A} ^{\top} \widehat\Sigma \widehat{A} /(1-r)^2$ to estimate $\widehat{A} ^{\top}  \Sigma \widehat{A} $ and use $\widehat{A} ^{\top} \hat{\bm{\mu}}^0 - \sqrt{\frac{n}{n_0}} \frac{r}{1-r} {\bf v}_1^{\top} {\bf e}_0$ to approximate $\widehat{A} ^{\top} {\bm{\mu}}^0$. Therefore,  we construct $\widehat{F}(\widehat\Sigma, \hat{\bm{\mu}}^0)$, whose definition is explicitly given in (\ref{def:whF}). Moreover, when $p$ is fixed, i.e., $r=O(1/n)$, we get the following simplified estimates
\begin{align}
&\widehat{A} ^{\top} \Sigma \widehat{A} = \varDelta_d+  O_\prec(n^{-\frac 12}\varDelta_d)\, ,\quad \widehat{A} ^{\top} \widehat\Sigma \widehat{A} =   \varDelta_d  + O_\prec(n^{-\frac 12}\varDelta_d)\, , \label{est:hAhShA_s}\\
&\widehat{A}^{\top}  {\bm{\mu}}_d  =   \varDelta_d + O_\prec(n^{-\frac 12} \varDelta_d)\,  ,
\quad \widehat{A} ^{\top} \hat{\bm{\mu}}^0 - \widehat{A} ^{\top}  {\bm{\mu}}^0 = O_\prec ( n_0^{-\frac 12} \varDelta_d^{\frac 12}) \,. \label{est:hAhmu0_s}
\end{align}
\end{myRem}

We provide a proof sketch  of Lemma \ref{lem:ests}, while a formal proof is presented in the Supplementary Materials.  Our starting point is to expand  $\widehat{\Sigma}^{-1}$ in terms of Green function $\mathcal{G}_1(z)= (XX^\top -z)^{-1}$ at $z=0$ since all the quadratic forms in Lemma \ref{lem:ests} can be rewritten as certain quadratic forms of $\widehat{\Sigma}^{-1}$ according to the representations (\ref{def:hSigma_matrix})-(\ref{eq:repre_mu_d}). Working with Green functions makes the analysis much easier due to the useful estimates in local laws, i.e.,  Proposition \ref{thm:locallaw} and its variants (\ref{est.DG}), (\ref{081501}).  In this expansion, we will need some elementary linear algebra (e.g., Woodbury matrix identity) to compute matrix inverse and local laws (\ref{est_m12N}), (\ref{est.DG}) and (\ref{081501}) to estimate the error terms. Next, with the expansion of $\widehat{\Sigma}^{-1}$ plugged in, all the quadratic forms we want to study in Lemma \ref{lem:ests} can be further simplified  to linear combinations of  quadratic forms of $\mathcal{G}_1^a(0)$, $\mathcal{G}_1^a(0)X$, and $X^\top \mathcal{G}_1^a(0) X$, for $a=1,2$. Then, further derivations with the aid of  local laws (\ref{est_m12N}), (\ref{est.DG}) and (\ref{081501}) lead to the ultimate expressions. All these derivations only need the first order expansion since we focus on the leading terms.

Next, we describe the difference between $\widehat{F}(\widehat\Sigma, \hat{\bm{\mu}}^0)$ and $F(\Sigma, {\bm{\mu}}^0)$ by a quantitative CLT.

\begin{prop}\label{prop.asym_dist}
Let $F(\Sigma, {\bm{\mu}}^0)$ and $\widehat{F}(\widehat\Sigma, \hat{\bm{\mu}}^0)$   be defined in (\ref{def:F}) and (\ref{def:whF}),  respectively. Under Assumption \ref{asm}, we have
\begin{align}\label{repre:sF-F}
\quad \widehat{F}(\widehat\Sigma, \hat{\bm{\mu}}^0)- F(\Sigma, {\bm{\mu}}^0)
&= \frac {\sqrt{ (1-r)  \hat{\bm{\mu}}_d^{\top}  \widehat\Sigma^{-1} \hat{\bm{\mu}}_d- \frac{n^2r}{n_0n_1}}}{\sqrt n}\, \; {\varTheta_{\alpha}}+  O_\prec \Big(n^{-1} \big(r^{\frac 12} + \varDelta_d^{\frac 12}\,  \big)\Big) \,, 
\end{align}
and the random part ${\varTheta_{\alpha}}$ satisfies
\begin{align*}
{\varTheta_{\alpha}}\simeq \mathcal{N}(0, \widehat {V})\,,
\end{align*}
where  $\widehat {V}$ was defined in (\ref{def:V123}).
%where $\widehat {V}=\sum_{i=1}^3 \widehat {V}_i$ such that 
%\begin{align*}
%&\widehat{V}_1:= \mathsf{C}^2 \Phi_\alpha^2  \frac{2(1+y)}{(1-y)^7},, \notag\\
%&\widehat{V}_2:= \frac{1}{ \big((1-y)\Vert \hat{\bf u}_1\Vert^2 - y\Vert {\bf v}_1\Vert^2\big)} \Big(\mathsf{C}^2 \Phi_\alpha^2 \Vert{\bf v}_1\Vert^2 \, \frac{y(1+y)}{(1-y)^7}  + \frac{n}{n_0(1-y)^3}
%+2 \mathsf{C} \Phi_\alpha \Vert{\bf v}_1\Vert \, \sqrt{\frac{n}{n_0}}\, \frac{y}{(1-y)^5}  \Big)\notag\\
%&\widehat{V}_3:= \frac{{V}_3}{ \big((1-y)\Vert \hat{\bf u}_1\Vert^2 - y\Vert {\bf v}_1\Vert^2\big)} .
%\end{align*}
Furthermore, the convergence rate of ${\varTheta_{\alpha}}$ to $\mathcal{N}(0, \widehat {V})$ is $O_\prec(n^{-1/2})$ in Kolmogorov-Smirnov distance, i.e., $\sup_{t}\Big|\p\big(\Theta_\alpha\leq t\big)-\p\big( \mathcal{N}(0, \widehat {V})\leq t\big)\Big|\prec  n^{-1/2}\,, $
%\begin{align*}
%\sup_{t}\Big|\p\big(\Theta_\alpha\leq t\big)-\p\big( \mathcal{N}(0, \widehat {V})\leq t\big)\Big|\prec  n^{-1/2}\,, 
%\end{align*} 
where we simply use $ \mathcal{N}(0, \widehat {V})$ to denote a random variable with distribution $ \mathcal{N}(0, \widehat {V})$. 

\end{prop}

We state the sketch of the proof of Proposition \ref{prop.asym_dist} as follows. First, we express $\widehat{F}(\widehat\Sigma, \hat{\bm{\mu}}^0)- F(\Sigma, {\bm{\mu}}^0)$ in terms of Green functions $\mathcal{G}_1(z)= (XX^\top -z)^{-1}$ and $(z\mathcal{G}_2(z))= z(X^\top X-z)^{-1}$ at $z=0$ (Lemma \ref{lem:green_diff} in Appendix \ref{sec:pf_LemProp}).  
%The latter is well-defined since $z=0$ is a removable singular point of $(z\mathcal{G}_2(z))$.
 Different from the derivations of the expansions of the quadratic forms in Lemma \ref{lem:ests}, here we need to do second order expansions for  $\widehat{\Sigma}^{-1}$ and quadratic forms of $\mathcal{G}_1^a(0)$, $\mathcal{G}_1^a(0)X$ and $X^\top \mathcal{G}_1^a(0) X$, for $a=1,2$. Because the leading terms of $\widehat{F}(\widehat\Sigma$, $\hat{\bm{\mu}}^0)$ and $F(\Sigma, {\bm{\mu}}^0)$ cancel out with each other due to  their definitions and Lemma \ref{lem:ests}, higher order terms are needed to study the asymptotic distribution. The error terms in the expansions can be estimated with the help of local laws (\ref{est_m12N}), (\ref{est.DG}) and (\ref{081501}).  It turns out that the leading terms of $\widehat{F}(\widehat\Sigma, \hat{\bm{\mu}}^0)- F(\Sigma, {\bm{\mu}}^0)$ in Lemma \ref{lem:green_diff} are given by linear combinations of certain quadratic forms of  $\mathcal{G}_1^{(\ell)} - m_1^{(\ell)}$ , $(z\mathcal{G}_2)^{(\ell)} -(zm_2)^{(\ell)}$ and $\mathcal{G}_1^{(\ell)}X$ where we omit the argument $z$ in $\mathcal{G}_1$, $\mathcal{G}_2$ at $z=0$. This inspires us to study the joint asymptotic distribution of these quadratic forms. To derive a multivariate Gaussian distribution, it is equivalent to show  the asymptotically Gaussian distribution for a generic linear combination $\mathcal{P}$ of the quadratic forms appeared in the Green function representation formula; see equation (\ref{def:mP_ori}) in  Appendix \ref{sec:pf_LemProp} for the specific expression of $\mathcal{P}$. Next, we aim to derive a differential equation of the characteristic function of $\mathcal{P}$, denoted by $\phi_n(\cdot)$. Concretely, we show that for $|t|\ll n^{\frac 12}$, $\varphi_n'(t) = - V t \varphi_n(t)  + O_\prec((|t|+1) n^{-\frac 12})\,$,
% \begin{align*}
%\varphi_n'(t) = - V t \varphi_n(t)  + O_\prec((|t|+1) n^{-\frac 12})\,.
%\end{align*}
where $V$ is some deterministic constant that indicates the variance of $\mathcal{P}$. The above estimate has two implications. First, it indicates the Gaussianity of $\mathcal{P}$. Second, applying Esseen's inequality, we can obtain its convergence rate as well. The proof of the above estimate relies on the technique of integration by parts and local laws. More details can be found in the proof of  Proposition  \ref{prop:rmeP} in Appendix \ref{sec:pf_LemProp} .

\begin{myRem} \label{rmk:simplified SF-F}
In the case that $p$ is fixed, or  $r\equiv r_n=O_\prec(1/n)$, we have the simplified version of Proposition \ref{prop.asym_dist} where $\widetilde{F}(\widehat\Sigma, \hat{\bm{\mu}}^0)$ defined in \eqref{def:wtF} is involved: 
\begin{align}\label{repre:sF-F_s}
\quad \widetilde{F}(\widehat\Sigma, \hat{\bm{\mu}}^0)- F(\Sigma, {\bm{\mu}}^0)
&=\frac {1}{\sqrt n}\, \sqrt{ \hat{\bm{\mu}}_d^{\top}  \widehat\Sigma^{-1} \hat{\bm{\mu}}_d} \, \widetilde {\varTheta}_{\alpha} + O_\prec \big(n^{-1} \varDelta_d^{\frac 12} \big)\,, 
\end{align}
and the random part $\widetilde{\varTheta}_{\alpha}$ satisfies $\widetilde{\varTheta}_{\alpha}\simeq \mathcal{N}(0, \widetilde  {V})$
%\begin{align*}
%\widetilde{\varTheta}_{\alpha}\simeq \mathcal{N}(0, \widetilde  {V})
%\end{align*}
with rate $O_\prec(n^{-1/2})$. We also remark that the proof of this simplified version  is similar to that of Proposition \ref{prop.asym_dist} by absorbing some terms containing $r$ into the error thanks to $r=O(1/n)$. Hence, we will omit the proof. 
%but discuss on the convergence rate $O_\prec(n^{-1/2})$ which differs from $O_\prec(p^{-\frac12})$ in Proposition \ref{prop.asym_dist} after we exhibit the convergence rate $O_\prec(p^{-\frac12})$ of Proposition \ref{prop.asym_dist} in Section \ref{sec:pf_LemProp}.

\end{myRem}

With the help of Lemma \ref{lem:ests} and Proposition \ref{prop.asym_dist}, we are now ready to prove the main theorem (c.f. Theorem \ref{mainthm}).

\begin{myPro} [Proof of  Theorem \ref{mainthm}]
Recall that $\hat \phi_\alpha(x)= \1 \Big( \widehat{A} ^{\top} x > \widehat C_\alpha^p\Big) $. If we can  claim  that 
\begin{align}\label{claim:hC>0}
\widehat C_\alpha^p\geq F(\Sigma, {\bm{\mu}}^0)
\end{align}
 with high probability,  then immediately, we can conclude that with high probability,
\begin{align*}
R_0(\hat \phi_\alpha) = \p \Big( \widehat{A} ^{\top} {\bf x} > \widehat C_\alpha^p\Big| {\bf x}\sim \mathcal{N}({\bm{\mu}}^0, \Sigma)\Big)\leq   \p \Big( \widehat{A} ^{\top} {\bf x} >F(\Sigma, {\bm{\mu}}^0) \Big| {\bf x}\sim \mathcal{N}({\bm{\mu}}^0, \Sigma)\Big)= R_0(\phi^*_\alpha)= \alpha.
\end{align*}
In the sequel, we establish inequality (\ref{claim:hC>0}) with high probability.
 By the definition of $\widehat C_\alpha^p$ in (\ref{def:whC}) and the representation (\ref{repre:sF-F}), we have  
 \begin{align*}
 \widehat C_\alpha^p- F(\Sigma, {\bm{\mu}}^0)
&= \widehat{F}(\widehat\Sigma, \hat{\bm{\mu}}^0)- F(\Sigma, {\bm{\mu}}^0) +\sqrt{\big((1-r) \widehat A^{\top} \widehat \Sigma\widehat A - r\Vert {\bf v}_1\Vert^2\big) }\sqrt{\frac{\widehat{V}}{n}} \, \Phi^{-1}(1-\delta)\notag\\
&=\frac {1}{\sqrt n}\,\sqrt{ (1-r)  \widehat{A} ^{\top}\widehat\Sigma \widehat{A} - r\Vert {\bf v}_1\Vert^2} \, \Big( {\varTheta_{\alpha}}- \sqrt{\widehat {V}} \, \Phi^{-1}(\delta) \Big)+ O_\prec(n^{-1} \varDelta_d^{\frac 12})\,.
\end{align*}
By Proposition \ref{prop.asym_dist}, ${\varTheta_{\alpha}}$ is asymptotically $\mathcal{N}(0, \widehat{V})$ distributed with convergence rate $O_\prec(n^{-1/2})$.
We then have  for any  constant $\varepsilon\in (0,\frac 12)$,
\begin{align*}
    \p \Big({\varTheta_{\alpha}} - \sqrt{\widehat {V}} \, \Phi^{-1}(\delta)>  n^{-\frac 12+ \varepsilon} \Big)&= \p \Big({\varTheta_{\alpha}} /\sqrt{\widehat {V}}>  \Phi^{-1}(\delta)+  n^{-\frac 12+ \varepsilon}/\sqrt{\widehat{V}} \, \Big) \notag\\
    &\geq  \p\Big(\mathcal{N}(0,1)>  \Phi^{-1}(\delta)+  n^{-\frac 12+ \varepsilon} /\sqrt{\widehat{V}} \, \Big) - n^{-\frac 12 + \varepsilon}\notag\\
    &= 1-
    \Phi\big(\Phi^{-1}(\delta)+  n^{-\frac 12+ \varepsilon}/\sqrt{\widehat{V}} \,\big) - n^{-\frac 12 + \varepsilon}\notag\\
    &\geq 1- \delta- C_1n^{-\frac 12 + \varepsilon}
\end{align*}
for some $C_1>0$ and $n>n(\varepsilon)$. Here  the second step is due to the convergence rate $O_\prec(n^{-1/2})$ of $\Theta_\alpha$;  And for the last step, we used the continuity of $\Phi(\cdot)$ together  with $\widehat{V}>c$ for some constant $c>0$ following from the definition (\ref{def:V123}). Further we have the estimate $\sqrt{ (1-r)  \widehat{A} ^{\top}\widehat\Sigma \widehat{A} - r\Vert {\bf v}_1\Vert^2}\asymp  \varDelta_d^{1/2}$ with probability at least $1-n^{-D}$ for any $D>0$ and $n>n(\varepsilon, D)$, which is  obtained from (\ref{est:hAhShA}). Thereby, we get that 
\begin{align*}
    \frac {1}{\sqrt {n}}\,\sqrt{ (1-r)  \widehat{A} ^{\top}\widehat\Sigma \widehat{A} - r\Vert {\bf v}_1\Vert^2 } \, \Big( {\varTheta_{\alpha}}- \sqrt{\widehat {V}} \, \Phi^{-1}(\delta) \Big)\geq c n^{-1+\varepsilon}\varDelta_d^{\frac 12}
\end{align*}
for some $c>0$,
with probability at least $1- \delta - C_1 n^{-\frac 12+ \varepsilon} -  n^{-D}$  when $n>n(\varepsilon, D)$.  As a consequence, there exist some $C_1, C_2>0$ such that
\begin{align*}
 \widehat C_\alpha^p- F(\Sigma, {\bm{\mu}}^0)>c n^{-1+\varepsilon}\big( r^{\frac12}+ \sqrt{ {\bm{\mu}}_d^{\top}  \Sigma^{-1}{\bm{\mu}}_d} \, \big)+   O_\prec \Big(n^{-1} \big(r^{\frac 12} + \sqrt{ {\bm{\mu}}_d^{\top}  \Sigma^{-1}{\bm{\mu}}_d}\,\big)\big(1+ \sqrt{\frac{n}{n_0}} \,\,  r^{\frac 12} \big)\Big) >0 
\end{align*}
with probability at least $1- \delta - C_1 n^{-\frac 12+ \varepsilon}  - C_2n^{-D}$ for  any $ \varepsilon \in (0, 1/2)$ and $D>0$, when $n>n(\varepsilon, D)$. 
%Here we used one derivation from (\ref{est:hAhShA}) that 
%$$ \big((1-r)  \widehat{A} ^{\top}\widehat\Sigma \widehat{A} - r\Vert {\bf v}_1\Vert^2={\bm{\mu}}_d^{\top} \Sigma^{-1}{\bm{\mu}}_d   + O_\prec(n^{-\frac 12}\Vert {\bm{\mu}}_d\Vert^2)= O(\Vert {\bm{\mu}}_d \Vert^2)\,,$$
%where the last bound holds with probability at least $1-n^{-D}$.

{\color{black}
In the sequel, we proceed to prove statement (ii) regarding the type II error. Note that by definition,
\begin{align}\label{steps:mainthm1}
R_1(\hat{\phi}_\alpha) & = \p (\hat{\phi}_\alpha({\bf x})\neq Y\big|Y=1)
= \p \Big( \widehat A^{\top} {\bf x}< \widehat C_\alpha^p\Big| {\bf x}\sim \mathcal{N}({\bm{\mu}}^1, \Sigma)\Big) \notag\\
&= \Phi\Big(( \widehat A^{\top}\Sigma \widehat A )^{-\frac 12}\big (\widehat C_\alpha^p- \widehat A^{\top}{\bm{\mu}}^1\big)\Big)
= \Phi\Big( \Phi^{-1}(1-\alpha) -\frac{\widehat A^{\top}{\bm{\mu}}_d }{\sqrt{\widehat A^{\top}\Sigma \widehat A}} + O_\prec(n^{-\frac12}) \Big).
\end{align}
Using the estimates in Lemma \ref{lem:ests}, if $p/n\to 0$, we further have 
\begin{align*}
R_1(\hat{\phi}_\alpha) &=  \Phi\Big(\Phi^{-1}(1-\alpha) -\varDelta_d^{\frac 12}  + O_\prec\big(n^{-\frac12}\varDelta_d^{\frac 12} \big) + O\big(r\varDelta_d^{\frac 12} \big)\, \Big)\,.
\end{align*}
Then, compared with $R_1({\phi}_\alpha^*)= \Phi\Big(\Phi^{-1}(1-\alpha) - \varDelta_d^{\frac 12} \, \Big)$,
%\begin{align} \label{steps:mainthm2}
%R_1({\phi}_\alpha^*)= \Phi\Big(\Phi^{-1}(1-\alpha) - \varDelta_d^{\frac 12} \, \Big)\,,
%\end{align}
it is not hard to deduce that in the case of $p/n\to 0$, (\ref{IIerror:p/n-0}) holds.
%\begin{align*}
%R_1(\hat{\phi}_\alpha) - R_1({\phi}_\alpha^*) \leq C\Big(r + n^{-\frac12+ \varepsilon}\Big) \varDelta_d^{\frac 12}\exp\Big(- \frac{c\varDelta_d}{2}\Big)
%\end{align*}
%with probability at least $1- n^{-D}$ for any small $ 0<\varepsilon<1/2$ and any large $D>0$, when $n>n(\varepsilon, D)$.  Here $C$ is some constant which may depend on $c_0, c_1, c_2$ and $\alpha$.

In the case that $p/n\to r_0\in (0,1)$, continuing with (\ref{steps:mainthm1}),  we arrive at 
\begin{align*}
R_1(\hat{\phi}_\alpha) &= \Phi\Big( \Phi^{-1}(1-\alpha) -\frac{(1-r)^{-1}\varDelta_d }{\sqrt{\frac{r}{(1-r)^2}\Vert {\bf v}_1\Vert^2 + \frac{1}{(1-r)^3} \varDelta_d}} + O_\prec\big(n^{-\frac 12} \varDelta_d^{\frac 12} \big)\Big).
\end{align*}
However, in this case, 
 $$\frac{\sqrt{1-r}}{\sigma} \, \varDelta_d^{\frac 12} < \frac{(1-r)^{-1}\varDelta_d }{\sqrt{\frac{r}{(1-r)^2}\Vert {\bf v}_1\Vert^2 + \frac{1}{(1-r)^3} \varDelta_d}}  <\sqrt{1-r}  \, \varDelta_d^{\frac 12}\, , $$
  for some $\sigma>1$ which depends on $r(1-r)\Vert {\bf v}_1\Vert^2/ \varDelta_d$. Thereby, by some elementary computations, one shall obtain that with probability at least  $1-n^{-D}$ for $D>0$ and $\varepsilon\in (0,1/2)$, when $n>n(\varepsilon, D)$, 
 \begin{align*}
&R_1(\hat{\phi}_\alpha) - R_1({\phi}_\alpha^*)\geq \frac{1}{\sqrt {2\pi }} \exp\Big( -\frac 12 \big(\Phi_\alpha - \delta_1 \sqrt{\varDelta_d}\; \big)^2\Big)  (1-\sqrt{1-r} - n^{-\frac{1}{2}+ \varepsilon}\, )\sqrt{\varDelta_d} \,, \notag\\
&R_1(\hat{\phi}_\alpha) - R_1({\phi}_\alpha^*) \leq \frac{1}{\sqrt {2\pi }} \exp\Big( -\frac 12 \big(\Phi_\alpha - \delta_2 \sqrt{\varDelta_d}\, \big)^2\Big)  \Big(1-\frac{\sqrt{1-r}}{\sigma}+ n^{-\frac{1}{2}+ \varepsilon}\, \Big)\sqrt{\varDelta_d} \,,
 \end{align*}
for some $\delta_1\in (\sqrt{1-r}, 1)$ and $\delta_2\in (\sqrt{1-r}/\sigma, 1)$.

Combining the loss of probability for both statements together, eventually we see that (i) and (ii) hold with probability at least $1- \delta - C_1 n^{-\frac 12+ \varepsilon}  - C_2 n^{-D}$ and hence we finished the proof of Theorem \ref{mainthm}.

}
\end{myPro}

\section{Numerical Analysis}\label{sec: numerical}
\subsection{Simulation Studies}
\begin{figure}[h]
\caption{Examples 1a,  type I and type II errors for competing methods with increasing balanced sample sizes.  \label{fig::ex 1a}}

 \begin{subfigure}[t]{0.5\textwidth}
        \centering
        \includegraphics[scale=0.24]{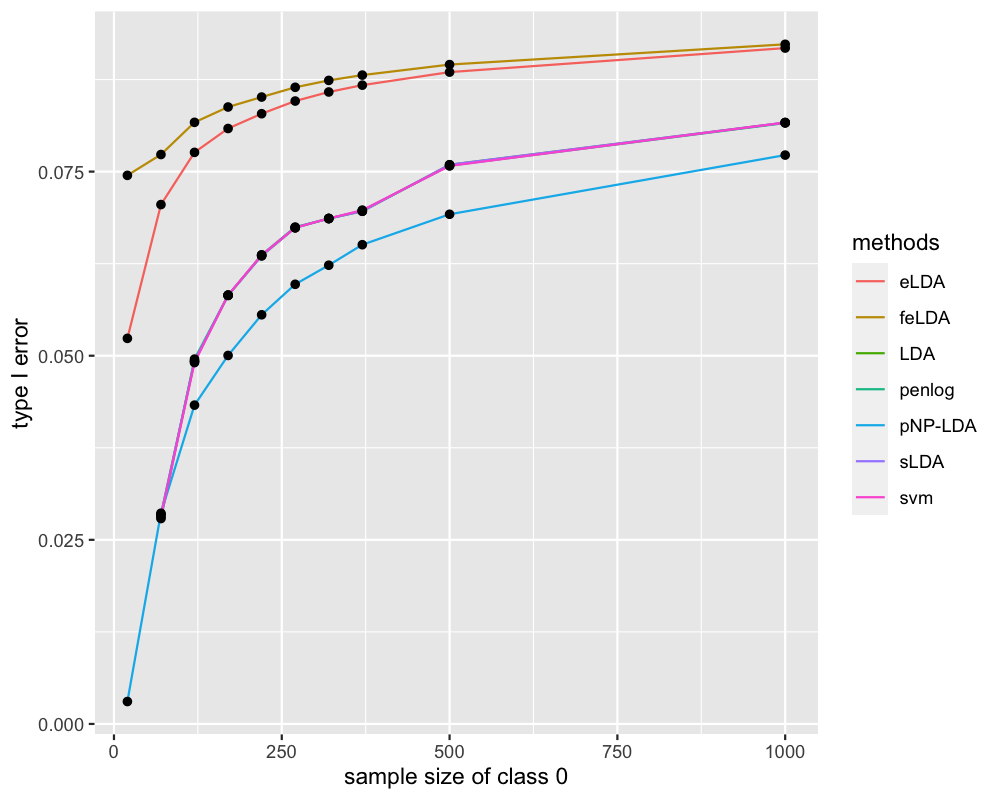}
        \caption{Example 1a, type I error}
    \end{subfigure}%
    \hspace{+0.1cm}
        \begin{subfigure}[t]{0.5\textwidth}
        \centering
        \includegraphics[scale=0.24]{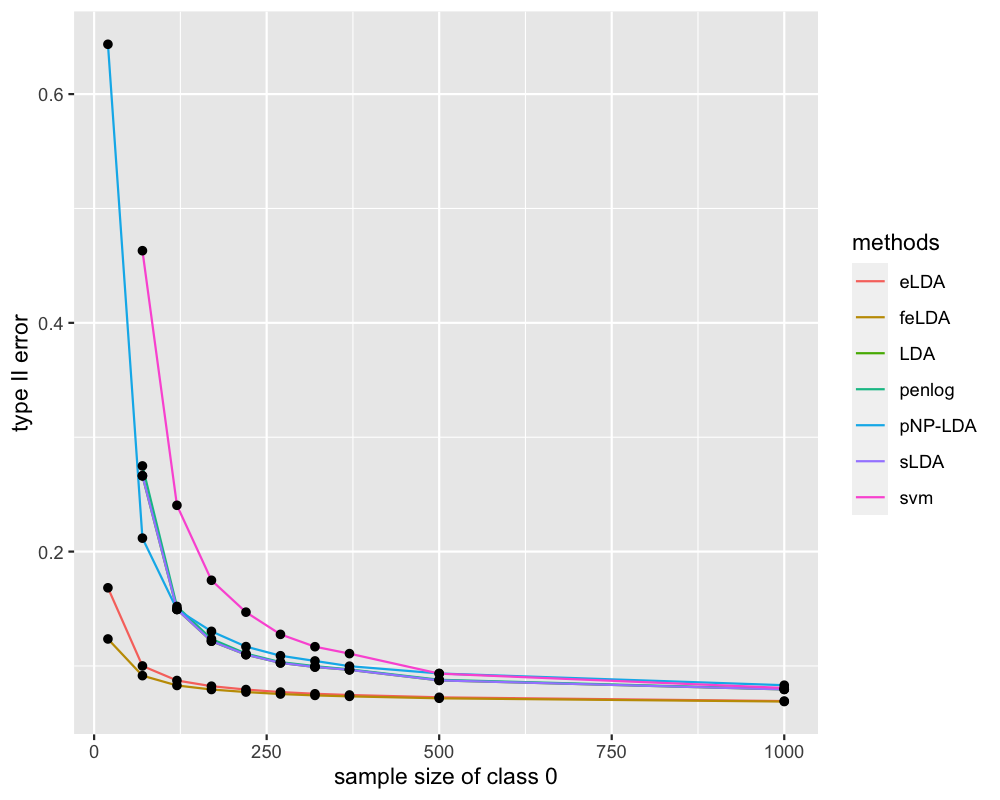}
        \caption{Example 1a, type II error}
    \end{subfigure}%    \begin{subfigure}[t]{0.5\textwidth}
%    
%    \begin{subfigure}[t]{0.5\textwidth}
%        \centering
%        \includegraphics[scale=0.24]{ex1b_type_I_error.png}
%        \caption{Example 1b, type I error}
%    \end{subfigure}%    \begin{subfigure}[t]{0.5\textwidth}
%    \begin{subfigure}[t]{0.5\textwidth}
%        \centering
%        \includegraphics[scale=0.24]{ex1b_type_II_error.png}
%        \caption{Example 1b, type II error}
%    \end{subfigure}%

\end{figure}

\begin{figure}[htbp]
\caption{Examples 1c, type I and type II errors for competing methods with increasing dimension $p$, $\delta=0.1$.  \label{fig::ex 1c}}

 \begin{subfigure}[t]{0.5\textwidth}
        \centering
        \includegraphics[scale=0.24]{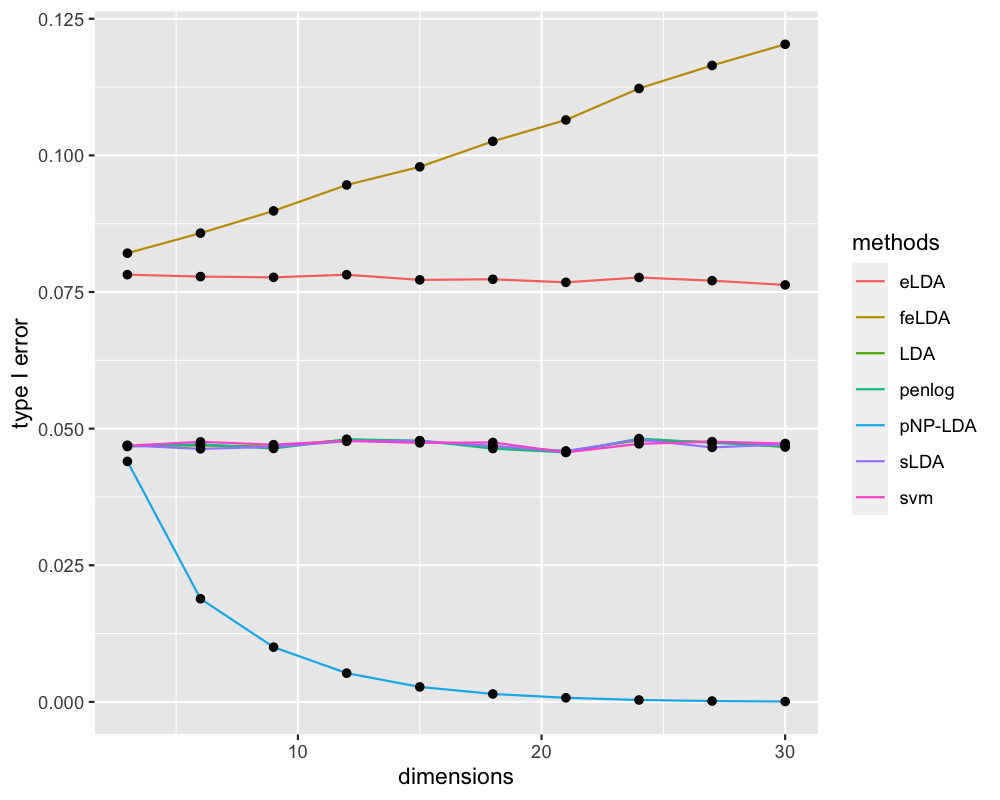}
        \caption{Example 1c, type I error}
    \end{subfigure}%
    \hspace{+0.1cm}
    \begin{subfigure}[t]{0.5\textwidth}
        \centering
        \includegraphics[scale=0.24]{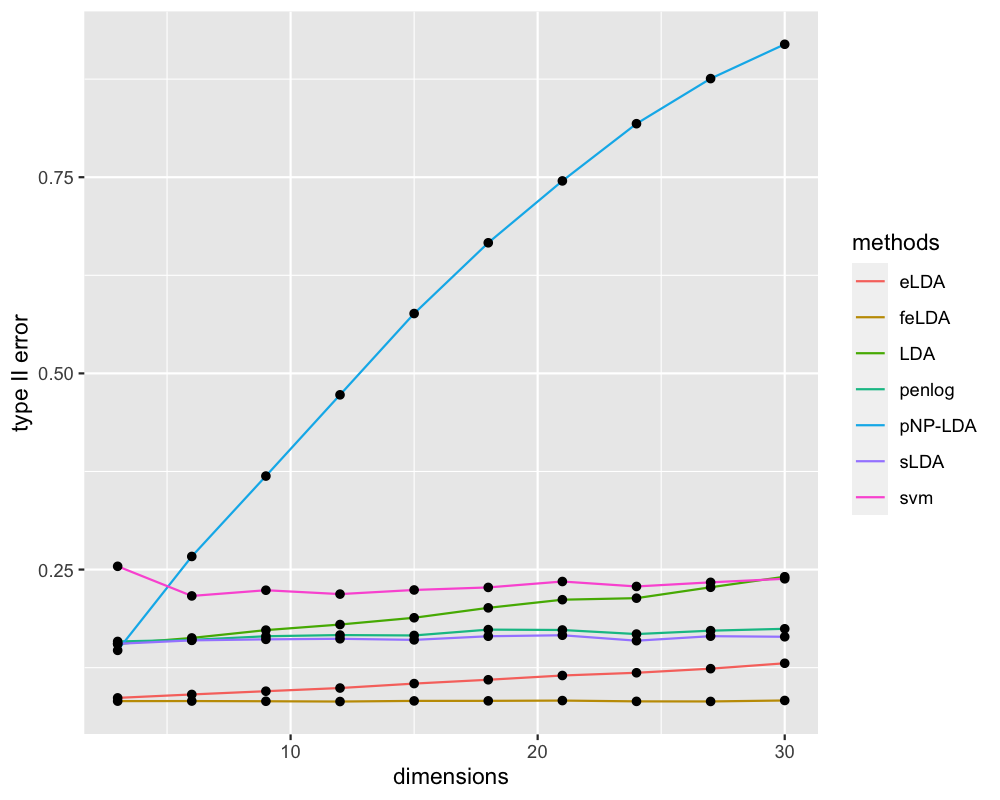}
        \caption{Example 1c, type II error}
    \end{subfigure}%    \begin{subfigure}[t]{0.5\textwidth}
    
%    \begin{subfigure}[t]{0.5\textwidth}
%        \centering
%        \includegraphics[scale=0.24]{ex1c_extra_type_I_error.png}
%        \caption{Example 1c', type I error}
%    \end{subfigure}%    \begin{subfigure}[t]{0.5\textwidth}
%    \begin{subfigure}[t]{0.5\textwidth}
%        \centering
%        \includegraphics[scale=0.24]{ex1c_extra_type_II_error.png}
%        \caption{Example 1c', type II error}
%    \end{subfigure}%
%    
%\begin{subfigure}[t]{0.5\textwidth}
%        \centering
%        \includegraphics[scale=0.24]{ex1c_star_type_I_error.png}
%        \caption{Example 1c* type I error}
%    \end{subfigure}%    \begin{subfigure}[t]{0.5\textwidth}
%    \begin{subfigure}[t]{0.5\textwidth}
%        \centering
%        \includegraphics[scale=0.24]{ex1c_star_type_II_error.png}
%        \caption{Example 1c*, type II error}
%    \end{subfigure}%
\end{figure}

In this section, we compare the performance of the two newly proposed classifiers \verb+eLDA+ and \verb+feLDA+ with that of five existing splitting NP methods: \verb+pNP-LDA+, \verb+NP-LDA+, \verb+NP-sLDA+, \verb+NP-svm+, and \verb+NP-penlog+. Here \verb+pNP-LDA+ is the parametric NP classifier as discussed in Section \ref{sec:intro}, where the threshold is constructed parametrically and the base algorithm is linear discriminant analysis (LDA). The latter four methods with \verb+NP+ as the prefix use the NP umbrella algorithm to select the threshold, and the base algorithms for scoring functions are LDA, sparse linear discriminant analysis (sLDA), svm and penalized logistic regression (penlog), respectively. In figures,  we omit the \verb+NP+ for these four methods for concise presentation.  Among the five existing methods, only \verb+pNP-LDA+ does not have sample size requirement on $n_0$. Thus for small $n_0$,  we can only compare our new methods with \verb+pNP-LDA+. For all five splitting NP classifiers, $\tau$, the class $0$ split proportion, is fixed at $0.5$, and the each experiment is repeated $1{,}000$ times. 
\begin{example}
The data are generated from an LDA model with common covariance matrix $\Sigma$, where $\Sigma$ is set to be an AR(1) covariance matrix with $\Sigma_{ij}=0.5^{|i-j|}$ for all $i$ and $j$. $\bm{\beta}^{\text{Bayes}}=\Sigma^{-1} \bm{\mu}_d= 1.2\times(\bm{1}_{p_0}, \bm{0}_{p - p_0})^{\top}$, $\bm{\mu}^0=\bm{0}_p$, $p_0 = 3$. We set $\pi_0 = \pi_1 = 0.5$ and $\alpha=0.1$.  Type I and type II errors are evaluated on a test set that contains $30{,}000$ observations from each class, and then we report the average over the $1{,}000$ repetitions.
\begin{enumerate}
\item[(1a)]    $\delta = 0.1$, $p = 3$, varying $n_0 = n_1 \in \{20, 70, 120, 170, 220, 270, 320, 370,500,1000\}$
  \item[(1b)] $\delta = 0.1$, $p = 3$, $n_1 = 500$,  varying $n_0 \in \{20, 70, 120, 170, 220, 270, 320, 370,500,1000\}$
  \item[(1c)] $\delta = 0.1$,  $n_0 = n_1 = 125$, varying $p \in \{3,6,9,12,15,18,21,24,27,30\}$
  \item[(1c')] $\delta = 0.05$, $n_0 = n_1 = 125$, varying $p \in \{3,6,9,12,15,18,21,24,27,30\}$
   \item[(1c*)] $\delta = 0.01$, $n_0 = n_1 = 125$, varying $p \in \{3,6,9,12,15,18,21,24,27,30\}$
  \item[(1d)]  $\delta = 0.1$, $n_0 = 125$, $n_1 = 500$, varying $p \in \{3,6,9,12,15,18,21,24,27,30\}$
  \item[(1d')] $\delta = 0.05$, $n_0 = 125$, $n_1 = 500$, varying $p \in \{3,6,9,12,15,18,21,24,27,30\}$
\end{enumerate}
\end{example}
We summarize the results for Example 1 in Figure \ref{fig::ex 1a},  Figure \ref{fig::ex 1c},  Appendix Figures \ref{fig::ex 1a 1b},  \ref{fig::ex 1c 1c'}, and \ref{fig::ex 1d 1d'},  Appendix Tables \ref{tb::simu1ab} and \ref{tb::simu1cc'}.  We discuss our findings in order.

Examples 1a and 1b share the common violation rate target $\delta=0.1$ and low dimension $p=3$. Their distinction comes from the two class sample sizes; Example 1a has balanced increasing sample sizes, i.e., $n_0=n_1$,  while Example 1b keeps $n_1$ fixed at 500, and only increases $n_0$.  Due to space limitations,  we only demonstrate the performance of Example 1a in Figure \ref{fig::ex 1a},  in terms of type I and type II errors.  We leave the comparison between Example 1a and Example 1b to Appendix Figure \ref{fig::ex 1a 1b}.  Notice that,  for very small class 0 sample sizes $n_0 = 20$,  all NP umbrella algorithm based methods (\verb+NP-LDA+, \verb+NP-sLDA+, \verb+NP-svm+, and \verb+NP-penlog+) fail their minimum class 0 sample size requirement and are not implementable, thus only the performances of \verb+eLDA+, \verb+feLDA+ and \verb+pNP-LDA+ are available in Figure \ref{fig::ex 1a}.  Consistently across Example 1a and Example 1b, we see that 1) as $n_0$ increases,  for all methods, the type I errors increase (but bounded above by $\alpha$), and the type II errors decrease.  Nevertheless,  the five existing NP methods present type I errors mostly below 0.08,  and are much more conservative compared to \verb+eLDA+ and \verb+feLDA+, whose type I errors closer to 0.1;  2) in terms of type II errors, \verb+eLDA+ and \verb+feLDA+ significantly outperform the other five methods across all $n_0$'s.   Comparing Example 1b to Example 1a,  keeping $n_1=500$ does not affect much the performance of \verb+eLDA+ and \verb+feLDA+.  However, Example 1b has aggravated the type I error performance of \verb+pNP-LDA+ for small $n_0$,  and also the type II error performance of \verb+NP-svm+.   

We further summarize the observed (type I error) violation rate\footnote{Strictly speaking, the observed violation rate on type I error is only an approximation to the real violation rate.  The approximation is two-fold: 1). in each repetition of an experiment, the population type I error is approximated by the empirical type I error on a large test set; 2). the violation rate should be calculated based on infinite repetitions of the experiment, but we only calculate it based on a finite number of repetitions. However, such approximation is unavoidable in numerical studies.} in Appendix Table \ref{tb::simu1ab}.  The five splitting NP classifiers all have violation rates smaller than targeted $\delta=0.1$,  and share a common increasing trend as $n_0$ increases.  In particular,  \verb+pNP-LDA+ is the most conservative one with the largest violation rate being 0.007 in Example 1a and 0.028 in Example 1b.  In contrast,  \verb+eLDA+ exhibits a much more accurate targeting at the violation rates, with all the observed violation rates around  $\delta=0.1$.  Theorem \ref{mainthm}  indicates that the type I error upper bound of \verb+eLDA+ might be violated with probability at most $\delta+C_1 n^{-1/2+\varepsilon}+ C_2 n^{-D}$.  As the sample size increases, this quantity gets closer to $\delta$.  The control of violation rates for \verb+feLDA+ is not desirable for small $n_0$. However, we observe a decreasing pattern as $n_0$ increases,  which agrees with Corollary \ref{thm:prop}.  When $n_0=1000$, for Example 1a, the violation rate of \verb+feLDA+ reaches the targeted level $\delta=0.1$. 

The common setting shared by Examples 1c,  1c' and 1c* includes balanced and fixed sample sizes, and increasing dimension $p$.  Similarly,  in the main text, we only present performance of Example 1c in Figure \ref{fig::ex 1c} and leave the comparison across Examples 1c,  1c' and 1c* to Appendix Figure \ref{fig::ex 1c 1c'}.  First,  we observe from Figure \ref{fig::ex 1c} that both \verb+eLDA+ and \verb+feLDA+ dominate existing methods in terms of type II errors.  Nevertheless,  Example 1c shows that when $p$ gets to 20 and beyond,  type I error of \verb+feLDA+ is no longer bounded by $\alpha=0.1$.  Changing the violation rate $\delta$ from 0.1 to 0.05 and further to 0.01 hinders the growth of type I error of \verb+feLDA+ as $p$ increases, but does not solve the problem ultimately as illustrated in Figure \ref{fig::ex 1c 1c'} panel (c) and (e).  This is due to the construction of \verb+feLDA+ which is specifically designed for small $p$; when $p$ gets large,  \verb+eLDA+ outperforms \verb+feLDA+. Therefore, considering the performance across different $p$'s,  \verb+eLDA+ performs the best among the seven methods. Second,  as dimension $p$ increases, all of the type II errors slightly increase or remain stable as expected,  except for that of \verb+pNP-LDA+.  This is due to a technical bound in the construction of the threshold of \verb+pNP-LDA+, which becomes loose when $p$ is large.  %comparing to its non-parametric NP umbrella algorithm competitors.

Appendix Table \ref{tb::simu1cc'} presents the violation rates from Examples 1c,  1c', and 1c*. Similar to what we have observed earlier, the five existing NP classifiers are relatively conservative and the observed violation rates of \verb+eLDA+ are mostly around the targeted $\delta$ in all the three sub-examples,  while that of \verb+feLDA+ goes beyond the targeted $\delta$ as $p$ increases.  When we decrease $\delta$ from 0.1 to 0.05 and further to 0.01,  we have the following two observations: 1) the violation rates of the four NP umbrella algorithm based classifiers \verb+NP-LDA+, \verb+NP-sLDA+, \verb+NP-penlog+ and \verb+NP-svm+ stay the same in Examples 1c and 1c'.  The violation rates decrease as we move to Example 1c*.  This is due to the discrete combinatorial construction of the thresholds in umbrella algorithms and thus the observed violation rates present discrete changes in terms of $\delta$.  In other words,  not necessarily small changes in $\delta$ will lead to a change in the constructed classifier and the observed violation rates.  For example,  for NP umbrella algorithm based methods, the number of left-out class 0 observations is 63, and the threshold is constructed as the $k^*$-th order statistics of the classification scores of the left-out class 0 sample, where $k^*=\min \{ k\in\{1,\cdots, 63\}: \nu(k)< \delta\}$, and $\nu (k)=\sum_{j=k}^{63}  {63 \choose j} (1-\alpha)^j \alpha^{63-j}$.  Plugging in $\alpha=0.1$,  we could easily calculate that $k^*=61$ for both $\delta=0.1$ and $\delta=0.05$, since $\nu (61)=\sum_{j=61}^{63}  {63 \choose j} (1-0.1)^j 0.1^{63-j}= 0.042$ and $\nu (60)= 0.113$. Furthermore, for $\delta = 0.01$,  the threshold changes as $k^*$ changes, since $0.042>0.01$; 2) \verb+pNP-LDA+, \verb+eLDA+, and \verb+feLDA+ have the parametric construction of the threshold and the observed violation rates of these methods respond to changes in $\delta$ more smoothly.  Nevertheless,  \verb+pNP-LDA+ is overly conservative,  with the observed violation rate almost all 0. %, so the resulted change is not obvious across Examples 1c, 1c' and 1c*.  

Examples 1d and 1d' also demonstrate the performances when dimension $p$ increases,  but with unequal class sizes.  %A similar message has been delivered that consistently \verb+eLDA+ and \verb+feLDA+ outperform the other competing methods. A smaller violation rate $\delta=0.05$ better controls the upper bound on the observed type I errors than with $\delta=0.1$. 
We omit the details in the main due to similar messages, and refer interested readers to Appendix Figure \ref{fig::ex 1d 1d'}. 

\begin{example}
The data are generated from an LDA model with common covariance matrix $\Sigma$, where $\Sigma$ is set to be an AR(1) covariance matrix with $\Sigma_{ij}=0.5^{|i-j|}$ for all $i$ and $j$. $\bm{\beta}^{\text{Bayes}}=\Sigma^{-1}\bm{\mu}_d = C_p \cdot \bm{1}_p^{\top}$, $\bm{\mu}^0=\bm{0}_p$. Here, $C_p$ is a constant depending on $p$, such that the NP oracle classifier always has type II error $0.236$ for any choice of $p$ when $\alpha = 0.1$. We set $\pi_0 = \pi_1 = 0.5$ and $\alpha=\delta=0.1$. 
\begin{enumerate}
  \item[(2a)] $n_0 = n_1 = 125$, varying $p \in \{3,6,9,12,15,18,21,24,27,30\}$
  \item[(2b)] $n_0 = 125, n_1 = 500$, varying $p \in \{3,6,9,12,15,18,21,24,27,30\}$
\end{enumerate}
\end{example}

% \begin{figure}[h]
% \caption{Examples 2a and 2b,  type I and type II error for competing methods with increasing balanced sample sizes. $p=3$ in Example 2a and $p=6$ in Example 2b.  \label{fig::ex 2a 2b}}

%  \begin{subfigure}[t]{0.5\textwidth}
%         \centering
%         \includegraphics[scale=0.23]{ex2a_type_I_error.png}
%         \caption{Example 2a, type I error}
%     \end{subfigure}%
%     \begin{subfigure}[t]{0.5\textwidth}
%         \centering
%         \includegraphics[scale=0.23]{ex2a_type_II_error.png}
%         \caption{Example 2a type II error}
%     \end{subfigure}%    \begin{subfigure}[t]{0.5\textwidth}
    
%     \begin{subfigure}[t]{0.5\textwidth}
%         \centering
%         \includegraphics[scale=0.23]{ex2b_type_I_error.png}
%         \caption{Example 2b, type I error}
%     \end{subfigure}%    \begin{subfigure}[t]{0.5\textwidth}
%     \begin{subfigure}[t]{0.5\textwidth}
%         \centering
%         \includegraphics[scale=0.23]{ex2b_type_II_error.png}
%         \caption{Example 2b, type II error}
%     \end{subfigure}%

% \end{figure}

Examples 2a and 2b are similar to Examples 1c and 1d, but their oracle projection direction $\bm{\beta}^{\text{Bayes}}$ is not sparse.  Appendix Figure \ref{fig::ex 2a 2b} summarizes the results on type I and type II errors. The delivered messages are similar to those of Examples 1c and 1d: 1) while \verb+eLDA+ enjoys controlled type I errors under $\alpha=0.1$ for all $p$ in both Examples 2a and 2b,  the type I errors of \verb+feLDA+ deteriorate above the target  for large $p$;  2) \verb+eLDA+ and  \verb+feLDA+ dominate all other competing methods in terms of type II errors.  Observed violation rates from Examples 2a and 2b present similar messages as in Examples 1c and 1d, so we omit the table for those results.

We have also conducted experiments under non-Gaussian settings.  In short, we observe that when sample size of class 0 is small,  \verb+eLDA+ and \verb+feLDA+ clearly outperform all their competitors.  As the sample size increases,  the performances of most umbrella algorithm based classifiers begin to catch up and eventually outperform  \verb+eLDA+ and \verb+feLDA+.  We believe this phenomenon is due to the fine calibration of the LDA model in the development of \verb+eLDA+ and \verb+feLDA+, which leads to conservative results in heavy-tail distribution settings.  Set-up of experiments and detail discussions are included in Appendix \ref{sec:heavy-tail}.

\subsection{Real Data Analysis}
We analyze two real datasets.  
The first one is a lung cancer dataset \citep{gordon2002translation,jin2016influential} that consists of gene expression measurements from 181 tissue samples.  Among them, 31 are malignant pleural mesothelioma (MPM) samples and 150 are adenocarinoma (ADCA) samples.   As MPM is known to be highly lethal pleural malignant and rare (in contrast to ADCA which is more common),  misclassifying MPM as ADCA would incur more severe consequences. Therefore,  we code MPM as class 0, and ADCA as class 1.  The feature dimension of this dataset is  $p =12{,}533$.  First, we set $\alpha=0.01$ and $\delta=0.05$.  Since the class 0 sample size is very small,  none of the umbrella algorithm based NP classifiers are implementable. Hence, we only compare the performance of  \verb+pNP-LDA+ with that of  \verb+eLDA+.  We choose to omit \verb+feLDA+ here because we have found from the simulation studies that \verb+feLDA+ outperforms \verb+eLDA+ only when the dimension is extremely small (e.g., $p\leq 3$).  On the other hand,  since \verb+eLDA+ is designed for $p<n$ settings and \verb+pNP-LDA+ usually works poorly for large $p$,  we first reduce the feature dimensionality to 40 by conducting two-sample t-test and selecting the 40 genes with smallest p-values.  To provide a more complete story,  we implemented further analysis with larger parameters ($\alpha = 0.1$ and $\delta = 0.4$) so that \verb+NP-sLDA+,  \verb+NP-penlog+,  \verb+NP-svm+ are also implementable. Those results are presented in Appendix Table \ref{tb::realdata1-2}.

The experiment is repeated 100 times and the type I and type II errors are the averages over these 100 replications.  In each replication, we randomly split the full dataset (class 0 and class 1 separately) into a training set (composed of 70\% of the data), and a test set (composed of 30\% of the data).  We train the classifiers on the training set,  with the feature selection step added before implementing \verb+eLDA+ and \verb+pNP-LDA+. Then we apply the classifiers to the test set to compute the empirical type I and type II errors.  Table \ref{tb::reaLDAta1} presents results from the parameter set $\alpha=0.01$ and $\delta=0.05$.  We observe that while both \verb+eLDA+ and \verb+pNP-LDA+ achieve type I errors smaller than the targeted $\alpha=0.01$,  \verb+pNP-LDA+ is overly conservative and has a type II error of 1.  In contrast,  \verb+eLDA+ provides a more reasonable type II error of 0.104, and the observed violation rate is 0.03 ($<0.05$). 
\begin{table}[t]
\caption{Lung cancer dataset  \label{tb::reaLDAta1}}
\centering
\renewcommand{\arraystretch}{0.6}
\begin{tabular}{r|rrr}
\hline
&&\texttt{pNP-LDA}&\texttt{eLDA}\\
\hline
\multirow{2}{2cm}{$\alpha=0.01$ \\ $\delta=0.05$}
&\texttt{type I error}&.000&.003\\
&\texttt{type II error}&1&.104\\
&\texttt{observed violation rate}&0&.03\\
\hline
\end{tabular}
\end{table}

The second dataset was originally studied in \cite{su2001molecular}. It contains microarray data from 11 different tumor cells,  including 27 serous papillary ovarian adenocarcinomas, 8 bladder/ureter carcinomas, 26 infiltrating ductal breast adenocarcinomas, 23 colorectal adenocarcinomas, 12 gastroesophageal adenocarcinomas, 11 clear cell carcinomas of the kidney,  7 hepatocellular carcinomas, 26 prostate adenocarcinomas, 6 pancreatic adenocarcinomas, 14 lung adenocarcinomas carcinomas, and 14 lung squamous carcinomas.  In more recent studies \citep{ jin2016influential,yousefi2010reporting},  the 11 different tumor cell types were aggregated into two classes, where class 0 contains bladder/ureter, breast,  colorectal and prostate tumor cells,  and class 1 contains the remaining groups.  We follow \cite{yousefi2010reporting} in determining the binary class labels, and we work on the modified dataset with $n_0=83$, $n_1=91$ and $p=12{,}533$.  

We repeat the data processing procedure as in the lung dataset, and report results from the parameter set $\alpha=0.01$ and $\delta = 0.05$ in Table \ref{tb::reaLDAta2}. While the sample size is too small for other umbrella algorithm based NP classifiers to work, the advantage of \verb+eLDA+ over \verb+pNP-LDA+ is obvious.  The observed violation rate 0.15 is larger than the targeted $\delta=0.05$.  However, we would like to emphasize that the observed violation rate in a real data study should not be interpreted as a close proxy to the true violation rate. First, the previous discussion on observed violation rate for simulation in the footnote also applies to the real data studies. Moreover, in simulations, samples are generated from population many times; however, in real data analysis, the one sample we have plays the role of population for repetitive sampling. Such substitute can be particularly inaccurate when the sample size is small.    

%which is different from the underlying population, and in each replication, we randomly split the full dataset into training and testing set; therefore, the observed violation rate can only be thought of as a rough approximation to the targeted $\delta$.

\begin{table}[t] 
\caption{Cancer dataset in \cite{su2001molecular}  \label{tb::reaLDAta2}}
\centering
\renewcommand{\arraystretch}{0.6}
\begin{tabular}{r|rrr}
\hline
&&\texttt{pNP-LDA}&\texttt{eLDA}\\
\hline
\multirow{2}{2cm}{$\alpha=0.01$ \\ $\delta=0.05$}
&\texttt{type I error}&.000&.008\\
&\texttt{type II error}&1&.437\\
&\texttt{observed violation rate}&0&.15\\
\hline
\end{tabular}
\end{table}

\section{Discussion} \label{sec:discussion}

%In this work, we propose \verb+eLDA+, the first non-splitting Neyman-Pearson classifier based on the LDA model, and its fixed dimensional variant $\verb+feLDA+$.  In practice, our non-splitting strategy is most beneficial when the sample sizes are limited, including situations where the more general nonparametric NP umbrella algorithm does not apply. A few directions, including the $p>n$ settings and more general parametric model settings, are worthy for future investigation. 

Our current work initiates  %{\color{red}[``only the beginning" kind of downplays our contribution. We may simply say that our current work initiates the investigations on ....]} of 
the investigations on non-splitting strategies under the NP paradigm.  For future works, we can work in settings where $p$ is larger than $n$ by selecting features via various marginal screening methods \citep{fan2010sure, li2012feature} and/or may add structural assumptions to the LDA model.  To accommodate diverse applications, one might also construct classifiers based on more complicated models, such as the quadratic discriminant analysis (QDA) model \citep{fan2015innovated, li2015sparse, yang2018quadratic,  pan2020efficient, wang2021phase, cai2021convex}.

%\vspace{-0.2in}

\appendix
\begin{appendices}

\section{Further remark on Assumption \ref{asm} } \label{rem:assump}
Previously, margin assumption and detection condition were assumed in \cite{tong2013plug} and subsequent works \cite{zhao2016neyman, Tong.Xia.Wang.Feng.2020} for an NP classifier to achieve a diminishing excess type II error. Concretely, write the level-$\alpha$ NP oracle as $\1(f_1({\bf x})/f_0({\bf x})> C^*_{\alpha})$, where $f_1$ and $f_0$ are class-conditional densities of the features,  then the margin assumption assumes that 
 \begin{align*}
  \p (|f_1({\bf x})/f_0({\bf x})-C^*_{\alpha}|\leq \delta | Y=0)\leq C_0  \delta^{\bar\gamma}\,,
 \end{align*}
for any $\delta>0$ and some positive constant $\bar\gamma$ and $C_0$. This is a low-noise condition around the oracle decision boundary that has roots in \cite{Polonik.1995, Mammen.Tsybakov.1999}. On the other hand, the detection condition, which was coined in \cite{tong2013plug} and refined in \cite{zhao2016neyman}, requires a lower bound:
\begin{align*}
     \p(C^*_{\alpha}\leq f_1({\bf x})/f_0({\bf x})\leq C^*_{\alpha}+ \delta | Y=0) \geq C_1 \delta^{\uderbar{\gamma}}\,,
\end{align*}
for small $\delta$ and some positive constant $\uderbar{\gamma}$. In fact, $\delta^{\uderbar{\gamma}}$ can be generalized to $u(\delta)$, where $u(\cdot)$ is any increasing function on $R^+$ that might be $(f_0, f_1)$-dependent and $\lim_{\delta\to 0^+}u(\delta)= 0$. The necessity of the detection condition under general models for achieving a diminishing excess type II error was also demonstrated in \cite{zhao2016neyman} by showing a counterexample  that has fixed $f_1$ and $f_0$, i.e., when $p$ does not grow with $n$. Note that although the feature dimension $p$ considered in \cite{zhao2016neyman, Tong.Xia.Wang.Feng.2020} can grow with $n$, both impose sparsity assumptions, and the ``effective" dimensionality $s$ has the property that $s/n\to 0$. Hence previously, there were no theoretical results regarding the excess type II error when the effective feature dimensionality and the sample size are comparable. 

Under Assumption \ref{asm}, the marginal assumption and detection condition hold automatically. To see this, recall the level-$\alpha$ NP oracle classifier defined in  (\ref{def:class_NP}), we can directly derive that for any $\delta>0$,
\begin{align*}
&\quad \p (C^*_{\alpha}\leq f_1({\bf x})/f_0({\bf x}) \leq C^*_{\alpha} + \delta| Y=0) \notag\\
    &= \p (F\leq (\Sigma^{-1} {\bm \mu}_d)^\top {\bf x} \leq F + \delta| Y=0) \notag\\
    &= \p \big(F - {\bm \mu}_d^\top \Sigma^{-1} {\bm \mu}^0 \leq (\Sigma^{-1} {\bm \mu}_d)^\top ({\bf x}-{\bm \mu}^0) \leq F - {\bm \mu}_d^\top \Sigma^{-1} {\bm \mu}^0+ \delta |Y=0\big)\notag\\
    &= \p\Big(\frac{F - {\bm \mu}_d^\top \Sigma^{-1} {\bm \mu}^0}{\sqrt{\varDelta_d}}\leq \mathcal{N}(0,1) \leq \frac{F - {\bm \mu}_d^\top \Sigma^{-1} {\bm \mu}^0 + \delta}{\sqrt{\varDelta_d}}\Big)\,,
\end{align*}
with the shorthand notation $F:=\sqrt{\varDelta_d} \, \Phi^{-1}(1-\alpha) + {\bm{\mu}}_d^{\top} \Sigma^{-1} {\bm{\mu}}^0$. The RHS above can be further simplified to get
\begin{align*}
    \p (F\leq (\Sigma^{-1} {\bm \mu}_d)^\top {\bf x} \leq F + \delta| Y=0)=\Phi\Big(\Phi^{-1}(1-\alpha) + {\delta}/{\sqrt{\varDelta_d} }\, \Big) - (1-\alpha). 
\end{align*}
Thereby, using mean value theorem, we simply bound the above probability from above and below as
\begin{align*}
    \p (F\leq (\Sigma^{-1} {\bm \mu}_d)^\top {\bf x} \leq F + \delta| Y=0)&\leq \frac{1}{\sqrt{2\pi}} \exp{(-\frac 12 \Phi_\alpha^2)} \frac{\delta}{\sqrt{\varDelta_d} }\,, \notag\\
    \p (F\leq (\Sigma^{-1} {\bm \mu}_d)^\top {\bf x} \leq F + \delta| Y=0)&\geq\frac{1}{\sqrt{2\pi}} \exp{\bigg(-\frac 12 \Big(\Phi_\alpha + \frac{\delta}{\sqrt{\varDelta_d} } \Big)^2 \bigg)} \frac{\delta}{\sqrt{\varDelta_d} }\,.
\end{align*}
where we recall $\Phi_\alpha= \Phi^{-1} (1-\alpha)$.
A similar upper bound can also be derived for $\p (F-\delta\leq (\Sigma^{-1} {\bm \mu}_d)^\top {\bf x} \leq F | Y=0)$. These coincide with the aforementioned marginal assumption and detection condition. %Hence, Theorem \ref{mainthm} indicates that under the LDA model,  when $p/n\to (0,1)$, addition to the margin assumption and detection condition, $\bm{\mu}_d$ also plays a role regarding the excess type II error. 

%For fixed dimensionality $p$, clearly these inequalities already imply the the margin assumption and detection condition.  For diverging $p$, when $\|\bm{\mu}_d\|$ is bounded, then the two conditions are still satisfied.

\section{Proofs of Lemma \ref{thm:1} and Corollary \ref{thm:prop} } \label{sec:appen_main-results}
We first show the proof of Lemma \ref{thm:1} below.
\begin{myPro}[Proof of Lemma \ref{thm:1}]
The statement (i) is easy to obtain by the definition of $\tilde \phi_\alpha^*(\cdot)$ in (\ref{def:inter_NP}) and the definition of the type I error. Specifically, 
\begin{align*}
R_0(\tilde \phi_\alpha^*) &= \p \Big(\widehat A^{\top} ({\bf x} -{\bm{\mu}}^0) > \sqrt{\widehat A^{\top} \Sigma \widehat A}\, \Phi^{-1}(1-\alpha) \Big| {\bf x}\sim \mathcal{N}({\bm{\mu}}^0, \Sigma) \Big)\notag\\
& = 1- \Phi(\Phi^{-1}(1-\alpha)) = \alpha\,.
\end{align*}
Next, we establish statement (ii). By definition, we have
\begin{align}
R_1(\tilde \phi_\alpha^*) &=\p \Big(\widehat A^{\top} ({\bf x} -{\bm{\mu}}^1) \leq \sqrt{\widehat A^{\top} \Sigma \widehat A}\, \Phi^{-1}(1-\alpha)  - \widehat A^{\top} {\bm{\mu}}_d\Big| {\bf x}\sim \mathcal{N}({\bm{\mu}}^1, \Sigma) \Big)\notag\\
&= \Phi\Big( \Phi^{-1}(1-\alpha)   - \frac{\widehat A^{\top} {\bm{\mu}}_d}{\sqrt{\widehat A^{\top} \Sigma \widehat A}}\, \Big)\,, \notag\\
R_1({\phi}_\alpha^*) &=\p \Big((\Sigma^{-1}{\bm{\mu}}_d)^{\top}{\bf x}< \sqrt{\varDelta_d} \, \Phi^{-1}(1-\alpha) + {\bm{\mu}}_d^{\top} \Sigma^{-1} {\bm{\mu}}^0 \Big|{\bf x}\sim \mathcal{N}({\bm{\mu}}^1, \Sigma)\Big) \notag\\
%&=  \Phi\bigg(( {\bm{\mu}}_d^*\Sigma^{-1}{\bm{\mu}}_d)^{-\frac 12} \Big[\sqrt{{\bm{\mu}}_d^* \Sigma^{-1}{\bm{\mu}}_d} \, \Phi^{-1}(1-\alpha) + {\bm{\mu}}_d^* \Sigma^{-1} {\bm{\mu}}^0- (\Sigma^{-1}{\bm{\mu}}_d)^*{\bm{\mu}}^1\Big]\bigg) \notag\\
&= \Phi\Big(\Phi^{-1}(1-\alpha) - \sqrt{\varDelta_d} \, \Big)\,. \label{21102901}
\end{align}
Lemma \ref{lem:ests} and some elementary calculations lead to the conclusion:  for any $\varepsilon \in (0,1/2)$ and $D>0$, when $n> n(\epsilon, D)$,  with probability at least $1- n^{-D}$ we have,
\begin{align*}
 \varDelta_d^{1/2}>\frac{\widehat A^{\top} {\bm{\mu}}_d}{\sqrt{\widehat A^{\top} \Sigma \widehat A}} =  \varDelta_d^{1/2}+ O\big( r \varDelta_d^{1/2} \big) + O\big(n^{-\frac 12 +\varepsilon} (1+  \varDelta_d^{1/2})\big)\,.
\end{align*}
Moreover, it is straightforward to check
\begin{align*}
 \exp\Big(-\frac12 {\Big(\Phi^{-1}(1-\alpha) -  \varDelta_d^{1/2}\, \Big)^2}\Big) \asymp\exp\Big(- \frac{c \varDelta_d}{2}\Big)\,.
\end{align*}
Thus, we conclude that there exists some fixed constant C which may depend on $c_0, c_1, c_2$ and $\alpha$ such that for any $\varepsilon \in (0, 1/2)$ and $D>0$, when $n\geq n(\varepsilon, D)$, with probability at least $1- n^{-D}$, we have
\begin{align*}
R_1(\tilde \phi_\alpha^*)- R_1({\phi}_\alpha^*) \leq  C \big(  r + n^{-\frac 12 +\varepsilon}  \big) \varDelta_d^{1/2} \exp\Big(- \frac{c\varDelta_d}{2}\Big)\, .
%\leq  C \big(r + n^{-\frac 12 +\varepsilon} \big){\color{blue}\Vert {\bm \mu}_d\Vert^{-1}}\,.
\end{align*}
This finished our proof. 
\end{myPro}

At the end of this section, we sketch the proof of Corollary \ref{thm:prop}.  %which is nearly the same as that of  Theorem \ref{mainthm}.

\begin{myPro} [Proof of Corollary \ref{thm:prop}]
By the definition of $\widetilde{F}(\widehat\Sigma, \hat{\bm{\mu}}^0)$ and $ \widetilde C_\alpha^p$ in  (\ref{def:wtF}), (\ref{def:wtC}), under the setting of $p= O(1)$, we observe that 
\begin{align*}
&\widetilde{F}(\widehat\Sigma, \hat{\bm{\mu}}^0) = \widehat{F}(\widehat\Sigma, \hat{\bm{\mu}}^0) + O_\prec \big(n^{-1} \varDelta_d^{\frac 12} \big)\,,\notag\\
&\widetilde C_\alpha^p=  \widehat C_\alpha^p + O_\prec \big(n^{-1} \varDelta_d^{\frac 12}\big)\,.
\end{align*} 
Then, similarly to the proof of Theorem \ref{mainthm}, with the aid of Remark \ref{rmk:ests} and Remark \ref{rmk:simplified SF-F}, we conclude the results in the same manner; hence we omit the details.

\end{myPro}

\section{Proofs for Section \ref{sec:Pre}} \label{sec:appen_pf_tech}

\subsection{Proof of Lemma \ref{prop_prec}}
%\begin{myPro}[Proof of Lemma \ref{prop_prec}]
	Part (i) is obvious from Definition \ref{def.sd}. For any fixed $\varrho>0$, we have
\begin{align*}
	|\E \mathsf{X}_1| &\leq \E |\mathsf{X}_1\1(| \mathsf{X}_1|\leq  N^{\varrho}\Phi)|+\E |\mathsf{X}_1\1(|\mathsf{X}_1|\geq  N^{\varrho}\Phi)|\nonumber\\
	&\leq N^{\varrho}\Phi+ N^C \p(|\mathsf{X}_1|\geq N^{\varrho}\Phi)=O(N^{\varrho}\Phi)
\end{align*}
	for for sufficiently large $N\geq N_0(\varrho)$. This proves part (ii). 
%\end{myPro}

\subsection{Proof of Proposition \ref{thm:locallaw}}
%\begin{myPro}[Proof of Proposition \ref{thm:locallaw}]
Define
\begin{align}\label{19071810}
\mathcal{D}\equiv\mathcal{D}(\tau):= \{ z\in \mathbb{C}^+: -\frac{\lambda_-}{2} <\Re{z} <\frac{\lambda_-}{2} , 0< \Im{z}\leq \tau^{-1}\}\,.
\end{align}
 All the estimates in Proposition  \ref{thm:locallaw} can be separately shown for the case of $p>n^{\epsilon}$ for some fixed small $\epsilon>0$ and the case  of $p< n^{\epsilon}$. We first show all the estimates hold for the case $l=0$ and then proceed to the case of $l\geq 1$.

\vspace{0.2cm}
\noindent$\bullet$ For the case of $l=0$.
\vspace{0.2cm}

 In the regime that $p\geq n^{\epsilon}$ for some fixed small  $\epsilon>0$, (\ref{est:locallaw1})  can be derived from the entrywise  local Marchenko-Pastur law for extended spectral domain in Theorem 4.1 of  \cite{bloemendal2014isotropic}. We emphasize that originally in \cite{bloemendal2014isotropic} the results are not provided for extended spectral domain
  one only need to adapt the arguments in Proposition 3.8 of \cite{bloemendal2016principal} to extend the results.  
  
  The estimates of (\ref{est_m12N})  can be obtained by the rigidity estimates of eigenvalues in \cite[Theorem 2.10]{bloemendal2014isotropic}.  We remark that we get the improved version in the second estimate of (\ref{2021051301}) due to the trivial bound $z= O(1)$, for $z\in \mathcal{D}^0$, while for $z\in \mathcal{D}$, we crudely bound $|z|$ by $r^{-\frac 12}$. For (\ref{est:locallaw2}), by noticing that ${X}^{\top} \mathcal{G}_1= \mathcal{G}_2 {X}$, one only needs to show the first estimate of (\ref{est:locallaw2}). Using singular value decomposition (SVD) of ${X}$, i,e., ${X}= U^{\top} (\Lambda^{\frac12}, 0) V $, {\color{black} where the diagonal matrix $\Lambda^{\frac12}$ collects the singular values of $X$ in a descending order,}  we arrive at 
 \begin{align*}
 \big( {X}^{\top}\mathcal{G}_1(z)\big)_{i'i} = V_{i'}^{\top}  \left(
 \begin{array}{c}
 \Lambda^{\frac12}(\Lambda-z)^{-1} \\
 {\bf 0}
 \end{array}
 \right)
 U_i, \qquad \Lambda:=\text{diag}(\lambda_1, \ldots, \lambda_p)
 \end{align*}
and $U_i$, $V_{i'}$ are independent and  uniformly distributed on $\mathbf{S}^{p-1}$ and $\mathbf{S}^{n-1}$, respectively, thanks  to  the fact that  ${X}$ is a GOE matrix. Here we abbreviate $\lambda_i(H)$ by $\lambda_i$. Then we can further write 
\begin{align}\label{2020120901}
\big( {X}^{\top}\mathcal{G}_1(z)\big)_{i'i} &\overset{\rm d}{=} \sum_{i=1}^p g_i \tilde{g}_i \frac{\sqrt \lambda_i}{\lambda_i-z} \frac{1}{\Vert {\bf g}\Vert\Vert {\bf \tilde{g}}\Vert} \notag\\
&=  \sum_{i=1}^p g_i \tilde{g}_i \frac{\sqrt \lambda_i}{\lambda_i-z} \Big(1- \frac{\Vert {\bf g} \Vert^2-1 }{2} + O_\prec(n^{-1})\Big) \Big(1-  \frac{\Vert {\bf \tilde{g}} \Vert^2-1 }{2} + O_\prec(p^{-1}) \Big),
\end{align}
where ${\bf g}:=(g_1, \cdots, g_p) \sim \mathcal{N}(0, \frac 1p I_p)$,  ${\bf \tilde{g}}:=(\tilde{g}_1, \cdots, \tilde{g}_n) \sim \mathcal{N}(0, \frac 1n I_n)$ and they are independent. The leading term on the RHS of (\ref{2020120901}) is $ \sum_{i=1}^p g_i \tilde{g}_i \frac{\sqrt \lambda_i}{\lambda_i-z}$. By the rigidity of eigenvalues, we easily get that $\sqrt \lambda_i /(\lambda_i-z)\asymp  r^{\frac 14}$ uniformly for $z\in \mathcal{D}$ with high probability. Further applying the randomness of $g_i$'s and $\tilde{g}_i$'s, it is easy to conclude the first estimate in (\ref{est:locallaw2}). The second estimate with the extension in (\ref{2021051301}) holds naturally from 
${X}^{\top} \mathcal{G}_1= \mathcal{G}_2 {X}$ and the facts that $|z|\leq r^{-\frac 12}$ for $z\in \mathcal{D}$,  $|z|=O(1)$ for $z\in \mathcal{D}^0$.

In the regime that $p< n^{\epsilon}$ for sufficiently small $\epsilon$. We first write 
\begin{align*}
{X}{X}^{\top} = r^{-\frac 12} I_p +{G}, 
\end{align*}
where ${G}$ is a $p$ by $p$ matrix defined entrywise by ${G}_{ij} = {\bf x}_i^{\top} {\bf x}_j - \mathbb{E}  {\bf x}_i^{\top} {\bf x}_j  $ and ${\bf x}_i$ represents the i-th row of ${X}$. One can easily see that ${G}_{ij}$ is asymptotically centred Gaussian with variance $1/p$  by CLT. Thus we can crudely estimate ${G}_{ij}= O_\prec(p^{-1/2})$ and  $\Vert {G}\Vert\leq \Vert G\Vert_{\text{HS}} = O_\prec(\sqrt p\, )$. Then, for $\mathcal{G}_1$, we can obtain that for $z\in \mathcal{D}$,
\begin{align*}
\mathcal{G}_1= (r^{-\frac 12} - z)^{-1} \Big(I_p +  (r^{-\frac 12} - z)^{-1} {G}\Big)^{-1}
=  (r^{-\frac 12} - z)^{-1}I_p  -  (r^{-\frac 12} - z)^{-2} G + O_\prec\big(r^{\frac 32} p\big)
\end{align*}
here with a little abuse of notation, we used $O_\prec\big(r^{\frac 32} p\big)$ to represent the higher order term of matrix form whose operator norm is $O_\prec\big(r^{\frac 32} p\big)$.  Choosing $\epsilon$ sufficiently small so that $p^{3}n^{-\frac 12} = o(1)$. After elementary calculation, we further have that 
\begin{align}
&(\mathcal{G}_1)_{ij} =  (r^{-\frac 12} - z)^{-1}  \delta_{ij}  -  (r^{-\frac 12} - z)^{-2}  {G}_{ij} + O_\prec(n^{-1})=  (r^{-\frac 12} - z)^{-1}  \delta_{ij}  + O_\prec(n^{-\frac 12}r^{\frac 12}),\notag\\
    & m_1(z) - (r^{-\frac 12} - z)^{-1} = O_\prec(r^{\frac 32}) , \label{2020121101}
\end{align}
which by the fact that $r^{\frac 32}\ll n^{-\frac 12}r^{\frac 12}$ indeed imply the first estimate in (\ref{est:locallaw1}) for the case $l=0$. By using the identity $z\mathcal{G}_2(z) = {X}^{\top} \mathcal{G}_1(z) {X} - I_p$, we also have that 
\begin{align}
&(z\mathcal{G}_2(z) )_{i'j'} = -\delta_{i'j'} + (r^{-\frac 12} - z)^{-1} ({X}^{\top} {X}) _{i'j'} -   (r^{-\frac 12} - z)^{-2} ( {X}^{\top} {G} {X})_{i'j'} + O_\prec(n^{-1}), \notag\\
&zm_2(z) = -1 + r(1+ zm_1(z)) = -1 + r^{\frac 12}(r^{-\frac 12} - z)^{-1} +  O_\prec(r^{\frac 32}). \label{2020121102}
\end{align}
It is easy to see that $ ({X}^{\top} {X}) _{i'j'} =( {\bf x}^{i'})^{\top}  {\bf x}^{j'} =  r^{\frac 12}\delta_{i'j'} + O_\prec(n^{-1/2})$, where ${\bf x}^{i'}$ is the $i'$-th column of ${X}$. Furthermore, $|( {X}^{\top} G {X})_{i'j'}|= |( {\bf x}^{i'})^{\top} G {\bf x}^{j'}| \leq \Vert G \Vert \Vert {\bf x}^{i'}\Vert \Vert {\bf x}^{j'}\Vert =   O_\prec(n^{-1/2}p)$. We then see that 
\begin{align*}
(z\mathcal{G}_2(z) )_{i'j'}- zm_2(z)\delta_{i'j'}= O_\prec(n^{-\frac 12} r^{\frac12}).
\end{align*}
Thus, we can conclude the second estimate in (\ref{est:locallaw1}).  Next,  for the two estimates in (\ref{est:locallaw2}), we only need to focus on the former one in light of ${X}^{\top} \mathcal{G}_1= \mathcal{G}_2 {X}$ and the facts $|z|\leq r^{-\frac 12}$ for $z\in \mathcal{D}$,  $|z|=O(1)$ for $z\in \mathcal{D}^0$. Similarly to the above discussion, we have 
\begin{align} \label{20201216008}
( {X}^{\top} \mathcal{G}_1(z))_{i'i} =  (r^{-\frac 12} - z)^{-1} X_{ii'} - (r^{-\frac 12} - z)^{-2} ({X}^{\top} G)_{i'i} + O_\prec(n^{-1}p^2r^{\frac 14}) = O_\prec(n^{-\frac 12} r^{\frac 14})
\end{align} 
following from the facts that $X_{ii'} = O_\prec(n^{-\frac 12} r^{-\frac 14})$, $|({X}^{\top} G)_{i'i}|\leq \Vert G\Vert  |({X}^{\top} {X}) _{i'i'}|^{1/2}= O_\prec(r^{1/4}\sqrt p)$, and $p^{2}n^{-\frac 12} = o(1)$. This proved (\ref{est:locallaw2}). We then turn to the estimates in (\ref{est_m12N}). Note that $ G_{ii} $ are i.i.d. random variables of order $O_\prec\big(p^{-\frac 12}\big)$, for $1\leq i \leq p$. Hence by CLT, $p^{-1}\sum_{i=1}^p G_{ii} $ is crudely of order $O_\prec(p^{-1})$. Applying the first estimate in (\ref{2020121101}), we  have 
\begin{align*}
m_{1n}(z) &= \frac{1}{p} \sum_{i=1}^p (\mathcal{G}_1)_{ii} = (r^{-\frac 12} - z)^{-1}+ (r^{-\frac 12} - z)^{-2} \frac{1}{p} \sum_{i=1}^p G_{ii}  + O_\prec(n^{-1}) \notag\\
&=(r^{-\frac 12} - z)^{-1} + O_\prec(n^{-1}).
\end{align*}
 The above estimate,
together with the second equation in  (\ref{2020121101}) and the estimate $r^{\frac 32}\ll n^{-1}$, yields the first estimate in  (\ref{est_m12N}). The second estimate in  (\ref{est_m12N}) can be concluded simply by using  the identity $zm_{2n}(z)= -1+ r(1+ zm_{1n}(z))$ and $zm_{2}(z)= -1+ r(1+ zm_{1}(z))$, since 
\begin{align*}
|rzm_{1n}(z)- rz m_{1}(z)|\prec r |z| |m_{1n}(z) - m_1(z)|\prec n^{-1}r^{\frac 12}
\end{align*}
 uniformly for $z\in \mathcal{D}$. Particularly for $z\in \mathcal{D}^0$, since $|z|= O(1)$, the bound above can be further improved to $n^{-1}r$.
 
 Therefore, we proved the estimates (\ref{est:locallaw1})-(\ref{est_m12N}) uniformly for   $z\in \mathcal{D}$ in the case of $l=0$. Since $\mathcal{D}^0$ is simply a subset of 
$\mathcal{D}$, we trivially have the results uniformly for   $z\in \mathcal{D}^0$. Now, we will proceed to the case that $l\geq 1$ by using the estimates for   $z\in \mathcal{D}$.

\vspace{0.2cm}
\noindent $\bullet$ For the case of $l\geq 1$.
\vspace{0.2cm}

We can derive the estimates easily from the case $l=0$ by using Cauchy integral with the radius of the contour taking value $|z-\lambda_-|/4\asymp r^{-\frac 12}$. Note that for any $z\in \mathcal{D}^0$, the contour $\Gamma$ centred at $z$ with radius $|z-\lambda_-|/4$ still lies in the regime $\mathcal{D}$, hence all the estimates (\ref{est:locallaw1})-(\ref{est_m12N}) hold uniformly on the contour.  Moreover, we shall see that 
\begin{align*}
\Big|\big( \mathcal{G}_1^{(l)}(z)\big)_{ij} - m_1^{(l)}(z) \delta_{ij} \bigg| \asymp\Big|\oint_{\Gamma} \frac{ \big( \mathcal{G}_1(\tilde z)\big)_{ij} - m_1(\tilde z) \delta_{ij} }{(\tilde z- z)^{l+1}} {\rm d} \tilde z\bigg| \prec  \frac{n^{-\frac 12 } r^{\frac 12}}{|z-\lambda_-|^l} = n^{-\frac 12} r^{\frac{1+l}{2}}.
\end{align*}
Similarly, we can show the error bounds for the other terms stated in  (\ref{est:locallaw1})-(\ref{est_m12N}).

%\end{myPro}

\section{Proofs of Lemma  \ref{lem:ests} and Proposition \ref{prop.asym_dist} } \label{sec:pf_LemProp}

In this section, we prove Lemma \ref{lem:ests} and Proposition \ref{prop.asym_dist}, which are the key technical ingredients of the proofs of our main theorem. We separate the discussion into three subsections: in the first subsection we will 
show the  proof of Lemma \ref{lem:ests}; then followed by the proof of Proposition \ref{prop.asym_dist}  in the second subsection;  in the last subsection, we provide the proofs for some technical results in the first two  subsections. In advance of the proofs, we discuss some identities regarding Stieltjes transforms $m_1(z), m_2(z)$ (see (\ref{m1m2}) for definitions) and list some basic identities of Green functions which will be used frequently throughout this section.

Using \eqref{m1m2} and \eqref{selfconeqt}, one can easily derive the following identities  
\begin{align}
m_1=-\frac{1}{z(1+r^{-1/2}m_2)}\,, \quad 1+zm_1=\frac{1+zm_2}{r}\,, \quad r^{-1/2}(zm_2)'+1=\frac{m_1'}{m_1^2}\,. \label{identitym1m2}
\end{align}

We remark that since our discussion is based on the assumption $r\equiv r_n\to r_0 \in [0,1)$, then by definition,  $\lambda_- =r^{1/2}+r^{-1/2} -2 = O(r^{-1/2})$. This implies the  support of $\nu_{\text{MP},a}({\rm d}x)$ for $a=1,2$   stays away from $0$ by  $O(r^{-1/2})$  distance. For the special case $z=0$, $m_1(z)$ is well-defined and analytic at $z=0$ since $r<1$. More specifically,  $m_1(0)= \sqrt r/(1-r)$ by the first equation of (\ref{selfconeqt}). In contrast, $z=0$  is a pole of  $m_2(z)$ due to the $(1-r)$ point mass at $0$ (see MP law $\nu_{\text{MP},2}({\rm d}x)$ in (\ref{19071801})). However, the singularity at $z=0$ is removable for $zm_2(z)$. We can get  $zm_2(z)|_{z=0} = r-1$ by simple calculations of the second equation of (\ref{m1m2}).We write $\widehat m_2(z):= zm_2(z)$ for simplicity. Let us simply list several results of functions in terms of $m_{1,2}$ at $z=0$ which can checked easily from either (\ref{m1m2}) or (\ref{identitym1m2}).
\begin{align}
&m_1(0)= \frac{\sqrt r}{1-r}\,,\quad m_1'(0)= \frac{r}{(1-r)^3}\,; \label{values:m_1s} \\
&\widehat m_2(0)= r-1\,, \quad \widehat m_2'(0) = \frac{r^{3/2}}{1-r} \label{values:zm_2s}\,.
\end{align}

Next for the Green functions $\mathcal{G}_1$, $\mathcal{G}_2$, we have some basic and useful identities which can be easily checked by some elementary computations. 
\begin{align}
&\mathcal{G}_1^{l}=\frac{1}{(l-1)!}\frac{\partial^{l-1}\mathcal{G}_1}{\partial z^{l-1}} = \frac{1}{(l-1)!} \mathcal{G}_1^{(l-1)}\,,\label{eq:basicG}\\   
&\mathcal{G}_1^l{X}{X}^{\top}=\mathcal{G}_1^{l-1}+z\mathcal{G}_1^{l}, \quad {X}^{\top}\mathcal{G}_1^l{X}=\mathcal{G}_2^l{X}^{\top} {X}=\mathcal{G}_2^{l-1}+z\mathcal{G}_2^{l}\,. \label{relationXG}
\end{align}
%In the end, let us proceed to the proof of Proposition \ref{thm:prop}

\subsection{Proof of Lemma \ref{lem:ests}}
%With the preparations above, in this subsection, we compute the green function representation forms of the approximations of the  two key quantities $\widehat A^{\top} \Sigma \widehat A$, $\widehat A^{\top}{\bm{\mu}}^0$ in terms of   their completely sample-based statistics $\widehat A^{\top}\widehat \Sigma \widehat A$ and $\widehat A^{\top}\hat {\bm{\mu}}^0$, respectively.
We start with the proof of (\ref{est:hAhShA}). 
Applying Woodbury matrix identity, from (\ref{def:hSigma_matrix}),  we see that 
\begin{align} \label{inverseSigma}
\widehat \Sigma ^{-1}&= \frac{n-2}{n\sqrt r} \Sigma^{-\frac 12} \Big(\mathcal{G}_1(0) + \mathcal{G}_1(0) X  {E}  \mathcal{I}_{2}^{-1} {E}^{\top} {X}^{\top}  \mathcal{G}_1(0) \Big) \Sigma^{-\frac 12}\,,
\end{align}
where we introduced the notation 
\begin{align*}
\mathcal{I}_{2}:=I_2 - {E}^{\top} {X}^{\top}    \mathcal{G}_1(0)  X{E}\,. 
\end{align*}
Recall the definition ${E}= ({\bf e}_0, {\bf e}_1)$. 
 By the second identity in (\ref{relationXG}) and the second  estimate of (\ref{est.DG}), we have the estimate
 \begin{align}\label{2021051401}
{\bf u}^{\top} {X}^{\top}  \mathcal{G}_1^{a}(z){X} {\bf v}= (1+zm_2(z))^{(a-1)}
{\bf u}^{\top}  {\bf v} + O_\prec(n^{-\frac 12} r^{\frac {a}{2} })
 \end{align} 
 for arbitrary unit vectors ${\bf u}, {\bf v}$ and any integer $a\geq 1$. Further by $\widehat m_2(0) =zm_2(z)\Big|_{z=0}=r-1 $,  we obtain
\begin{align*}
{\bf e}_0^{\top}{{X}}^{\top}    \mathcal{G}_1(0)  {X}{\bf e}_1= O_\prec(n^{-1/2}r^{1/2}),\quad 
1- {\bf e}_i^{\top} {{X}}^{\top}   \mathcal{G}_1(0)  {X}  {\bf e}_i  = 1- r + O_\prec(n^{-1/2}r^{1/2})\,, \; i=1,2\,.
\end{align*}
Then, 
\begin{align} \label{inverM_dim2}
 \mathcal{I}_2^{-1} = 
\frac{1}{1-r} \, I_2 + \Delta\,,
\end{align}
where $ \Delta$ represents a $2\times 2$ matrix with $\Vert \Delta \Vert =O_\prec(n^{-1/2}r^{1/2}) $. Plugging \eqref{inverM_dim2} into (\ref{inverseSigma}), we can write 
\begin{align}
\widehat \Sigma^{-1} = \frac{n-2}{n\sqrt r}  \Sigma^{-\frac 12}\mathcal{G}_1(0)  \Sigma^{-\frac 12}+ \frac{n-2}{n(1-r)\sqrt r} \sum_{i=1,2}  \Sigma^{-\frac 12} \mathcal{G}_1(0){X}{\bf e}_i {\bf e}_i^{\top} {X}^{\top} \mathcal{G}_1(0)  \Sigma^{-\frac 12} +\widehat \Delta\,, \label{031501}
\end{align}
where  
$$\widehat \Delta=\frac{n-2}{n\sqrt{r}} \Sigma^{-\frac12}\mathcal{G}_1(0)XE\Delta E^{\top}X^{\top}\mathcal{G}_1(0)\Sigma^{-\frac12}\,, $$
and it is easy to check $\|\widehat \Delta \|\prec n^{-1/2}r^{1/2} $.
%\begin{align*}
%\widehat \Sigma ^{-1}&= \Big(\frac{1}{n-2}  {{X}} {{X}}^* \Big)^{-1}  \notag\\
%& + 
% \Big(\frac{1}{n-2}  {{X}} {{X}}^* \Big)^{-1} \frac{1}{n-2} {X}{E}  \Big(I_2 - \frac{1}{n-2} {E}^{\top} {{X}}^*   \Big(\frac{1}{n-2}  {{X}} {{X}}^* \Big)^{-1}  {X}{E}\Big)^{-1} {E}^{\top} {{X}}^*   \Big(\frac{1}{n-2}  {{X}} {{X}}^* \Big)^{-1} 
%\end{align*}

With the above preparation, we now compute the leading term of $\widehat A^{\top} \widehat \Sigma \widehat A$. Recall $\widehat A= \widehat \Sigma^{-1}\hat {\bm{\mu}}_d$.  We have
\begin{align}\label{def:rewrite AhSigmaA}
\widehat A^{\top} \widehat \Sigma \widehat A = \hat {\bm{\mu}}_d^{\top}  \widehat \Sigma^{-1}  \hat {\bm{\mu}}_d  &= \sqrt r{\bf v}_1^{\top}  {X}^{\top} \Sigma^{\frac 12} \widehat \Sigma^{-1} \Sigma^{\frac 12}   {X} {\bf v}_1 +{\bm{\mu}}_d^{\top}  \widehat \Sigma^{-1}   {\bm{\mu}}_d  + 2  r^{\frac 14} {\bf v}_1^{\top} {X}^{\top} \Sigma^{\frac 12}\widehat \Sigma^{-1}   {\bm{\mu}}_d \notag\\
&=:T_1+ T_2+T_3\,.
\end{align}
For $T_1$, with (\ref{031501}), we have
\begin{align}\label{eq:est_T1}
T_1 
& = \frac {n-2}{n}\,   {\bf v}_1^{\top}{X}^{\top} \mathcal{G}_1(0){X} {\bf v}_1
+  \frac {n-2}{n(1-r)}\,   {\bf v}_1^{\top}{X}^{\top} \mathcal{G}_1(0){X}\Big( \sum_{i=0,1} {\bf e}_i {\bf e}_i^{\top}  \Big) {X}^{\top} \mathcal{G}_1(0){X} {\bf v}_1 \notag\\
&\qquad + \sqrt r {\bf v}_1^{\top}{X}^{\top} \Sigma^{\frac 12} \widehat \Delta \Sigma^{\frac 12}  {X} {\bf v}_1 \notag\\
&= r \Vert {\bf v}_1\Vert^2  + \frac{r^2}{1-r}\Big(\big({\bf v}_1^{\top} {\bf e}_0\big)^2 + \big({\bf v}_1^{\top} {\bf e}_1\big)^2 \Big) + O_\prec(n^{-\frac 12}r^{\frac 12})\,.
 \end{align}
 Here in the last step, we repeatedly used the estimate (\ref{2021051401}) and $1+ zm_2(z)|_{z=0}= r $. In addition, for the last term of the second line of (\ref{eq:est_T1}), we trivially bound it by 
 \begin{align*}
 \sqrt r {\bf v}_1^{\top}{X}^{\top} \Sigma^{\frac 12} \widehat \Delta \Sigma^{\frac 12}  {X} {\bf v}_1\leq \Vert \widehat \Delta \Vert \Vert \Sigma\Vert   (\sqrt r {\bf v}_1^{\top}{X}^{\top}{X} {\bf v}_1) = O_\prec (n^{-\frac 12}r^{\frac 12})\,.
 \end{align*}
% suppose we assume that $n_0, n_1$ are both comparable to $n$ so that $\Vert {\bf v}_1\Vert = O(1)$. 
 Similarly, for $T_2$, we have 
 \begin{align}
 T_2 
 & = \frac {n-2}{n\sqrt r}\,   {\bf u}_1^{\top} \ \mathcal{G}_1(0){\bf u}_1
+  \frac {n-2}{n(1-r)\sqrt r}\,   {\bf {u}}_1^{\top} \mathcal{G}_1(0){X}\Big( \sum_{i=0,1} {\bf e}_i {\bf e}_i^{\top}  \Big) {X}^{\top} \mathcal{G}_1(0) {\bf u}_1 +  {\bf u}_1^{\top} \Sigma^{\frac 12} \widehat \Delta \Sigma^{\frac  12}  {\bf u}_1 \notag\\
&=\frac{ \Vert {\bf u}_1\Vert^2 }{1-r}  + O_\prec(n^{-\frac 12}\Vert {\bf u}_1\Vert^2)\,, \label{031505}
 \end{align}
 where we employed the shorthand notation 
 \begin{align}
 {\bf u}_1:= \Sigma^{-\frac 12} {\bm{\mu}}_d\,. \label{def of mu1}
 \end{align} 
 Here in  (\ref{031505}), we applied the estimates
 \begin{align}
 & {\bf u}_1^{\top} \ \mathcal{G}_1(0){\bf u}_1 = m_1(0) \Vert {\bf u}_1\Vert^2 + O_\prec (n^{-\frac 12}r^{\frac 12}\Vert {\bf u}_1 \Vert^2)  \label{eq:est(1)}\\
 & {\bf u}_1^{\top} \mathcal{G}_1(z){X} {\bf e}_i =  O_\prec(n^{-\frac 12}r^{\frac 14}\Vert {\bf u}_1  \Vert ) = O_\prec \big(n^{-\frac 12} r^{\frac 14}\Vert {\bf u}_1\Vert \big),\quad i=0,1\,.   \label{eq:est(2)}
 \end{align}
 with the fact $m_1(0)= 1/(1-r)$. 
%Here we make assumption that  $\Vert \Sigma\Vert$ is bounded below. And we keep this $\Vert {\bm{\mu}}_d\Vert^2$ factor because we are not sure that it is of order $1$  or higher order depending on $n$, $\sqrt{n}$, say. However, in any sense the leading term on RHS of (\ref{eq:est(1)}) is always $ m_1(0) \Vert {\bf u}_1\Vert^2$. 
Next, we turn to estimate $T_3$. Similarly, we have
\begin{align*}
T_3
&= \frac{2(n-2)}{nr^{\frac 14}}  {\bf v}_1^{\top} {X}^{\top} \mathcal{G}_1(0){\bf u}_1 +  \frac{2(n-2)}{n(1-r)r^{\frac 14}}  {\bf v}_1^{\top} {X}^{\top} \mathcal{G}_1(0){X}\Big(\sum_{i=0,1} {\bf e}_i{\bf e}_i^{\top} \Big){X}^{\top} \mathcal{G}_1(0){\bf u}_1  + O_\prec(n^{-\frac 12}r^{\frac 12}\Vert {\bf u}_1\Vert) \notag\\
&= O_\prec(n^{-\frac 12} \Vert {\bf u}_1\Vert)\, .
\end{align*}
 Therefore, we arrive at 
 \begin{align*} %\label{est:qua_hSigma}
\widehat A^{\top} \widehat \Sigma \widehat A 
%&=   r\Vert {\bf v}_1\Vert^2  + \frac{r^2}{1-r}\Big(\big({\bf v}_1^{\top} {\bf e}_0\big)^2 + \big({\bf v}_1^{\top} {\bf e}_1\big)^2  \Big) +  \frac{1}{1-r}\Vert {\bf u}_1\Vert^2     + O_\prec(n^{-\frac 12}\Vert {\bm{\mu}}_d\Vert^2)\notag\\
&=  \frac{r}{1-r} \Vert {\bf v}_1\Vert^2 +  \frac{1}{1-r}\Vert {\bf u}_1\Vert^2     + O_\prec \big(n^{-\frac 12}(r^{\frac 12} + \Vert {\bf u}_1\Vert^2 + \Vert {\bf u}_1\Vert )  \big)\, .
\end{align*}
This proved  (\ref{est:hAhShA}).

To proceed, we  estimate $\widehat A^{\top} \Sigma \widehat A $. By definition, 
\begin{align*}
\widehat A^{\top} \Sigma \widehat A =\Big(\frac {n-2}{n\sqrt r}\Big)^2\hat {\bm{\mu}}_d^{\top} \Sigma^{-\frac12} H_E^{-2}  \Sigma^{-\frac12}\hat {\bm{\mu}}_d\, ,
\end{align*}
where we introduced the notation 
\begin{align*}
H_E:={{X}}( I_n - {{E}}{{E}}^{\top}) {{X}}^{\top}.
\end{align*}
Applying Woodbury matrix identity again, we have 
\begin{align}\label{eq:X_XT-2}
H_E^{-2} &= \mathcal{G}_1^2(0) + \mathcal{G}_1^2(0)  {X}{E}  \mathcal{I}_2^{-1} {E}^{\top} {{X}}^{\top}  \mathcal{G}_1(0) \notag\\
&\quad +  \mathcal{G}_1(0)  {X}{E}  \mathcal{I}_2^{-1} {E}^{\top} {{X}}^{\top}  \mathcal{G}_1^2(0)+   \Big(\mathcal{G}_1(0)  {X}{E}  \mathcal{I}_2^{-1} {E}^{\top} {{X}}^{\top}  \mathcal{G}_1(0)\Big)^2. 
\end{align}
Analogously to the way we deal with $\widehat A^{\top} \widehat \Sigma \widehat A $, applying the representation of  $\hat {\bm{\mu}}_d$ in (\ref{eq:repre_mu_d}) and also the notation in (\ref{def of mu1}), we can write
\begin{align}\label{def:rewrite ASigmaA}
\widehat A^{\top} \Sigma \widehat A &=\Big(\frac {n-2}{n}\Big)^2r^{-\frac 12}{\bf v}_1^{\top} {X}^{\top} H_E^{-2} {X}{\bf v}_1
+ \Big(\frac {n-2}{n}\Big)^2 r^{-1}{\bf u}_1^{\top}  H_E^{-2}{\bf u}_1 \notag\\
&\qquad +2 \Big(\frac {n-2}{n}\Big)^2 r^{-\frac 34}{\bf v}_1^{\top} {X}^{\top} H_E^{-2} {\bf u}_1=: \mathcal{T}_1 + \mathcal{T}_2 + \mathcal{T}_3\,,
\end{align}
and we analyse  the RHS of the above equation term by term. First, for $\mathcal{T}_1$, substituting (\ref{eq:X_XT-2}) and (\ref{inverM_dim2}), we have 
\begin{align} \label{20201230002}
\mathcal{T}_1 &= \Big(\frac{n-2}{n}\Big)^2 r^{-\frac 12}\Big( {\bf v}_1^{\top} {X}^{\top} \mathcal{G}_1^2(0){X}{\bf v}_1 + \frac {2}{1-r}\sum_{i=0,1} \big({\bf v}_1^{\top} {X}^{\top} \mathcal{G}_1^2(0){X}{\bf e}_i \big)\big({\bf e}_i^{\top} {X}^{\top} \mathcal{G}_1(0){X}{\bf v}_1\big) \notag\\
&\quad + \frac{1}{(1-r)^2} \sum_{i,j=0,1} \big({\bf v}_1^{\top} {X}^{\top} \mathcal{G}_1(0){X}{\bf e}_i \big) \big({\bf e}_i^{\top} {X}^{\top} \mathcal{G}_1^2(0){X}{\bf e}_j\big)
\big({\bf e}_j^{\top} {X}^{\top} \mathcal{G}_1(0){X}{\bf v}_1\big)\Big) + O_\prec(n^{-\frac 12}) \notag\\
&=\bigg[r^{-\frac 12} (zm_2(z))' \Vert {\bf v}_1\Vert^2 + \frac{2r^{-\frac 12}}{1-r} (zm_2(z))' (1+zm_2(z))\Big( \sum_{i=0,1}({\bf v}_1^{\top} {\bf e}_i)^2\Big)  \notag\\
&\quad + \frac{r^{-\frac 12}}{(1-r)^2} (zm_2(z))'\Big( (1+ zm_2(z))\Big)^2 \Big( \sum_{i=0,1}({\bf v}_1^{\top} {\bf e}_i)^2\Big)\bigg]\bigg|_{z=0}+ O_\prec(n^{-\frac 12}r^{\frac 12}) \notag\\
&= \frac{r}{(1-r)^3}\Vert {\bf v}_1\Vert^2 +  O_\prec(n^{-\frac 12}r^{\frac 12})\,. 
\end{align}
Here we used the estimate (\ref{2021051401}) and the facts that $(zm_2(z))'\big|_{z=0} = r^{3/2}/(1-r) $, $(1+ zm_2(z))\big|_{z=0}= r$ and $\sum_{i=0,1}({\bf v}_1^{\top} {\bf e}_i)^2= \Vert{\bf v}_1\Vert^2$ according to the definition of ${\bf v}_1$ in (\ref{eq:repre_mu_d}). 

Next, similarly to $\mathcal{T}_1$, for $\mathcal{T}_2$, we have the estimates 
\begin{align*}
\mathcal{T}_2&= \Big(\frac{n-2}{n}\Big)^2 r^{-1} \Big( {\bf u}_1^{\top}  \mathcal{G}_1^2(0){\bf u}_1 + \frac {2}{1-r}\sum_{i=0,1} \big({\bf u}_1^{\top}  \mathcal{G}_1^2(0){X}{\bf e}_i \big)\big({\bf e}_i^{\top} {X}^{\top} \mathcal{G}_1(0){\bf u}_1\big) \notag\\
&\quad + \frac{1}{(1-r)^2} \sum_{i,j=0,1} \big({\bf u}_1^{\top} \mathcal{G}_1(0){X}{\bf e}_i \big) \big({\bf e}_i^{\top} {X}^{\top} \mathcal{G}_1^2(0){X}{\bf e}_j\big)
\big({\bf e}_j^{\top} {X}^{\top} \mathcal{G}_1(0){\bf u}_1\big)\Big) + O_\prec(n^{-\frac 12}\Vert {\bf u}_1\Vert^2) \notag\\
& =  \frac{1}{(1-r)^3} \Vert {\bf u}_1\Vert^2 +  O_\prec(n^{-\frac 12}\Vert {\bf u}_1\Vert^2 )\,.
\end{align*}
In the last step, we applied (\ref{est.DG}), the second estimate of (\ref{081501}) and (\ref{2021051401}).
Further for $\mathcal{T}_3$, we have the following estimate 
\begin{align*}
\mathcal{T}_3 &=2\Big(\frac{n-2}{n}\Big)^2 r^{-\frac 34} \Big( {\bf v}_1^{\top} {X}^{\top}  \mathcal{G}_1^2(0){\bf u}_1 + \frac {1}{1-r}\sum_{i=0,1} \big({\bf v}_1^{\top} {X}^{\top}  \mathcal{G}_1^2(0){X}{\bf e}_i \big)\big({\bf e}_i^{\top} {X}^{\top} \mathcal{G}_1(0){\bf u}_1\big)  \notag\\
&\quad 
+  \frac {1}{1-r}\sum_{i=0,1}\big({\bf v}_1^{\top} {X}^{\top}  \mathcal{G}_1(0){X}{\bf e}_i \big)\big({\bf e}_i^{\top} {X}^{\top} \mathcal{G}_1^2(0){\bf u}_1\big)  \notag\\
&\quad + \frac{1}{(1-r)^2} \sum_{i,j=0}^1 \big({\bf v}_1^{\top} {X}^{\top} \mathcal{G}_1(0){X}{\bf e}_i \big) \big({\bf e}_i^{\top} {X}^{\top} \mathcal{G}_1^2(0){X}{\bf e}_j\big)
\big({\bf e}_j^{\top} {X}^{\top} \mathcal{G}_1(0){\bf u}_1\big)\Big) + O_\prec(n^{-\frac 12}\Vert {\bf u}_1 \Vert) \notag\\
&=  O_\prec(n^{-\frac 12}\Vert {\bf u}_1 \Vert)\,. 
\end{align*}
Here all the summands above contain  quadratic forms of $({X}\mathcal{G}_1^{a})$, and by (\ref{081501}), we see such quadratic forms are of order $O_\prec(n^{-\frac12}r^{1/4+(a-1)/2}\Vert {\bf u}_1 \Vert)$. Further with  the estimate (\ref{2021051401}) and identities (\ref{values:zm_2s}), we shall get the estimate $O_\prec(n^{-\frac 12}\Vert {\bf u}_1 \Vert) $ for $\mathcal{T}_3$.
According to the above estimates of $\mathcal{T}_1, \mathcal{T}_2, \mathcal{T}_3 $, we now  see that 
\begin{align*}%\label{est:qua_Sigma}
\widehat A^{\top} \Sigma \widehat A 
%&= \frac{r}{1-r}\Vert {\bf v}_1\Vert^2 + \Big(\frac{2r^2}{(1-r)^2} + \frac{r^3}{(1-r)^3}  \Big)\Big( \sum_{i=0,1}({\bf v}_1^{\top} {\bf e}_i)^2\Big) +  \frac{1}{(1-r)^3} \Vert {\bf u}_1\Vert^2 +  O_\prec(n^{-\frac 12}\Vert {\bm{\mu}}_d\Vert^2)\notag\\
&=  \frac{r}{(1-r)^3}\Vert {\bf v}_1\Vert^2 + \frac{1}{(1-r)^3} \Vert {\bf u}_1\Vert^2 +  O_\prec \big(n^{-\frac 12}(r^{\frac 12} + \Vert {\bf u}_1\Vert^2 + \Vert {\bf u}_1\Vert )  \big)\,.
\end{align*}
Thus we completed the proof of  (\ref{est:hAShA}) by the fact that $\Vert {\bf u}_1\Vert^2 = \varDelta_d$.

Next, we turn to prove the estimates (\ref{est:hAmu0}) and (\ref{est:hAhmu0}). 
Recall the  representations of $\hat {\bm{\mu}}^0$ and $\hat {\bm{\mu}}_d$ in (\ref{eq:repre_mu_0}) and  (\ref{eq:repre_mu_d}), and also the notation in (\ref{def of mu1}). Applying Woodbury matrix identity to $H_E^{-1}$, we can write
\begin{align*}
\widehat{A} ^{\top} {\bm{\mu}}_d &=  \frac{n-2}{n\sqrt r} \big( r^{\frac 14}{\bf v}_1^{\top} {X}^{\top} + {\bf u}_1^{\top}\big) H_E^{-1} {\bf u}_1  \notag\\
& =\frac{n-2}{n} r^{-\frac 14} \Big( {\bf v}_1^{\top} {X}^{\top} \mathcal{G}_1(0) {\bf u}_1+ {\bf v}_1^{\top} {X}^{\top} \mathcal{G}_1(0){X}{E} \mathcal{I}_2^{-1}{E}^{\top} {X}^{\top} \mathcal{G}_1(0) {\bf u}_1\Big)\notag\\
&\quad  +   \frac{n-2}{n\sqrt r} \Big( {\bf u}_1^{\top} \mathcal{G}_1(0)  {\bf u}_1 + {\bf u}_1^{\top}  \mathcal{G}_1(0){X}{E} \mathcal{I}_2^{-1}{E}^{\top} {X}^{\top} \mathcal{G}_1(0){\bf u}_1 \Big)\,, 
\end{align*}
and 
\begin{align*}
\widehat{A} ^{\top} \hat{\bm{\mu}}^0 - \widehat{A} ^{\top} {\bm{\mu}}^0 &= \hat {\bm{\mu}}_d^{\top} \widehat \Sigma^{-1}(\hat{\bm{\mu}}^0 -  \hat{\bm{\mu}}^0)  = \frac{n-2}{n\sqrt r} \big( r^{\frac 14}{\bf v}_1^{\top} {X}^{\top} + {\bf u}_1^{\top}\big) H_E^{-1}\Big(\sqrt{\frac{n}{n_0}} \, r^{\frac 14}{X}{\bf e}_0\Big)\notag\\
&= \frac{n-2}{\sqrt{nn_0}}  \Big( {\bf v}_1^{\top} {X}^{\top} \mathcal{G}_1(0)  {X}{\bf e}_0+ {\bf v}_1^{\top} {X}^{\top} \mathcal{G}_1(0){X}{E} \mathcal{I}_2^{-1}{E}^{\top} {X}^{\top} \mathcal{G}_1(0) {X}{\bf e}_0\Big)\notag\\
& \quad + \frac{n-2}{\sqrt{nn_0}} r^{-\frac 14} \Big( {\bf u}_1^{\top}  \mathcal{G}_1(0)  {X}{\bf e}_0+ {\bf u}_1^{\top} \mathcal{G}_1(0){X}{E} \mathcal{I}_2^{-1}{E}^{\top} {X}^{\top} \mathcal{G}_1(0) {X}{\bf e}_0\Big)\, .
\end{align*}
%
% \begin{align*}
%\widehat{A} ^{\top} \hat{\bm{\mu}}^0 &= \hat {\bm{\mu}}_d^{\top} \widehat \Sigma^{-1}\hat{\bm{\mu}}^0 = \frac{n-2}{n\sqrt r} \big( r^{\frac 14}{\bf v}_1^{\top} {X}^{\top} + {\bf u}_1^{\top}\big) H_E^{-1}\Big(\sqrt{\frac{n}{n_0}} \, r^{\frac 14}{X}{\bf e}_0+ \Sigma^{-\frac12}{\bm{\mu}}^0\Big)\notag\\
%&= \frac{n-2}{\sqrt{nn_0}}  \Big( {\bf v}_1^{\top} {X}^{\top} \mathcal{G}_1(0)  {X}{\bf e}_0+ {\bf v}_1^{\top} {X}^{\top} \mathcal{G}_1(0){X}{E} \mathcal{I}_2^{-1}{E}^{\top} {X}^{\top} \mathcal{G}_1(0) {X}{\bf e}_0\Big)\notag\\
%&\quad + \frac{n-2}{n} r^{-\frac 14} \Big( {\bf v}_1^{\top} {X}^{\top} \mathcal{G}_1(0)  \Sigma^{-\frac 12} {\bm{\mu}}^0+ {\bf v}_1^{\top} {X}^{\top} \mathcal{G}_1(0){X}{E} \mathcal{I}_2^{-1}{E}^{\top} {X}^{\top} \mathcal{G}_1(0) \Sigma^{-\frac 12} {\bm{\mu}}^0\Big)\notag\\
%&\quad +  \frac{n-2}{\sqrt{nn_0}} r^{-\frac 14} \Big( {\bf u}_1^{\top}  \mathcal{G}_1(0)  {X}{\bf e}_0+ {\bf u}_1^{\top} \mathcal{G}_1(0){X}{E} \mathcal{I}_2^{-1}{E}^{\top} {X}^{\top} \mathcal{G}_1(0) {X}{\bf e}_0\Big)\notag\\
%&\quad  +   \frac{n-2}{n\sqrt r} \Big( {\bf u}_1^{\top} \mathcal{G}_1(0)  \Sigma^{-\frac 12} {\bm{\mu}}^0+ {\bf u}_1^{\top}  \mathcal{G}_1(0){X}{E} \mathcal{I}_2^{-1}{E}^{\top} {X}^{\top} \mathcal{G}_1(0) \Sigma^{-\frac 12} {\bm{\mu}}^0\Big)\,.
%\end{align*}
Similarly to the derivation of the leading term of $\widehat A^{\top} \widehat\Sigma^{-1} \widehat A$, by (\ref{est.DG}),  (\ref{081501}) and (\ref{2021051401}), after elementary calculation, we arrive at 
\begin{align*}
\widehat{A} ^{\top} {\bm{\mu}}_d = \frac{1}{1-r} {\bm {\mu}}_d^{\top} \Sigma^{-1} {\bm {\mu}}_d +  O_\prec \big(n^{-\frac 12} (\Vert {\bf u}_1\Vert^2 + \Vert {\bf u}_1\Vert )\big)
\end{align*}
and 
 \begin{align*}
\widehat{A} ^{\top} \hat{\bm{\mu}}^0 - \widehat{A} ^{\top} {\bm{\mu}}^0  =\sqrt{\frac{n}{n_0}} \frac{r}{1-r} {\bf v}_1^{\top} {\bf e}_0  + O_\prec\big(n_0^{-\frac 12} (r^{\frac 12}  + \Vert {\bf u}_1\Vert)\big)\,.
\end{align*}
%For $\widehat{A} ^{\top} {\bm{\mu}}^0 $, which  can  be viewed as a term in $\widehat{A} ^{\top} \hat{\bm{\mu}}^0$ (c.f.(\ref{eq:repre_mu_0})), we have 
%\begin{align*}%\label{2020101301}
%\widehat{A} ^{\top} {\bm{\mu}}^0  &=  \frac{n-2}{n\sqrt r} \big(r^{\frac 14}{\bf v}_1^{\top} {X}^{\top} + {\bf u}_1^{\top}\big)H_E^{-1}\Sigma^{-\frac12}{\bm{\mu}}^0=  \frac{1}{1-r} {\bm{\mu}}_d^{\top} \Sigma^{-1}{\bm{\mu}}^0+ O_\prec(n^{-\frac 12} \Vert{\bm{\mu}}_d\Vert \Vert {\bm{\mu}}^0\Vert )\,.
%\end{align*}
Finally,  analogously to $\widehat{A} ^{\top} \hat{\bm{\mu}}^0 - \widehat{A} ^{\top} {\bm{\mu}}^0  $, the estimates  with the triple $({\bm{\mu}}^0, \hat{{\bm{\mu}}}_0, \sqrt{n/n_0} \, {\bf e}_0)$ replaced by $({\bm{\mu}}^1, \hat{{\bm{\mu}}}^1, \sqrt{n/n_1} \, {\bf e}_1)$ or $({\bm{\mu}}_d, \hat{{\bm{\mu}}}_d , {\bf v}_1)$ can be derived similarly.  Hence we skip the details and conclude the proof of Lemma \ref{lem:ests}. 

%We remark here that for both (\ref{est:qua_hSigma}) and (\ref{est:qua_Sigma}), if $\Vert{\bm{\mu}}_d\Vert$ is of size $o(1)$ then the error bounds on both RHS should change $n^{-1/2} \Vert {\bm{\mu}}_d\Vert$. But in this classification problem, ${\bm{\mu}}^0, {\bm{\mu}}^1$ shall be distinguishable so that $\Vert{\bm{\mu}}_d\Vert$ should be at least of size $O(1)$.

\subsection{Proof of Proposition \ref{prop.asym_dist} } \label{subsec:21051501}

In this part, we show the proof of Proposition \ref{prop.asym_dist}. First, we introduce the Green function representation of $\widehat{F}(\widehat\Sigma, \hat{\bm{\mu}}^0)- F(\Sigma, {\bm{\mu}}^0)$ based on Lemma \ref{lem:ests} and Remark \ref{rmk:ests}.

\begin{lem} \label{lem:green_diff}
Let $\widehat{F}(\widehat\Sigma, \hat{\bm{\mu}}^0)$ and $F(\Sigma, {\bm{\mu}}^0)$  be defined in (\ref{def:whF}) and (\ref{def:F}), respectively. Suppose that Assumption \ref{asm} holds. Then, 
\begin{align}\label{eq:green_diff}
 \widehat{F}(\widehat\Sigma, \hat{\bm{\mu}}^0)- F(\Sigma, {\bm{\mu}}^0) &=\Bigg[\frac{1-r}{2\sqrt{   \hat {\bm{\mu}}_d^{\top} \widehat\Sigma^{-1} \hat {\bm{\mu}}_d}}\bigg(\frac{1-2r}{(1-r)^4} {\bf v}_1^{\top} \big( z\mathcal{G}_2-zm_2\big){\bf v}_1  - \frac{r^{-\frac 12}}{(1-r)^2} {\bf v}_1^{\top}\Big((z\mathcal{G}_2)'-(zm_2)'\Big){\bf v}_1 \notag\\
& \quad  +  \frac{r^{-\frac 12}}{(1-r)^2} {\bf u}_1^{\top}(\mathcal{G}_1-m_1){\bf u}_1 - r^{-1}{\bf u}_1^{\top}(\mathcal{G}_1^2- m_1'){\bf u}_1 \notag\\
&\quad +  \frac{2r^{-\frac 14}}{(1-r)^2} {\bf u}_1^{\top}\mathcal{G}_1 {X}{\bf v}_1 - \frac{2r^{-\frac 34}}{1-r}  {\bf u}_1^{\top} \mathcal{G}_1^2 {X}{\bf v}_1
 \bigg) \, \Phi^{-1}(1-\alpha) \notag\\
& \quad +  \sqrt{\frac{n}{n_0}} \Big( \frac{1}{(1-r)^2} {\bf v}_1^{\top}(z\mathcal{G}_2-zm_2){\bf e}_0+ \frac{r^{-\frac 14}}{1-r} {\bf u}_1^{\top}\mathcal{G}_1 {X}{\bf e}_0\Big)\Bigg]\Bigg|_{z=0}\notag\\
&\quad+   O_\prec \big(n^{-1}(r^{\frac 12} +\varDelta_d^{\frac 12} )\big)  \,.
%(1+ \sqrt{n/n_0} \,\,  r^{\frac 12} )\big) } \,. 
\end{align}

%{\color{blue}{If we allow the case $\Vert {\bm{\mu}}_d\Vert$ is small, the error is $ O_\prec\big(n^{-1}(r^{1/2}+\Vert {\bm{\mu}}_d\Vert)\big) $}}
\end{lem}

\begin{myRem} \label{rmk:green_diff}
Here we emphasize again that  $z=0$ is a removable singularity of $z\mathcal{G}_2(z)$ and  $zm_2(z)$. Additionally,  $z\mathcal{G}_2(z)\neq 0$ and  $zm_2(z)\neq 0$ when $z=0$ (see (\ref{values:zm_2s})). By (\ref{est:hAhShA}), (\ref{est.DG})  and (\ref{081501}), it is not hard to see that the factor before $ \Phi^{-1}(1-\alpha)$ on the RHS of (\ref{eq:green_diff}) is of order $O_\prec(n^{-1/2} \varDelta_d^{1/2})$. Similarly,  the term in the fourth line of (\ref{eq:green_diff}) is also crudely bounded  by $O_\prec(n^{-1/2} \varDelta_d^{1/2})$. 
\end{myRem}

Here to the rest of this subsection, we will adopt the notation $(M)_{{\bf u} {\bf v}}$ as the quadratic form ${\bf u}^* M {\bf v}$ for arbitrary two column vectors ${\bf u}, {\bf v}$ of dimension $a, b$, respectively, and any $a\times b$ matrix $M$.
In light of Lemma \ref{lem:green_diff} and Remark \ref{rmk:green_diff}, it suffices to study the joint distribution of the following terms with appropriate scalings which make them order one random variables,
 \begin{align}\label{randomterms}
& \frac{\sqrt{n}}{\sqrt r}\big( z\mathcal{G}_2-zm_2\big)_{\bar{\bf v}_1\bar{\bf v}_1}, \, \frac{\sqrt{n}}{r}\Big((z\mathcal{G}_2)'-(zm_2)'\Big)_{\bar{\bf v}_1\bar{\bf v}_1}, \,
\frac{\sqrt n}{\sqrt r} (\mathcal{G}_1-m_1)_{\bar{\bf u}_1\bar{\bf u}_1} ,\,  \frac{\sqrt n}{r}(\mathcal{G}_1^2- m_1')_{\bar{\bf u}_1\bar{\bf u}_1}, \notag\\
&
\sqrt n r^{-\frac 14}( \mathcal{G}_1 {X})_{\bar{\bf u}_1\bar{\bf v}_1}, \, \sqrt n r^{-\frac 34}( \mathcal{G}_1^2 {X})_{\bar{\bf u}_1\bar{\bf v}_1},\,  \frac{\sqrt{n}}{\sqrt r}(z\mathcal{G}_2-zm_2)_{\bar{\bf v}_1{\bf e}_0},\, \sqrt n r^{-\frac 14}(\mathcal{G}_1 {X})_{\bar{\bf u}_1{\bf e}_0}\,.
 \end{align} 
 Here we adopt the notation $\bar{\bf u}$ to denote the normalized version of a generic vector ${\bf u}$, i.e. 
 \begin{align*}
 \bar{\bf u} = \left\{
 \begin{array}{cc}
\frac{{\bf u}}{\Vert {\bf u}\Vert}, & \text{ if $\Vert {\bf u}\Vert \neq 0$\,;}\\
0, & \text{ otherwise\,.}
 \end{array}
 \right.
 \end{align*}
 And for a fixed deterministic column vector ${\bf c}:=\big(c_{10}, \cdots, c_{13}, \,c_{20},\, c_{21}\big)^{\top}\in \mathbb{R}^{8}$, we define for $z\in \mathcal{D}$
 \begin{align}\label{def:mP_ori}
 \mathcal{P} &\equiv \mathcal{P}(\mathbf{c},z):= \frac{\sqrt{n}}{\sqrt{r}}  c_{10} (\mathcal{G}_1-m_1)_{\bar{\bf u}_1\bar{\bf u}_1} +  \frac{\sqrt {n}}{r}  c_{11}(\mathcal{G}_1^2- m_1')_{\bar{\bf u}_1\bar{\bf u}_1}  \notag\\
 &
  + \frac{\sqrt n}{r^{\frac14}}  c_{12}( \mathcal{G}_1 {X})_{\bar{\bf u}_1\bar{\bf v}_1} + \frac{\sqrt n}{r^{\frac14}}  c_{13}(\mathcal{G}_1 {X})_{\bar{\bf u}_1{\bf e}_0} +\frac{\sqrt n}{r^{\frac34}} c_{14}( \mathcal{G}_1^2 {X})_{\bar{\bf u}_1\bar{\bf v}_1} \notag\\
 &+  \frac{\sqrt{n}}{\sqrt r} c_{20}\big( z\mathcal{G}_2-zm_2\big)_{\bar{\bf v}_1\bar{\bf v}_1} + \frac{\sqrt{n}}{\sqrt r}c_{21}(z\mathcal{G}_2-zm_2)_{\bar{\bf v}_1{\bf e}_0} + \frac{\sqrt{n}}{ r} c_{22}\Big((z\mathcal{G}_2)'-(zm_2)'\Big)_{\bar{\bf v}_1\bar{\bf v}_1}.
 \end{align}
% Furthermore, by introducing 
% \begin{align}\label{def:y_ty_eta}
%& {\bf y}_0:= c_{10}\bar{\bf u}_1, \quad {\bf y}_1:= c_{11} \bar{\bf u}_1, \quad \tilde{{\bf y}}_0:=c_{12} \bar{\bf v}_1 + c_{13}{\bf e}_0,\quad  \tilde{{\bf y}}_1:= c_{14} \bar{\bf v}_1,\notag\\
% & {\bm \eta}_0:= c_{20}{\bf v}_1^0+ c_{21} {\bf e}_0,\quad {\bm \eta}_1:= c_{22}{\bf v}_1^0, 
% \end{align}
% together with the trivial identity $z\mathcal{G}_2= {X}^{\top} \mathcal{G}_1{X}- I_n$,
% we can simplify the expression of $\mathcal{P}$ by 
% \begin{align*}
% \mathcal{P}= \sqrt n \sum_{t=0}^1 \Big( (\mathcal{G}_1^{(t)}-m_1^{(t)})_{\bar{\bf u}_1{\bf y}_t} + ( \mathcal{G}_1^{(t)} {X})_{\bar{\bf u}_1\tilde{{\bf y}}_t} + \Big(({X}^{\top} \mathcal{G}_1{X})^{(t)}-(1+zm_2)^{(t)}\Big)_{{\bf v}_1^0{\bm \eta}_t}\Big).
% \end{align*}
% Further, by isotropic local laws, it is easy to see that 
% \begin{align}\label{est:P}
% \mathcal{P} = O_\prec(1).
% \end{align}
 
 Further we define  $\mathcal{M}\equiv\mathcal{M}(z) $ to be a $8$-by-$8$ block diagonal matrix such that $\mathcal{M}={\rm diag}(\mathcal{M}_1, \mathcal{M}_2, \mathcal{M}_3)$, and  the main-diagonal blocks $\mathcal{M}_1, \mathcal{M}_2, \mathcal{M}_3$ are all symmetric matrices with dimension $2,3,3$, respectively.  The entrywise definition of the diagonal blocks are given below.

With certain abuse of notation, in this part, let us use ${\mathcal{M}}_a(i,j)$ to denote the $(i,j)$-th entry of matrix $ \mathcal{M}_a, a=1,2,3$. For the  matrix $\mathcal{M}_1$, it is defined entrywise by 
\begin{align*}
&\mathcal{M}_1(1,1)=2r^{-\frac 32}m_1^2(zm_1)' \,,\qquad \mathcal{M}_1(1,2)=r^{-2}m_1^2(zm_1)''+ 2r^{-2}m_1m_1'(zm_1)'\,, \notag\\
&\mathcal{M}_1(2,2)= 2r^{-\frac 52}\Big( \frac{m_1^2(zm_1)'''}{3!} + m_1m_1'(zm_1)'' + (m_1')^2(zm_1)'\Big)\,.
\end{align*}
 The entries of $\mathcal{M}_2$ are given by 
 \begin{align*}
 &\mathcal{M}_2(1,1)=-\frac{m_1'(zm_2)}{r(1+\sqrt r m_1)}\,,\qquad \mathcal{M}_2(1,2)=\frac{m_1'(zm_2)}{r(1+\sqrt r m_1)} \sqrt{\frac{n_1}{n}}\,, \notag\\
&\mathcal{M}_2(1,3)=\frac{1}{2}\Big[ -\frac{m_1''(zm_2)}{r^{\frac 32}(1+\sqrt r m_1)} -\frac{m_1'(zm_2)'}{r^{\frac 32}(1+\sqrt r m_1)}+ \frac{(m_1')^2(zm_2)}{r(1+\sqrt r m_1)^2}\Big]\,,\notag\\
&\mathcal{M}_2(2,2)= -\frac{m_1'(zm_2)}{r(1+\sqrt r m_1)}\,,\notag\\
&\mathcal{M}_2(2,3)=- \frac{1}{2}\Big[ -\frac{m_1''(zm_2)}{r^{\frac 32}(1+\sqrt r m_1)} -\frac{m_1'(zm_2)'}{r^{\frac 32}(1+\sqrt r m_1)}+ \frac{(m_1')^2(zm_2)}{r(1+\sqrt r m_1)^2}\Big] \sqrt{\frac{n_1}{n}} \,,\notag\\
&\mathcal{M}_2(3,3)= -\frac{1}{r^2(1+\sqrt r m_1)} \Big( \frac{m_1'''(zm_2)}{3!} + \frac{m_1''(zm_2)'}{2}\Big) +  \frac{m_1'}{r^{\frac32}(1+\sqrt r m_1)^2} \Big( \frac{m_1''(zm_2)}{2} + m_1'(zm_2)' \Big)\,. 
 \end{align*}
Further, we define  $\mathcal{M}_3$ entrywise by 
 \begin{align*}
 &\mathcal{M}_3(1,1)=-\frac{2(zm_2)'(zm_2)}{r^{\frac 32}(1+\sqrt r m_1)}\,,\qquad\mathcal{M}_3(1,2)=\frac{2(zm_2)'(zm_2)}{r^{\frac 32}(1+\sqrt r m_1)}\sqrt{\frac{n_1}{n}} \,,\notag\\
&\mathcal{M}_3(1,3)= -\frac{(zm_2)''(zm_2)}{r^2(1+\sqrt r m_1)} + \frac{m_1'(zm_2)'(zm_2)}{r^{\frac 32}(1+\sqrt r m_1)^2} - \frac{\big((zm_2)'\big)^2}{r^2(1+\sqrt r m_1)}  \,,\notag\\
&\mathcal{M}_3(2,2)= -\frac{(zm_2)'(zm_2)}{r^{\frac 32}(1+\sqrt r m_1)}\Big(1+\frac{n_1}{n}\Big)\,,\notag\\
&\mathcal{M}_3(2,3)=\Big( -\frac{(zm_2)''(zm_2)}{r^2(1+r m_1)} + \frac{ m_1'(zm_2)'(zm_2)}{r^{\frac 32}(1+r m_1)^2} - \frac{\big((zm_2)'\big)^2}{r^2(1+r m_1)} \Big)\Big(- \sqrt{\frac{n_1}{n}}\, \Big)\,,\notag\\
&\mathcal{M}_3(3,3)=2\Big[  -\frac{1}{r^{\frac 52}(1+\sqrt r m_1)} \Big( \frac{(zm_2)'''(zm_2)}{3!} + \frac{(zm_2)''(zm_2)'}{2}\Big) \notag\\
&\qquad\qquad \qquad +  \frac{m_1'}{r^2(1+\sqrt r m_1)^2} \Big( \frac{(zm_2)''(zm_2)}{2} + \big((zm_2)' \big)^2\Big)\Big]\,. 
 \end{align*}

 Next, we set 
\begin{align}\label{def:fixed_z}
z:= {\rm i} n^{-K}
\end{align}
for some sufficiently large constant $K>0$. This setting allows us to use the high probability bounds for the quadratic forms of $\mathcal{G}_1^{a}, (z\mathcal{G}_2)^{(a)}, ({X}^{\top} \mathcal{G}_1^{a})$  for $a=0,1$,  even when we estimate their moments. To see this, first we can always bound those quadratic forms deterministically by $(\Im z)^{-s}$ for some fixed $s>0$, up to some constant. Then according to Lemma \ref{prop_prec} (ii) and  Proposition \ref{thm:locallaw} with Remark \ref{rmk:locallaw}, we get that the high probability bound in Remark \ref{rmk:locallaw} can be directly  applied in calculations of the expectations.

With all the above notations, we introduce the following proposition.
 \begin{prop}\label{prop:rmeP}
 Let $ \mathcal{P}$ be defined above and $z$ given in (\ref{def:fixed_z}).  Denote by $\varphi_n(\cdot)$  the characteristic function of $ \mathcal{P}$. Suppose that $p/n\to[0,1)$.
Then, for $|t|\ll n^{1/2}$,
\begin{align*}
\varphi_n'(t) = - \big({\bf c}^{\top} \mathcal{M} {\bf c} \big) t \varphi_n(t)  + O_\prec((|t|+1) n^{-\frac 12})\,.
\end{align*}

%  \begin{align}
%  &(i)\qquad  \E \mathcal{P} = O_\prec(n^{-\frac 12}) \notag\\
%  &(ii)\qquad  \E \mathcal{P}^l = (l-1){\bf c}^* \mathcal{M} {\bf c} \E \mathcal{P}^{l-2} + O_\prec(n^{-\frac 12}),
%  \end{align}
%  for any integer $l\geq 2$.
 \end{prop}

The proof of Proposition \ref{prop:rmeP} will be postponed. With the aid of Lemma \ref{lem:green_diff} and Proposition \ref{prop:rmeP},  we can now finish  the proof of Proposition \ref{prop.asym_dist}. 
 %Moreover, in the following proof, we also show the convergence rate of $\widetilde{\varTheta}_\alpha$ in Remark \ref{rmk:simplified SF-F}
 \begin{myPro}{(Proof of Proposition \ref{prop.asym_dist})}
 First by Proposition \ref{prop:rmeP}, we claim  that the random vector 
 \begin{align}\label{def:randomV}
& \Bigg( \frac{\sqrt{n}}{\sqrt{r}}  (\mathcal{G}_1-m_1)_{\bar{\bf u}_1\bar{\bf u}_1},\,   \frac{\sqrt {n}}{r} (\mathcal{G}_1^2- m_1')_{\bar{\bf u}_1\bar{\bf u}_1},\,  \frac{ \sqrt n}{r^{\frac14}} ( \mathcal{G}_1 {X})_{\bar{\bf u}_1\bar{\bf v}_1} ,\,  \frac{\sqrt n}{r^{\frac14}}(\mathcal{G}_1 {X})_{\bar{\bf u}_1{\bf e}_0},\, 
 \frac{\sqrt n}{r^{\frac34}} ( \mathcal{G}_1^2 {X})_{\bar{\bf u}_1\bar{\bf v}_1},\notag\\ 
  &\frac{\sqrt{n}}{\sqrt{r}} \big( z\mathcal{G}_2-zm_2\big)_{\bar{\bf v}_1\bar{\bf v}_1} ,\,  \frac{\sqrt{n}}{\sqrt{r}} (z\mathcal{G}_2-zm_2)_{\bar{\bf v}_1{\bf e}_0} ,\,  \frac{\sqrt{n}}{r} \Big((z\mathcal{G}_2)'-(zm_2)'\Big)_{\bar{\bf v}_1\bar{\bf v}_1}\Bigg)
 \end{align}
 is asymptotically Gaussian  with mean ${\bf 0}$ and covariance matrix $\mathcal{M}$ at $z=0$. To see this, we only need to claim that $\mathcal{P}$ is asymptotically normal with mean $0$ and variance ${\bf c}^{\top} \mathcal{M} {\bf c}$ due to the arbitrariness of the fixed vector $\mathbf{c}$. Let us denote by $\varphi_0(t)$ the characteristic function of standard normal distribution with mean $0$ and variance ${\bf c}^{\top} \mathcal{M} {\bf c}$ which takes the expression $\varphi_0(t)= \exp\{-\big({\bf c}^{\top} \mathcal{M} {\bf c} \big) t^2/2\}$. According to Proposition \ref{prop:rmeP}, for $|t|\ll n^{1/2}$, we have
 \begin{align*}
 \frac{\mathrm{d}}{\mathrm{d}t} \frac{\varphi_n(t)}{\varphi_0(t)} =  \frac{\varphi'_n(t)+ \big({\bf c}^{\top} \mathcal{M} {\bf c} \big) t\varphi_n(t) }{\varphi_0(t)} =O_\prec \Big(  (|t|+1) e^{\big({\bf c}^{\top} \mathcal{M} {\bf c} \big) t^2/2}n^{-\frac12}\Big)\,.
 \end{align*}
Notice  the fact ${\varphi(0)}/{\varphi_0(0)} =1$, we shall have 
 \begin{align*}
\frac {\varphi_n(t)}{\varphi_0(t)} -1= \left\{
\begin{array}{ll}
O_\prec \Big( e^{\big({\bf c}^{\top} \mathcal{M} {\bf c} \big) t^2/2}n^{-\frac12}\Big), &1< |t| \ll \sqrt n\,;\\
O_\prec (|t| n^{-\frac 12}), & |t|\leq 1\, .
\end{array}
\right. 
 \end{align*}
This further implies that 
 \begin{align}\label{est:varphi(t)}
  {\varphi_n(t)}= \varphi_0(t) + O_\prec(n^{-\frac 12}), \text{ for } 1<|t|\ll \sqrt n; \qquad 
  {\varphi_n(t)}= \varphi_0(t) + O_\prec(|t|n^{-\frac 12}), \text{ for } |t|\leq1\,.
 \end{align}
 We can then conclude the asymptotical distribution of $\mathcal{P}$.
 
Recall  the Green function representation in (\ref{eq:green_diff}). Set 
\begin{align*}
&{\varTheta_{\alpha}}:= \Bigg[\frac{(1-r)\sqrt n}{2\sqrt{   \hat {\bm{\mu}}_d^{\top} \widehat\Sigma^{-1} \hat {\bm{\mu}}_d}}\bigg(\frac{1-2r}{(1-r)^4}  \big( z\mathcal{G}_2-zm_2\big)_{{\bf v}_1{\bf v}_1}  - \frac{1}{r^{\frac 12}(1-r)^2} \Big((z\mathcal{G}_2)'-(zm_2)'\Big)_{{\bf v}_1{\bf v}_1}  \notag\\
& \qquad\qquad\qquad \qquad +  \frac{1}{r^{\frac 12}(1-r)^2} (\mathcal{G}_1-m_1)_{{\bf u}_1{\bf u}_1}- \frac{1}{r}(\mathcal{G}_1^2- m_1')_{{\bf u}_1{\bf u}_1} \notag\\
&\qquad\qquad\qquad \qquad +  \frac{2}{r^{\frac 14}(1-r)^2} ( \mathcal{G}_1 {X})_{{\bf u}_1{\bf v}_1} - \frac{2}{r^{\frac 34}(1-r)}  ( \mathcal{G}_1^2 {X})_{{\bf u}_1{\bf v}_1} \bigg) \, \Phi^{-1}(1-\alpha) \notag\\
& \qquad +  \frac{n}{\sqrt{n_0}} \Big( \frac{1}{(1-r)^2} (z\mathcal{G}_2-zm_2)_{{\bf v}_1{\bf e}_0} + \frac{1}{r^{\frac 14}(1-r)} (\mathcal{G}_1 {X})_{{\bf u}_1{\bf e}_0}\Big)\Bigg]
\Big/ \sqrt{ \Big((1-r)  \hat{\bm{\mu}}_d^{\top}  \widehat\Sigma^{-1} \hat{\bm{\mu}}_d- \frac{n^2r}{n_0n_1}\Big)} \,,
\end{align*}
which is a linear combination of the components of the  vector in (\ref{def:randomV}). Therefore by elementary calculations of the quadratic form of $\mathcal{M}$ with the identities 
\begin{align*}
&m_1(0)= \frac{\sqrt r }{1-r}, \quad m_1'(0) = \frac{r}{(1-r)^3},\quad  m_1''(0)=\frac{2r^{\frac 32}(1+r)}{(1-r)^5}, \quad m_1'''(0)= \frac{6r^2(1+3r+r^2)}{(1-r)^7}\notag\\
&\widehat m_2(0):= (zm_2(z))\Big|_{z=0}= r-1, \quad \widehat m_2'(0)= \frac{r^{\frac 32}}{1-r},\quad 
\widehat m_2''(0)= \frac{2r^2}{(1-r)^3},\quad \widehat m_2'''(0)= \frac{6r^{\frac 52}(1+r)}{(1-r)^5},
\end{align*}
together with the estimate 
\begin{align*}
\frac{{\bm{\mu}}_d^{\top} \Sigma^{-1}{\bm{\mu}}_d}{(1-r)  \hat{\bm{\mu}}_d^{\top} \widehat\Sigma^{-1} \hat{\bm{\mu}}_d- \frac{n^2r}{n_0n_1}} = 1+ O_\prec (n^{-\frac 12})
\end{align*}
which follows from Lemma \ref{lem:ests},
we can finally prove (\ref{repre:sF-F}) and the fact ${\varTheta_{\alpha}}\simeq \mathcal{N}(0, \widehat {V})$. 

In the end, we show the convergence rate of ${\varTheta_{\alpha}}$ again using Proposition \ref{prop:rmeP}. It suffices to obtain the convergence rate of the general form of linear combination, i.e. $\mathcal{P}$. We follow the derivations for  Berry-Esseen bound, more precisely, by Esseen's inequality, we have
\begin{align*}
\sup_{x\in \mathbf{R}} \big|F_n(x)- F_0(x) \big| \leq C_1 \int_{0}^T \frac{|\varphi_n(t)- \varphi_0(t)|}{t} \mathrm{d} t + \frac{C_2}{ T}
\end{align*}
for some fixed constants $C_1, C_2>0$. Here we use $F_n(x), F_0(x)$ to denote the distribution functions of $\mathcal{P}$ and centred normal distribution with variance ${\bf c}^{\top} \mathcal{M} {\bf c}$, respectively.  Applying (\ref{est:varphi(t)}), and choose $T=\sqrt n$, we then get 
\begin{align*}
\sup_{x\in \mathbf{R}} \big|F_n(x)- F_0(x) \big| \leq C_1 \int_{1}^T t^{-1} O_\prec(n^{-\frac12})  \mathrm{d} t +   C_1 \int_{0}^1 t^{-1} O_\prec(|t|n^{-\frac12})  \mathrm{d} t +  C_2 n^{-\frac 12}= O_\prec (n^{-\frac 12})\,.
\end{align*}
This indicates that the convergence rate of $\mathcal{P}$ is $O_\prec (n^{-\frac 12})$, and hence the same rate applies to ${\varTheta_{\alpha}}$.
 \end{myPro}
 
 \begin{myRem}
 The arguments of the  convergence rate of $\widetilde{\varTheta}_\alpha$ of  Remark  \ref{rmk:simplified SF-F}, which leads to the high probability bound in Corollary \ref{thm:prop} is actually the same, since $\widetilde{\varTheta}_\alpha$ again takes the form of $\mathcal{P}$ with appropriate $\bf c$.
 \end{myRem}

\subsection{Proofs of Lemma  \ref{lem:green_diff} and Proposition \ref{prop:rmeP}}
In the last subsection, we prove the technical results from Section \ref{subsec:21051501}, i.e.,  Lemma  \ref{lem:green_diff} and Proposition \ref{prop:rmeP}.

\begin{myPro}{(Proof of Lemma \ref{lem:green_diff})}

Recall the definitions of $\widehat{F}(\widehat\Sigma, \hat{\bm{\mu}}^0)$ and $F(\Sigma, {\bm{\mu}}^0)$  in (\ref{def:whF}) and (\ref{def:F}).  In light of Lemma \ref{lem:ests}, it suffices to further identify  the differences $\frac{1}{(1-r)^2} \widehat A^{\top}\widehat\Sigma \widehat A - \widehat A^{\top}\Sigma \widehat A$ and $\widehat{A} ^{\top} \hat{\bm{\mu}}^0- \sqrt{\frac{n}{n_0}} \frac{r}{1-r} {\bf v}_1^{\top} {\bf e}_0 - \widehat{A} ^{\top} {\bm{\mu}}^0$. We start with the first term.  We write
%By (\ref{def:rewrite AhSigmaA}) and (\ref{def:rewrite ASigmaA}),
%By the two  estimates (\ref{est:qua_Sigma}) and  (\ref{est:qua_hSigma}), together with the identities
%\begin{align*}
%\Vert {\bf v}_1\Vert^2= \sum_{i=0,1}({\bf v}_1^{\top} {\bf e}_i)^2= \frac{n}{n_0} + \frac{n}{n_1}
%\end{align*}
%which can be check by definitions of ${\bf v}_1, {\bf e}_0, {\bf e}_1$ in (\ref{def:v1 X^E}) and (\ref{def:e_0e_1}), we can come up with an approximation of $\widehat A^{\top}\Sigma \widehat A$ by 
%\begin{align*}
%\frac{1}{(1-y)^2}\widehat A^{\top}\widehat\Sigma \widehat A
%\end{align*}
%since it is not hard to check the leading term of $\widehat A^{\top}\Sigma \widehat A$ coincides to that of the proposed linear transformation of $\widehat A^{\top}\widehat\Sigma \widehat A$. In the sequel, we further capture their difference and determine its asymptotic distribution and even explicit order of the errors.
\begin{align*}
 \frac{1}{(1-r)^2} \widehat A^{\top}\widehat\Sigma \widehat A - \widehat A^{\top}\Sigma \widehat A
&=\Big[\frac{n-2}{n(1-r)^2}  {\bf v}_1^{\top} {X}^{\top} H_E^{-1}  {X}{\bf v}_1 - \Big(\frac{n-2}{n}\Big)^2  r^{-\frac 12} {\bf v}_1^{\top} {X}^{\top} H_E^{-2}  {X}{\bf v}_1\Big] \notag\\
&\quad + \Big[\frac{n-2}{n(1-r )^2\sqrt r}  {\bf u}_1^{\top} H_E^{-1} {\bf u}_1 - \Big(\frac{n-2}{n\sqrt r}\Big)^2  {\bf u}_1^{\top} H_E^{-2} {\bf u}_1 \Big]\notag\\
&\quad + 2\Big[\frac{n-2}{n(1-r)^2r^{\frac 14}}  {\bf v}_1^{\top} {X}^{\top} H_E^{-1} {\bf u}_1 - \Big(\frac{n-2}{n}\Big)^2 r^{-\frac 34}  {\bf v}_1^{\top} {X}^{\top} H_E^{-2} {\bf u}_1 \Big]\notag\\
&=: D_1+D_2+D_3\,,
\end{align*}
in which we used (\ref{def:hSigma_matrix}), (\ref{def:rewrite AhSigmaA}), (\ref{def:rewrite ASigmaA}), and the shorthand notation ${\bf u}_1=\Sigma^{-\frac 12} {\bm{\mu}}_d $. In the sequel, we estimate $D_1, D_2, D_3$ term by term. Before we commence the details,  we first continue with (\ref{inverM_dim2}) to seek for the explicit form of one higher order term by resolvent expansion formula, \begin{align}\label{2020092801}
 \mathcal{I}_2^{-1} = 
\frac{1}{1-r} \, I_2 + \frac{1}{(1-r)^2} {\bf \Delta} + O_\prec(n^{-1}r)\,,
\end{align}
where 
\begin{align*}
{\bf \Delta} = \Big({E}^{\top} \big(z\mathcal{G}_2(z)- zm_2(z)I_p\big) {E}\Big)\Big|_{z=0}\,,
\end{align*}
and $\Vert {\bf \Delta}\Vert= O_\prec(n^{-\frac 12}r^{\frac 12})$ by (\ref{est.DG}).
Here in (\ref{2020092801}) $O_\prec(n^{-1}r)$ represents an error matrix which  is stochastically bounded by $r/n$ in operator norm.  We remark here that the above estimate will be frequently used in the following calculations.

Let us start with $D_1$. Similarly to (\ref{inverseSigma}), by applying Woodbury matrix identity, we get 
\begin{align*}
D_1&=\frac{n-2}{n(1-r)^2} \Big( {\bf v}_1^{\top} {X}^{\top} \mathcal{G}_1{X}{\bf v}_1 + {\bf v}_1^{\top} {X}^{\top} \mathcal{G}_1{X}{E}\mathcal{I}_2^{-1}{E}^{\top} {X}^{\top} \mathcal{G}_1 {X}{\bf v}_1\Big) \notag\\
&\quad -\Big(\frac{n-2}{n}\Big)^2 r^{-\frac 12} \Big( {\bf v}_1^{\top} {X}^{\top} \mathcal{G}_1^2{X}{\bf v}_1 + 2{\bf v}_1^{\top} {X}^{\top} \mathcal{G}_1^2{X}{E} \mathcal{I}_2^{-1}{E}^{\top} {X}^{\top} \mathcal{G}_1{X}{\bf v}_1\Big) \notag\\
&\quad - \Big(\frac{n-2}{n}\Big)^2 r^{-\frac 12} {\bf v}_1^{\top} {X}^{\top} \mathcal{G}_1{X}{E} \mathcal{I}_2^{-1}{E}^{\top} {X}^{\top} \mathcal{G}_1^2 {X}{E} \mathcal{I}_2^{-1}{E}^{\top} {X}^{\top} \mathcal{G}_1{X}{\bf v}_1\,.
\end{align*} 
Hereafter, for brevity,  we drop the $z$-dependence  from the notations $\mathcal{G}_1(z)$, $\mathcal{G}_2(z)$ and $m_1(z), m_2 (z)$ and set $z=0$ but omit this fact from the notations. 
Recall (\ref{eq:est_T1}) and (\ref{20201230002}). Analogously, we can compute 
\begin{align} \label{est:D_1}
D_1&= \frac{1}{(1-r)^2} {\bf v}_1^{\top}\big( z\mathcal{G}_2-zm_2\big){\bf v}_1
-r^{-\frac 12}{\bf v}_1^{\top}\Big( \big(z\mathcal{G}_2\big)' - (zm_2)'\Big){\bf v}_1 + \frac{2(1+zm_2)}{(1-r)^3}{\bf v}_1^{\top} ( z\mathcal{G}_2- zm_2) {E}{E}^{\top} {\bf v}_1\notag\\
&\quad + \frac{(1+zm_2)^2}{(1-r)^4}{\bf v}_1^{\top} {E}{E}^{\top} (z\mathcal{G}_2 - zm_2(z)){E}{E}^{\top}{\bf v}_1 - \frac{2(1+zm_2)}{(1-r)\sqrt r}{\bf v}_1^{\top} \Big((z\mathcal{G}_2)'-(zm_2)'\Big) {E}{E}^{\top} {\bf v}_1 \notag\\
&\quad -\frac{2(zm_2)'}{(1-r)\sqrt r}{\bf v}_1^{\top}  {E}{E}^{\top} (z\mathcal{G}_2-zm_2){\bf v}_1 -  \frac{2(zm_2)'(1+zm_2)}{(1-r)^2\sqrt r}{\bf v}_1^{\top} {E}{E}^{\top} (z\mathcal{G}_2 - zm_2(z)){E}{E}^{\top}{\bf v}_1\notag\\
&\quad - \frac{2(1+zm_2)(zm_2)'}{(1-r)^2\sqrt r}{\bf v}_1^{\top} ( z\mathcal{G}_2-zm_2) {E}{E}^{\top} {\bf v}_1 -\frac{(1+zm_2)^2}{(1-r)^2\sqrt r}{\bf v}_1^{\top}  {E}{E}^{\top} \Big((z\mathcal{G}_2 )'- (zm_2)'\Big){E}{E}^{\top}{\bf v}_1 \notag\\
&\quad - \frac{2(1+zm_2)^2(zm_2)'}{(1-r)^3\sqrt r}  {\bf v}_1^{\top} {E}{E}^{\top} \big( z\mathcal{G}_2 -zm_2(z)\big)  {E}{E}^{\top}{\bf v}_1+ O_\prec(n^{-1}r)\notag\\
&=  \frac{1-2r}{(1-r)^4}{\bf v}_1^{\top} \big( z\mathcal{G}_2-zm_2\big){\bf v}_1
-\frac{1}{(1-r)^2\sqrt r}{\bf v}_1^{\top}\Big( \big(z\mathcal{G}_2\big)' - (zm_2)'\Big){\bf v}_1+O_\prec(n^{-1}r)\,.
\end{align}
%Here we emphasize that among the above derivations, we always omit the $z$-dependence for $\mathcal{G}_{1,2}$ and also $m_2$. Additionally, we take $z=0$. It is easy to observe that $D_1$ is represented by linear combinations of quadratic forms of Green function. And further implied by Lemma \ref{lem:ests} and Remark \ref{rmk:ests} the leading order of $D_1$ is roughly $O_\prec(n^{-1/2}r^{1/2})$.
%

Next, we turn to estimate $D_2$. Similarly to $D_1$, by Woodbury matrix identity, we have
\begin{align*}
D_2&= \frac{n-2}{n(1-r)^2\sqrt r} \Big( {\bf u}_1^{\top} \mathcal{G}_1 {\bf u}_1 + {\bf u}_1^{\top}  \mathcal{G}_1{X}{E} \mathcal{I}_2^{-1}{E}^{\top} {X}^{\top} \mathcal{G}_1 {\bf u}_1\Big) \notag\\
&\quad -\Big(\frac{n-2}{n\sqrt r}\Big)^2  \Big( {\bf u}_1^{\top}  \mathcal{G}_1^2{\bf u}_1 + 2{\bf u}_1^{\top}  \mathcal{G}_1^2{X}{E} \mathcal{I}_2^{-1}{E}^{\top} {X}^{\top} \mathcal{G}_1{\bf u}_1\Big) \notag\\
&\quad - \Big(\frac{n-2}{n\sqrt r}\Big)^2 {\bf u}_1^{\top} \mathcal{G}_1{X}{E} \mathcal{I}_2^{-1}{E}^{\top} {X}^{\top} \mathcal{G}_1^2 {X}{E} \mathcal{I}_2^{-1}{E}^{\top} {X}^{\top} \mathcal{G}_1{\bf u}_1\,.
\end{align*}
Then, by  (\ref{2020092801}), it is not hard to derive that {\small
\begin{align}\label{est:D_2}
&D_2= \frac{r^{-\frac 12}}{(1-r)^2} \Big( {\bf u}_1^{\top}\mathcal{G}_1{\bf u}_1 + \frac{1}{1-r} \sum_{i=0}^1  ({\bf u}_1^{\top}\mathcal{G}_1{X}{\bf e}_i) ^2 + \frac{1}{(1-r)^2} \sum_{i,j=0}^1  ({\bf u}_1^{\top}\mathcal{G}_1{X}{\bf e}_i) ({\bf u}_1^{\top}\mathcal{G}_1{X}{\bf e}_j)\big({\bf e}_i^{\top}(z\mathcal{G}_2-zm_2){\bf e}_j\big)\Big) \notag\\
&\quad - \frac 1r\Big( {\bf u}_1^{\top}\mathcal{G}_1^2{\bf u}_1 +  \frac{2}{1-r} \sum_{i=0}^1  ({\bf u}_1^{\top}\mathcal{G}_1{X}{\bf e}_i) ({\bf u}_1^{\top}\mathcal{G}_1^2{X}{\bf e}_i)+ \frac{1}{(1-r)^2} \sum_{i,j=0}^1  ({\bf u}_1^{\top}\mathcal{G}_1^2{X}{\bf e}_i) ({\bf u}_1^{\top}\mathcal{G}_1{X}{\bf e}_j)\big({\bf e}_i^{\top}(z\mathcal{G}_2-zm_2){\bf e}_j\big)\Big)\notag\\
&\quad -\frac{1}{(1-r)^2r} \sum_{i,j=0}^1  ({\bf u}_1^{\top}\mathcal{G}_1{X}{\bf e}_i) ({\bf u}_1^{\top}\mathcal{G}_1{X}{\bf e}_j)\big({\bf e}_i^{\top}(z\mathcal{G}_2)'{\bf e}_j\big)\notag\\
&\quad -
\frac{2}{(1-r)^3r}\sum_{i,j,k=0}^1  ({\bf u}_1^{\top}\mathcal{G}_1{X}{\bf e}_i) ({\bf u}_1^{\top}\mathcal{G}_1{X}{\bf e}_k) \big( {\bf e}_i^{\top}(z\mathcal{G}_2)'{\bf e}_j\big) \big({\bf e}_j^{\top}(z\mathcal{G}_2-zm_2){\bf e}_k\big) +O_\prec(n^{-1} \Vert {\bf u}_1 \Vert^2) \notag\\
&\quad = \frac{1}{(1-r)^2\sqrt r} {\bf u}_1^{\top}(\mathcal{G}_1-m_1){\bf u}_1 -r^{-1} {\bf u}_1^{\top}(\mathcal{G}_1^2- m_1'){\bf u}_1 +O_\prec(n^{-1}\Vert {\bf u}_1 \Vert^2)\,,
\end{align}
}
where in the last step we applied the estimate ${\bf u}_1^{\top}\mathcal{G}_1^a{X}{\bf e}_i = O_\prec(n^{-1/2}r^{1/4+(a-1)/2}\Vert {\bf u}_1\Vert), a=1,2$ and  ${\bf e}_i^{\top}(z\mathcal{G}_2)'{\bf e}_j = O_\prec(r^{3/2})$ which follow from (\ref{081501}) and (\ref{est.DG}). Further, we also used the trivial identity $m_1(0)/(1-r)^2= m_1'(0)$. 

Next, we estimate $D_3$ as follows,
\begin{align*}
D_3&=\frac{2(n-2)}{n(1-r)^2r^{\frac 14}} \Big( {\bf v}_1^{\top} {X}^{\top} \mathcal{G}_1{\bf u}_1 + {\bf v}_1^{\top} {X}^{\top} \mathcal{G}_1{X}{E} \mathcal{I}_2^{-1}{E}^{\top} {X}^{\top} \mathcal{G}_1 {\bf u}_1\Big) \notag\\
&\quad -2\Big(\frac{n-2}{n}\Big)^2  r^{-\frac 34}\Big( {\bf v}_1^{\top} {X}^{\top} \mathcal{G}_1^2{\bf u}_1 + {\bf v}_1^{\top} {X}^{\top} \mathcal{G}_1^2{X}{E} \mathcal{I}_2^{-1}{E}^{\top} {X}^{\top} \mathcal{G}_1{\bf u}_1 + {\bf v}_1^{\top} {X}^{\top} \mathcal{G}_1{X}{E} \mathcal{I}_2^{-1}{E}^{\top} {X}^{\top} \mathcal{G}_1^2{\bf u}_1\Big) \notag\\
&\quad -  2\Big(\frac{n-2}{n}\Big)^2 r^{-\frac 34} {\bf v}_1^{\top} {X}^{\top} \mathcal{G}_1{X}{E} \mathcal{I}_2^{-1}{E}^{\top} {X}^{\top} \mathcal{G}_1^2 {X}{E} \mathcal{I}_2^{-1}{E}^{\top} {X}^{\top} \mathcal{G}_1{\bf u}_1\notag\\
&= \frac{2}{(1-r)^2r^{\frac 14}} \Big(  {\bf v}_1^{\top} {X}^{\top} \mathcal{G}_1{\bf u}_1+ \frac{1}{1-r}{\bf v}_1^{\top} {X}^{\top} \mathcal{G}_1{X}{E} {E}^{\top} {X}^{\top} \mathcal{G}_1 {\bf u}_1 \Big) \notag\\
 &\quad -  2  r^{-\frac 34}\Big( {\bf v}_1^{\top} {X}^{\top} \mathcal{G}_1^2{\bf u}_1+ \frac{1}{1-r}{\bf v}_1^{\top} {X}^{\top} \mathcal{G}_1^2{X}{E} {E}^{\top} {X}^{\top} \mathcal{G}_1 {\bf u}_1+ \frac{1}{1-r}{\bf v}_1^{\top} {X}^{\top} \mathcal{G}_1{X}{E} {E}^{\top} {X}^{\top} \mathcal{G}_1 ^2{\bf u}_1 \Big) \notag\\
 &\quad - \frac{2}{(1-r)^2}  r^{-\frac 34}{\bf v}_1^{\top} {X}^{\top} \mathcal{G}_1{X}{E}{E}^{\top} {X}^{\top} \mathcal{G}_1^2 {X}{E} {E}^{\top} {X}^{\top} \mathcal{G}_1{\bf u}_1 
 + O_\prec(n^{-1}r^{\frac 12} \Vert {\bf u}_1 \Vert )\,.
 \end{align*}
Further, by   (\ref{est.DG}), (\ref{081501}), and (\ref{2021051401}), we have
 \begin{align}\label{est:D_3}
D_3
%&= \frac{2}{(1-r)^2r^{\frac 14}} \Big( (\mathcal{G}_1{X})_{{\bf u}_1{\bf v}_1}+ \frac{r}{1-r}\sum_{i=0}^1{\bf v}_1^{\top} {\bf e}_i ( \mathcal{G}_1 {X})_{{\bf u}_1{\bf e}_i} \Big)\notag\\
%&\quad  - 2 r^{-\frac 34} \Big(  (\mathcal{G}_1^2{X})_{{\bf u}_1{\bf v}_1}+ \frac{r^{\frac 32}}{(1-r)^2} \sum_{i=0}^1{\bf v}_1^{\top} {\bf e}_i ( \mathcal{G}_1 {X})_{{\bf u}_1{\bf e}_i} + \frac{r}{(1-r)} \sum_{i=0}^1{\bf v}_1^{\top} {\bf e}_i ( \mathcal{G}_1^2 {X})_{{\bf u}_1{\bf e}_i}\Big)
%  \notag\\
%  &\quad - \frac{2r^{\frac 74}}{(1-r)^3} \sum_{i=0}^1{\bf v}_1^{\top} {\bf e}_i ( \mathcal{G}_1 {X})_{{\bf u}_1{\bf e}_i}  + O_\prec(n^{-1}\Vert {\bm{\mu}}_d\Vert)\nonumber\\
  &= \frac{2r^{-\frac 14}}{(1-r)^2}  {\bf v}_1^{\top} {X}^{\top} \mathcal{G}_1{\bf u}_1 - \frac{2r^{-\frac 34}}{1-r}{\bf v}_1^{\top} {X}^{\top} \mathcal{G}_1^2{\bf u}_1  + O_\prec(n^{-1}r^{\frac 12} \Vert {\bf u}_1\Vert)\,.
 \end{align}
%Here in the last step, we used  ${\bf v}_1= -\sqrt{n/n_0} {\bf e}_0+ \sqrt{n/n_1} {\bf e}_1$ and ${\bf v}_1^{\top} {\bf e}_0= -\sqrt{n/n_0}, {\bf v}_1^{\top} {\bf e}_1=\sqrt{n/n_1}$ by definition. 
Combining (\ref{est:D_1}), (\ref{est:D_2}) and (\ref{est:D_3}), we  conclude that 
\begin{align}\label{est:Repre_1}
&\quad \frac{1}{(1-r)^2} \widehat A^{\top}\widehat\Sigma \widehat A - \widehat A^{\top}\Sigma \widehat A\notag\\
%&=  \frac{1}{(1-r)^2} \big( z\mathcal{G}_2-zm_2\big)_{{\bf v}_1{\bf v}_1} 
%-\Big( \big(z\mathcal{G}_2\big)' - (zm_2)'\Big)_{{\bf v}_1{\bf v}_1} -\frac{2r}{1-r} \Big((z\mathcal{G}_2)'-(zm_2)'\Big)_{{\bf v}_1{\bf v}_1} \notag\\
%&\quad - \frac{r^2}{(1-r)^4} \big(z\mathcal{G}_2-zm_2\big)_{{\bf v}_1{\bf v}_1} -\frac{r^2}{(1-r)^2}\Big((z\mathcal{G}_2)'-(zm_2)'\Big)_{{\bf v}_1{\bf v}_1} \notag\\
%& \quad +  \frac{1}{(1-r)^2} (\mathcal{G}_1-m_1)_{{\bf u}_1{\bf u}_1} - (\mathcal{G}_1^2- m_1')_{{\bf u}_1{\bf u}_1} +  \frac{2}{(1-r)^2} ( \mathcal{G}_1 {X})_{{\bf u}_1{\bf v}_1} - \frac{2}{1-r}  ( \mathcal{G}_1^2 {X})_{{\bf u}_1{\bf v}_1} +O_\prec(n^{-1}\Vert {\bm{\mu}}_d\Vert^2)\notag\\
&= \frac{1-2r}{(1-r)^4} {\bf v}_1^{\top} \big( z\mathcal{G}_2-zm_2\big){\bf v}_1  - \frac{r^{-\frac 12}}{(1-r)^2} {\bf v}_1^{\top}\Big((z\mathcal{G}_2)'-(zm_2)'\Big){\bf v}_1 +  \frac{r^{-\frac 12}}{(1-r)^2} {\bf u}_1^{\top}(\mathcal{G}_1-m_1){\bf u}_1  \notag\\
& \quad - r^{-1}{\bf u}_1^{\top}(\mathcal{G}_1^2- m_1'){\bf u}_1 +  \frac{2r^{-\frac 14}}{(1-r)^2}{\bf u}_1^{\top}  \mathcal{G}_1 {X}{\bf v}_1- \frac{2r^{-\frac 34}}{1-r}  {\bf u}_1^{\top} \mathcal{G}_1^2 {X}{\bf v}_1+ O_\prec\big(n^{-1}(\Vert {\bf u}_1 \Vert^2+r)\big)\, .
\end{align}
Then,  expanding  $\sqrt{\widehat A^{\top}\Sigma \widehat A}$ around $\sqrt{ \widehat A^{\top}\widehat\Sigma \widehat A }/(1-r)$, we  finally obtain 
\begin{align*}
&\frac{\sqrt{ \widehat A^{\top}\widehat\Sigma \widehat A }}{1-r} - \sqrt{\widehat A^{\top}\Sigma \widehat A} = \frac{1-r}{2\sqrt{\widehat A^{\top}\widehat\Sigma \widehat A }} \bigg( \frac{1-2r}{(1-r)^4}  {\bf v}_1^{\top}\big( z\mathcal{G}_2-zm_2\big){\bf v}_1  - \frac{r^{-\frac12}}{(1-r)^2} {\bf v}_1^{\top}\Big((z\mathcal{G}_2)'-(zm_2)'\Big){\bf v}_1 \notag\\
& \quad +  \frac{r^{-\frac12}}{(1-r)^2} {\bf u}_1^{\top}(\mathcal{G}_1-m_1){\bf u}_1 - r^{-1}{\bf u}_1^{\top}(\mathcal{G}_1^2- m_1'){\bf u}_1 +  \frac{2r^{-\frac14}}{(1-r)^2} {\bf u}_1^{\top}\mathcal{G}_1 {X}{\bf v}_1 - \frac{2r^{-\frac 34}}{1-r}  {\bf u}_1^{\top}\mathcal{G}_1^2 {X}{\bf v}_1\bigg) \notag\\
& \quad + O_\prec\big(n^{-1}( r^{\frac 12} + \Vert {\bf u}_1 \Vert )\big)\,.
\end{align*}

Next, analogously, we have 
\begin{align*}
&\quad \widehat{A} ^{\top} \hat{\bm{\mu}}^0- \sqrt{\frac{n}{n_0}} \frac{r}{1-r} {\bf v}_1^{\top} {\bf e}_0 - \widehat{A} ^{\top} {\bm{\mu}}^0\notag\\
&= \frac{n-2}{\sqrt{nn_0}} \Big( {\bf v}_1^{\top} {X}^{\top} \mathcal{G}_1  {X}{\bf e}_0+ {\bf v}_1^{\top} {X}^{\top} \mathcal{G}_1{X}{E} \mathcal{I}_2^{-1}{E}^{\top} {X}^{\top} \mathcal{G}_1 {X}{\bf e}_0\Big)\notag\\
&\quad +  \frac{n-2}{\sqrt{nn_0}} r^{-\frac 14}\Big( {\bf u}_1^{\top}  \mathcal{G}_1  {X}{\bf e}_0+ {\bf u}_1^{\top} \mathcal{G}_1{X}{E} \mathcal{I}_2^{-1}{E}^{\top} {X}^{\top} \mathcal{G}_1 {X}{\bf e}_0\Big) - \sqrt{\frac{n}{n_0}} \frac{r}{1-r} {\bf v}_1^{\top} {\bf e}_0\notag\\
&= \sqrt{\frac{n}{n_0}} \Big({\bf v}_1^{\top}(z\mathcal{G}_2-zm_2){\bf e}_0 + {\bf v}_1^{\top} {X}^{\top} \mathcal{G}_1{X}{E} \mathcal{I}_2^{-1}{E}^{\top} {X}^{\top} \mathcal{G}_1 {X}{\bf e}_0 - \frac{r^2}{1-r} {\bf v}_1^{\top} {\bf e}_0\Big)\notag\\
&\quad + \sqrt{\frac{n}{n_0}}r^{-\frac 14}\Big( {\bf u}_1^{\top}  \mathcal{G}_1  {X}{\bf e}_0+ {\bf u}_1^{\top} \mathcal{G}_1{X}{E}\mathcal{I}_2^{-1}{E}^{\top} {X}^{\top} \mathcal{G}_1 {X}{\bf e}_0\Big) + O_\prec(n^{-\frac 12} n_0^{-\frac 12}r^{\frac 12} ( r^{\frac 12} + \Vert {\bf u}_1\Vert) )\,.
\end{align*}
Employing (\ref{2020092801}) and the estimates  (\ref{est.DG}), (\ref{081501}) with (\ref{2021051401}), we can further get that 
%Then, by (\ref{2020092801}), we can further estimate 
%\begin{align*}
%&\quad{\bf v}_1^{\top} {X}^{\top} \mathcal{G}_1{X}{E} \big( I_2- {E}^{\top} {X}^{\top} \mathcal{G}_1{X}{E}\big)^{-1}{E}^{\top} {X}^{\top} \mathcal{G}_1 {X}{\bf e}_0 - \frac{r^2}{1-r} {\bf v}_1^{\top} {\bf e}_0 \notag\\
%&= \frac{1}{1-r} {\bf v}_1^{\top} {X}^{\top} \mathcal{G}_1{X}{E} {E}^{\top} {X}^{\top} \mathcal{G}_1 {X}{\bf e}_0  - \frac{r^2}{1-r} {\bf v}_1^{\top} {\bf e}_0 \notag\\
%&\quad + \frac{1}{(1-r)^2} {\bf v}_1^{\top} {X}^{\top} \mathcal{G}_1{X}{E} {E}^{\top}\big(z\mathcal{G}_2- zm_2\big){E} {E}^{\top} {X}^{\top} \mathcal{G}_1 {X}{\bf e}_0 + O_\prec(n^{-1}r) \notag\\
%&=\frac{2r-r^2}{(1-r)^2}(z\mathcal{G}_2-zm_2)_{{\bf v}_1{\bf e}_0} + O_\prec(n^{-1}r),
%\end{align*}
%and 
%\begin{align*}
%&\quad {\bf u}_1^{\top} \mathcal{G}_1{X}{E} \big( I_2- {E}^{\top} {X}^{\top} \mathcal{G}_1{X}{E}\big)^{-1}{E}^{\top} {X}^{\top} \mathcal{G}_1 {X}{\bf e}_0\notag\\
%&= \frac{1}{1-r}{\bf u}_1^{\top} \mathcal{G}_1{X}{E}{E}^{\top} {X}^{\top} \mathcal{G}_1 {X}{\bf e}_0 +\frac{1}{(1-r)^2} {\bf u}_1^{\top} \mathcal{G}_1{X}{E} {E}^{\top} (z\mathcal{G}_2-zm_2){E} {E}^{\top} {X}^{\top} \mathcal{G}_1 {X}{\bf e}_0 + O_\prec(n^{-1} \Vert {\bm{\mu}}_d\Vert r)\notag\\
%&= \frac{r}{1-r} (\mathcal{G}_1 {X})_{{\bf u}_1{\bf e}_0} + O_\prec(n^{-1}\Vert {\bm{\mu}}_d\Vert r^{\frac 34})
%\notag\\
%&=  \frac{r}{1-r} (\mathcal{G}_1 {X})_{{\bf u}_1{\bf e}_0} + O_\prec(n^{-1}\Vert {{\bm{\mu}}}_d\Vert  r^{\frac 34}).
%\end{align*}
%In a summary, we get 
\begin{align}\label{repre:Sampled-Amu}
&\quad \widehat{A} ^{\top} \hat{\bm{\mu}}^0- \sqrt{\frac{n}{n_0}} \frac{r}{1-r} {\bf v}_1^{\top} {\bf e}_0 - \widehat{A} ^{\top} {\bm{\mu}}^0\notag\\
&= \sqrt{\frac{n}{n_0}} \Big( \frac{1}{(1-r)^2} {\bf v}_1^{\top}(z\mathcal{G}_2-zm_2){\bf e}_0+ \frac{r^{-\frac 14}}{1-r} {\bf u}_1^{\top}\mathcal{G}_1 {X}{\bf e}_0\Big)+   O_\prec \big(n^{-\frac 12} n_0^{-\frac 12} r^{\frac 12} ( r^{\frac 12} + \Vert {\bf u}_1\Vert)\big)\,.
\end{align}
In light of   (\ref{repre:Sampled-Amu}) and (\ref{est:Repre_1}), together with the fact $\Vert {\bf u}_1 \Vert^2 = \varDelta_d$, we can now conclude the proof of Lemma \ref{lem:green_diff}.
\end{myPro}

In the sequel, we state the proof of Proposition \ref{prop:rmeP} which  will rely on  Gaussian integration by parts. For simplicity, we always drop $z$-dependence  from the notations $\mathcal{G}_1(z)$, $\mathcal{G}_2(z)$ and $m_1(z), m_2 (z)$. We also fix the choice of $z$ in  (\ref{def:fixed_z}) and omit this fact from the notations.

 Recall the definition of $\mathcal{P}$ in (\ref{def:mP_ori}). For brevity, we introduce the shorthand notation
 \begin{align}\label{def:y_ty_eta}
& {\bf y}_0:= c_{10}\bar{\bf u}_1, \quad {\bf y}_1:= c_{11} \bar{\bf u}_1, \quad \tilde{{\bf y}}_0:=c_{12} \bar{\bf v}_1 + c_{13}{\bf e}_0,\quad  \tilde{{\bf y}}_1:= c_{14} \bar{\bf v}_1,\notag\\
 & {\bm \eta}_0:= c_{20}\bar{\bf v}_1+ c_{21} {\bf e}_0,\quad {\bm \eta}_1:= c_{22}\bar{\bf v}_1\,.
 \end{align}
Using the basic identity $z\mathcal{G}_2= {X}^{\top} \mathcal{G}_1{X}- I_n$,
 we can simplify the expression of $\mathcal{P}$ in (\ref{def:mP_ori}) to 
 \begin{align}\label{20210205}
 \mathcal{P}=&\sqrt n \sum_{t=0}^1 \Big( r^{-\frac{1+t}{2}}(\mathcal{G}_1^{(t)}-m_1^{(t)})_{\bar{\bf u}_1{\bf y}_t} + r^{-\frac{1+2t}{4}}( \mathcal{G}_1^{(t)} {X})_{\bar{\bf u}_1\tilde{{\bf y}}_t} \notag\\
 &\qquad\qquad+  r^{-\frac{1+t}{2}}\Big(({X}^{\top} \mathcal{G}_1{X})^{(t)}-(1+zm_2)^{(t)}\Big)_{\bar{\bf v}_1{\bm \eta}_t}\Big)\,.
 \end{align}
 Further, by Proposition \ref{thm:locallaw} and Remark \ref{rmk:locallaw}, it is easy to see 
 \begin{align}\label{est:P}
 \mathcal{P} = O_\prec(1)\,.
 \end{align}

 Using the identity 
 \begin{align*}
 \mathcal{G}_1^t = z^{-1} (H\mathcal{G}_1^t - \mathcal{G}_1^{t-1})\,, \qquad t=1,2\,,
 \end{align*}

 we can rewrite
 \begin{align}\label{eq:Rewrite1}
  &\sqrt n \sum_{t=0,1}  r^{-\frac{1+t}{2}}  (\mathcal{G}_1^{(t)}-m_1^{(t)})_{\bar{\bf u}_1{\bf y}_t}  \notag\\
 &=\frac{\sqrt n}{ r}\Bigg( \frac{1}{(1+ r^{-\frac 12}m_2)z}({H}\mathcal{G}_1^2)_{\bar{\bf u}_1{\bf y}_1} + \frac{r^{-\frac 12}m_2}{1+ r^{-\frac 12}m_2}( \mathcal{G}_1^2)_{\bar{\bf u}_1{\bf y}_1} \notag\\
 &\qquad+ \Big(\frac{r^{-\frac 12}(zm_2)'}{(1+ r^{-\frac 12}m_2)z}-  \frac{r^{-\frac 12}(zm_2)'+1 }{(1+ r^{-\frac 12}m_2)z}\Big) (\mathcal{G}_1) _{\bar{\bf u}_1{\bf y}_1} - m_1' \big(\bar{\bf u}_1\big)^{\top} {\bf y}_1 \Bigg) \notag\\
 &\quad+ \frac{\sqrt n}{\sqrt r}\Bigg( \frac{1}{(1+ r^{-\frac 12}m_2)z}({H}\mathcal{G}_1)_{\bar{\bf u}_1{\bf y}_0} + \frac{r^{-\frac 12}m_2}{1+ r^{-\frac 12}m_2}( \mathcal{G}_1)_{\bar{\bf u}_1{\bf y}_0} - \frac{1}{(1+ r^{-\frac 12}m_2)z} \big(\bar{\bf u}_1\big)^{\top} {\bf y}_0 - m_1 \big(\bar{\bf u}_1\big)^{\top} {\bf y}_0\Bigg)\notag\\
 &= \frac{\sqrt n}{r}\Bigg( \frac{1}{(1+ r^{-\frac 12}m_2)z}({H}\mathcal{G}_1^2)_{\bar{\bf u}_1{\bf y}_1} + \frac{r^{-\frac 12}m_2}{1+ r^{-\frac 12}m_2}( \mathcal{G}_1^2)_{\bar{\bf u}_1{\bf y}_1} + \frac{r^{-\frac 12}(zm_2)'}{(1+ r^{-\frac 12}m_2)z} (\mathcal{G}_1) _{\bar{\bf u}_1{\bf y}_1} \Bigg)\notag\\
 &\quad +\frac{m_1'}{m_1}\frac{\sqrt n}{r} \Bigg( \frac{1}{(1+ r^{-\frac 12}m_2)z}({H}\mathcal{G}_1)_{\bar{\bf u}_1{\bf y}_1} + \frac{r^{-\frac 12}m_2}{1+ r^{-\frac 12}m_2}( \mathcal{G}_1)_{\bar{\bf u}_1{\bf y}_1} \Bigg)\notag\\
 &\quad 
 + \frac{\sqrt n}{\sqrt r} \Bigg( \frac{1}{(1+ r^{-\frac 12}m_2)z}({H}\mathcal{G}_1)_{\bar{\bf u}_1{\bf y}_0} + \frac{r^{-\frac 12}m_2}{1+ r^{-\frac 12}m_2}( \mathcal{G}_1)_{\bar{\bf u}_1{\bf y}_0} \Bigg)\notag\\
 &=: \mathds{T}_{11} + \mathds{T}_{12}  + \mathds{T}_{13}\,.
 \end{align}

 Here we used the first and last identities in (\ref{identitym1m2}) to gain some cancellations. Particularly, from first step to second step, we also do the derivation 
 \begin{align*}
 -  \frac{r^{-\frac 12}(zm_2)'+1 }{(1+ r^{-\frac 12}m_2)z}(\mathcal{G}_1) _{\bar{\bf u}_1{\bf y}_1} - m_1' \big(\bar{\bf u}_1\big)^{\top} {\bf y}_1 &= \frac{m_1'}{m_1} \Big((\mathcal{G}_1) _{\bar{\bf u}_1{\bf y}_1} -m_1\big(\bar{\bf u}_1\big)^{\top} {\bf y}_1 \Big)\notag\\
 &=\frac{m_1'}{m_1} \Bigg( \frac{1}{(1+ r^{-\frac 12}m_2)z}({H}\mathcal{G}_1)_{\bar{\bf u}_1{\bf y}_1} + \frac{r^{-\frac 12}m_2}{1+ r^{-\frac 12}m_2}( \mathcal{G}_1)_{\bar{\bf u}_1{\bf y}_1} \Bigg)\,.
 \end{align*}

  Next, we also rewrite 
 \begin{align}\label{eq:Rewrite2}
 &\sqrt n \sum_{t=0}^1 r^{-\frac{1+2t}{4}} ( \mathcal{G}_1^{(t)} {X})_{\bar{\bf u}_1\tilde{{\bf y}}_t} \notag\\
 &=\sqrt n r^{-\frac 34}\bigg( (\mathcal{G}_1^2{X})_{\bar{\bf u}_1\tilde{{\bf y}}_1}  +  \frac{r^{\frac 12}  m_1'}{1+r^{\frac 12} m_1}(\mathcal{G}_1{X})_{\bar{\bf u}_1\tilde{{\bf y}}_1}\bigg) - \frac{r^{-\frac14} m_1'}{1+r^{\frac 12}  m_1} \,\sqrt n (\mathcal{G}_1{X})_{\bar{\bf u}_1\tilde{{\bf y}}_1} + \sqrt n r^{-\frac 14} (\mathcal{G}_1{X})_{\bar{\bf u}_1\tilde{{\bf y}}_0} \notag\\
 &=: \mathds{T}_{21} + \mathds{T}_{22}+ \mathds{T}_{23}\,,
 \end{align}
 and 
  \begin{align}\label{eq:Rewrite3}
 &{\sqrt n} \sum_{t=0}^1  r^{-\frac{1+t}{2}}\Big(({X}^{\top} \mathcal{G}_1{X})^{(t)}-(1+zm_2)^{(t)}\Big)_{\bar{\bf v}_1{\bm \eta}_t} \notag\\
& =\frac{\sqrt n}{ r}\bigg( ({X}^{\top} \mathcal{G}_1^2{X})_{\bar{\bf v}_1{\bm \eta}_1} 
+ \frac{\sqrt r m_1'}{1+\sqrt r m_1}  ({X}^{\top} \mathcal{G}_1{X})_{\bar{\bf v}_1{\bm \eta}_1} -  \frac{\sqrt r m_1'}{1+ \sqrt r m_1} {\big(\bar{\bf v}_1\big)^{\top}{\bm \eta}_1} \bigg) \notag\\
&\quad - \frac{\sqrt r m_1'}{1+\sqrt r m_1} \frac{\sqrt n}{ r}  \bigg(({X}^{\top} \mathcal{G}_1{X})_{\bar{\bf v}_1{\bm \eta}_1} - \Big(1-\frac{1+\sqrt r m_1}{\sqrt r m_1'} (zm_2)' \Big) {\big(\bar{\bf v}_1\big)^{\top}{\bm \eta}_1} \bigg) \notag\\
& \quad + \frac{\sqrt n}{\sqrt r} \bigg( ({X}^{\top} \mathcal{G}_1{X})_{\bar{\bf v}_1{\bm \eta}_0} 
- (1+zm_2) {\big(\bar{\bf v}_1\big)^{\top}{\bm \eta}_0}\bigg) \notag\\
&=\frac{\sqrt n}{ r}\bigg( ({X}^{\top} \mathcal{G}_1^2{X})_{\bar{\bf v}_1{\bm \eta}_1} 
+ \frac{\sqrt r m_1'}{1+\sqrt r m_1}  (z \mathcal{G}_2)_{\bar{\bf v}_1{\bm \eta}_1} \bigg) \notag\\
&\quad - \frac{\sqrt r m_1'}{1+\sqrt r m_1} \frac{\sqrt n}{ r}\bigg( \frac{1}{1+ \sqrt r m_1} ({X}^{\top} \mathcal{G}_1{X})_{\bar{\bf v}_1{\bm \eta}_1} + \frac{\sqrt r m_1}{1+\sqrt r m_1}  (z\mathcal{G}_2)_{\bar{\bf v}_1{\bm \eta}_1} \bigg) \notag\\
& \quad +\frac{\sqrt n}{\sqrt r} \bigg(\frac{1}{1+ \sqrt r m_1} ({X}^{\top} \mathcal{G}_1{X})_{\bar{\bf v}_1{\bm \eta}_0} + \frac{\sqrt r m_1}{1+\sqrt r m_1}  (z \mathcal{G}_1)_{\bar{\bf v}_1{\bm \eta}_0} \bigg) \notag\\
&=:\mathds{T}_{31}+ \mathds{T}_{32}+ \mathds{T}_{33}\,,
 \end{align}
 where we used the second identity in (\ref{relationXG}) and the identities
 \begin{align}
 1-\frac{1+\sqrt r m_1}{\sqrt r m_1'} (zm_2)' = 1+ zm_2\,, \quad  \frac{\sqrt r m_1}{1+\sqrt r m_1} = 1+zm_2\,. \label{042101}
 \end{align}
 We remark here that (\ref{042101}) can be easily checked by applying the identities in (\ref{identitym1m2}),  the first equation in (\ref{selfconeqt}),  and also the identity obtained by taking derivative w.r.t $z$ for both sides of the first equation in (\ref{selfconeqt}), i.e.,
 \begin{align*}
\sqrt  r m_1^2 + 2z \sqrt r\,  m_1m_1' + m_1+ (z-1/\sqrt r+\sqrt r)m_1'=0\,.
 \end{align*}
% {\color{blue}[Think if it's better to do decomposition of  $T_{ab}$'s in a later stage]}
% {\color{red}And we also remark here, we did some seemingly artificial decompositions,  of the form $\mathcal{G}_1X=s\mathcal{G}_1X+(1-s)\mathcal{G}_1X$ for instance, in the terms $\mathds{T}_{ab}$'s, in order to facilitate our later derivations. More specifically, to prove Proposition  \ref{prop:rmeP}, we will  derive  a self-consistent equation  for the characteristic function of $\mathcal{P}$,  for which we will need to apply the basic integration by parts formula for Gaussian variables. In the sequel, very often, we will apply the integration by parts to a part such as $s\mathcal{G}_1X$ and meanwhile keep the other part $(1-s)\mathcal{G}_1X$ untouched. One will see that applying integration by parts only partially will help us gain some simple algebraic cancellations. }
 
 Before we commence the proof of Proposition  \ref{prop:rmeP},  let us first state below  the derivative of $\mathcal{P}$,  which follows from a direct calculation
 \begin{align}\label{eq:deri_P}
  \frac{\partial \mathcal{P}}{\partial x_{ij}}  
&= -\sqrt n\sum_{\begin{subarray}{c} a_1, a_2\geq 1\\ a=a_1+a_2\leq 3\end{subarray}}  r^{-\frac{a-1}{2}}\Big((\mathcal{G}_1^{a_1}\bar{\bf u}_1)_{i} ({X}^{\top}\mathcal{G}_1^{a_2} {\bf y}_{a-2})_j +  ({X}^{\top}\mathcal{G}_1^{a_1}\bar{\bf u}_1)_{j} (\mathcal{G}_1^{a_2} {\bf y}_{a-2})_i   \Big)\notag\\
&\quad - \sqrt n\sum_{\begin{subarray}{c} a_1, a_2\geq 1\\ a=a_1+a_2\leq3\end{subarray}}
 r^{-\frac{1+2(a-2)}{4}}\Big( (\mathcal{G}_1^{a_1}\bar{\bf u}_1)_{i}((z \mathcal{G}_2)^{(a_2-1)}\tilde{{\bf y}}_{a_2})_j +
  ({X}^{\top}\mathcal{G}_1^{a_1}\bar{\bf u}_1)_{j}( \mathcal{G}_1^{a_2}{X}\tilde{{\bf y}}_{a-2})_i\Big)\notag\\
&\quad - {\sqrt n} \sum_{\begin{subarray}{c} a_1, a_2\geq 1\\ a=a_1+a_2\leq3\end{subarray}} r^{-\frac{a-1}{2}}
 \Big( (\mathcal{G}_1^{a_1}{X}\bar{\bf v}_1)_i \big((z\mathcal{G}_2)^{(a_2-1)}{\bm \eta}_{a-2}\big)_j + (\mathcal{G}_1^{a_1}{X}{\bm \eta}_{a-2})_i \big((z\mathcal{G}_2)^{(a_2-1)}\bar{\bf v}_1\big)_j\Big).
 \end{align}

 Now, let us proceed to the proof of Proposition  \ref{prop:rmeP}.
   \begin{myPro}{(Proof of Proposition  \ref{prop:rmeP})}

 By the definition of characteristic function, we have, for $t\in \mathbf{R}$, 
  \begin{align*}
  \varphi_n(t) = \E e^{\mathrm{i} t \mathcal{P}}\,, \quad \varphi_n'(t) = \mathrm{i} \E \mathcal{P} e^{\mathrm{i} t \mathcal{P}}\,.
  \end{align*}
  
  We will estimate $\varphi_n'(t)$ via Gaussian integration by parts. Recall the representation of $\mathcal{P}$ in (\ref{20210205}) together with (\ref{eq:Rewrite1})-(\ref{eq:Rewrite3}), we may further express 
  \begin{align*}
  \varphi_n'(t)= \mathrm{i}\E  \sum_{i,j=1}^3 \mathds{T}_{ij} h(t)\,,\qquad h(t):= e^{\mathrm{i}t  \mathcal{P}}\,. 
  \end{align*}
  For convenience, we use the following shorthand notation for summation
  \begin{align*}
  \sum_{i,j}:=\sum_{i=1}^p \sum_{j=1}^n\,.
  \end{align*}
 Since all entries $x_{ij}$  are i.i.d $\mathcal{N}(0, 1/\sqrt{np})$, applying  Gaussian integration by parts leads to  
 \begin{align*}
 \mathrm{i} \E \frac{\sqrt n}{ r} ({H}\mathcal{G}_1^2)_{\bar{\bf u}_1{\bf y}_1}\, h(t) &=\mathrm{i} \frac{\sqrt n}{r} \sum_{i,j}  \E \bar{u}_{1i} x_{ij} ({X}^{\top}\mathcal{G}_1^2{\bf y}_1)_j \,  h(t)= \mathrm{i}\frac{r^{-\frac 32}}{\sqrt n}\, \E\sum_{i,j} \bar{u}_{1i} \frac{ \partial ({X}^{\top}\mathcal{G}_1^2{\bf y}_1)_jh(t)}{\partial x_{ij}}\notag\\
%  &=\mathrm{i} \frac{r^{-\frac 32}}{\sqrt n}\,  \E\sum_{i,j} \bar{u}_{1i} \Big((\mathcal{G}_1^2{\bf y}_1)_i - ({X}^{\top} \mathcal{G}_1)_{ji} ({X}^{\top} \mathcal{G}_1^2{\bf y}_1)_j- ({X}^{\top} \mathcal{G}_1^2)_{ji} ({X}^{\top} \mathcal{G}_1{\bf y}_1)_j \notag\\
%&\qquad   -  ({X}^{\top} \mathcal{G}_1{X})_{jj} ( \mathcal{G}_1^2{\bf y}_1)_i - ({X}^{\top} \mathcal{G}_1^2{X})_{jj} ( \mathcal{G}_1{\bf y}_1)_i\Big)h(t) \notag\\
%& \quad +  \frac{\mathrm{i}^2t}{\sqrt {nr^3}}\E\sum_{i,j} \bar{u}_{1i}  ({X}^{\top}\mathcal{G}_1^2{\bf y}_1)_j \frac{\partial \mathcal{P}}{\partial x_{ij}} h(t)\notag\\
&=  \mathrm{i}\E \Big( \sqrt n r^{-\frac 32} (\mathcal{G}_1^2)_{\bar{\bf u}_1{\bf y}_1} - \frac{r^{-\frac 32}}{\sqrt n} (\mathcal{G}_1 {H}\mathcal{G}_1^2)_{\bar{\bf u}_1{\bf y}_1}- \frac{r^{-\frac 32}}{\sqrt n} (\mathcal{G}_1^2 {H}\mathcal{G}_1)_{\bar{\bf u}_1{\bf y}_1} \notag\\
&\qquad - \sqrt n r^{-\frac 32} (\mathcal{G}_1^2)_{\bar{\bf u}_1{\bf y}_1} \frac{{\rm Tr} \,{X}^{\top} \mathcal{G}_1{X}}{n} - \sqrt n  r^{-\frac 32}(\mathcal{G}_1)_{\bar{\bf u}_1{\bf y}_1} \frac{{\rm Tr} \,{X}^{\top} \mathcal{G}_1^2{X}}{n}\Big)h(t)\notag\\
&  \quad +  \frac{\mathrm{i}^2t}{\sqrt {nr^3}}\E\sum_{i,j} \bar{u}_{1i}  ({X}^{\top}\mathcal{G}_1^2{\bf y}_1)_j \frac{\partial \mathcal{P}}{\partial x_{ij}} h(t)\,.
 \end{align*}
 Then, by  Proposition \ref{thm:locallaw}, Remark \ref{rmk:locallaw}, and the fact $m_1^{(a)}(z)= O(r^{(1+a)/2})$ for $a=0,1,2$ owing to the choice of $z$ in  (\ref{def:fixed_z}), we further have
 \begin{align}\label{2020100702}
  \mathrm{i}\E\frac{\sqrt n}{r} ({H}\mathcal{G}_1^2)_{\bar{\bf u}_1{\bf y}_1}h(t) &=\mathrm{i}  \E \bigg( - \sqrt n  r^{-\frac 32}(\mathcal{G}_1^2)_{\bar{\bf u}_1{\bf y}_1} \frac{{\rm Tr} \,z \mathcal{G}_2}{n} - \sqrt n r^{-\frac 32} (\mathcal{G}_1)_{\bar{\bf u}_1{\bf y}_1} \frac{{\rm Tr} \, (z \mathcal{G}_2)'}{n} \notag\\
  &\qquad - n^{-\frac 12} r^{-\frac 32} (z\mathcal{G}_1)''_{\bar{\bf u}_1{\bf y}_1} \bigg)h(t) +  \frac{\mathrm{i}^2t}{\sqrt {nr^3}}\E\sum_{i,j} \bar{u}_{1i}  ({X}^{\top}\mathcal{G}_1^2{\bf y}_1)_j \frac{\partial \mathcal{P}}{\partial x_{ij}} h(t) \notag\\
  &= -\mathrm{i}\sqrt n r^{-\frac 32}  \E \Big( zm_2 (\mathcal{G}_1^2)_{\bar{\bf u}_1{\bf y}_1}  + (zm_2 )'(\mathcal{G}_1)_{\bar{\bf u}_1{\bf y}_1}  \Big) h(t)\notag\\
  & \qquad +  \frac{\mathrm{i}^2t}{\sqrt {nr^3}}\E\sum_{i,j} \bar{u}_{1i}  ({X}^{\top}\mathcal{G}_1^2{\bf y}_1)_j \frac{\partial \mathcal{P}}{\partial x_{ij}} h(t)+ O_\prec(p^{-\frac 12})\,.
 \end{align}
 Next, plugging in (\ref{eq:deri_P}),   we have the term {\small
 \begin{align} \label{2020100701}
& \frac{\mathrm{i}^2 t}{\sqrt {nr^3}}\E\sum_{i,j} \bar{u}_{1i}  ({X}^{\top}\mathcal{G}_1^2{\bf y}_1)_j \frac{\partial \mathcal{P}}{\partial x_{ij}}  h(t)\notag\\
 & =
   -\mathrm{i}^2t r^{-\frac 32}\E\sum_{i,j} \bar{u}_{1i}({X}^{\top}\mathcal{G}_1^2{\bf y}_1)_j\sum_{\begin{subarray}{c} a_1, a_2\geq 1\\a= a_1+a_2\leq 3\end{subarray}} \Bigg[
   r^{-\frac{a-1}{2}}\Big((\mathcal{G}_1^{a_1}\bar{\bf u}_1)_{i} ({X}^{\top}\mathcal{G}_1 ^{a_2}{\bf y}_{a-2})_j +  ({X}^{\top}\mathcal{G}_1^{a_1}\bar{\bf u}_1)_{j} (\mathcal{G}_1^{a_2} {\bf y}_{a-2})_i  \Big) \notag\\
   & \qquad +
    r^{-\frac{1+2(a-2)}{4}}\Big((\mathcal{G}_1^{a_1}\bar{\bf u}_1)_{i} \big((z\mathcal{G}_2 )^{(a_2-1)}\tilde{\bf y}_{a-2}\big)_j +  ({X}^{\top}\mathcal{G}_1^{a_1}\bar{\bf u}_1)_{j} (\mathcal{G}_1^{a_2} {X}\tilde{\bf y}_{a-2})_i  \Big)  \notag\\
    & \qquad+   r^{-\frac{a-1}{2}}  \Big((\mathcal{G}_1^{a_1}{X}{\bm \eta}_{a-2})_{i} \big( (z\mathcal{G}_2)^{(a_2-1)}\bar{\bf v}_1\big)_j + (\mathcal{G}_1^{a_1}{X}\bar{\bf v}_1)_{i} \big( (z\mathcal{G}_2)^{(a_2-1)}{\bm \eta}_{a-2}\big)_j\Big)\Bigg]h(t)\,. 
 \end{align}
 }
 It is easy to see that the RHS of the above equation is a linear combination of the expectations of the terms taking the following forms 
% {\color{red}[I added $t$ to the following forms and use $O_\prec(tp^{-1/2})$ for the error bounds. For the convergence rate, we will need $n$ or $p$-dependent $t$.  check the other error bounds. One more question, if the error bounds is $tp^{-1/2}$,is it still correct that the final convergence rate is $n^{-1/2}$ rather than $p^{-1/2}$, say. Check again ]}
 \begin{align}\label{forms.1}
 &tr^{-\frac{b_1+b_2+b_3}{2}}\big({\bf {\vartheta}}_1^{\top} \mathcal{G}_1^{b_1} {\bf {\vartheta}}_2 \big) \big({\bf {\vartheta}}_3^{\top} \mathcal{G}_1^{b_2}{H}\mathcal{G}_1^{b_3} {\bf {\vartheta}}_4 \big)\,,\qquad 
 tr^{-\frac{2(b_1+b_2+b_3)-1}{4}} \big({\bf {\vartheta}}_1^{\top} \mathcal{G}_1^{b_1} {\bf {\vartheta}}_2 \big) \big({\bf {\vartheta}}_3^{\top} \mathcal{G}_1^{b_2}{X} (z\mathcal{G}_2)^{(b_3-1)} {\bf {\vartheta}}_4 \big)\,,\notag\\
  &
  tr^{-\frac{2(b_1+b_2+b_3)-1}{4}}  \big({\bf {\vartheta}}_1^{\top} \mathcal{G}_1^{b_1}{X} {\bf {\vartheta}}_2 \big) \big({\bf {\vartheta}}_3^{\top} \mathcal{G}_1^{b_2}{H}\mathcal{G}_1^{b_3} {\bf {\vartheta}}_4 \big)\,,
   \qquad  tr^{-\frac{b_1+b_2+b_3}{2}}\big({\bf {\vartheta}}_1^{\top} \mathcal{G}_1^{b_1} {X}{\bf {\vartheta}}_2 \big) \big({\bf {\vartheta}}_3^{\top} \mathcal{G}_1^{b_2}{X} (z\mathcal{G}_2)^{(b_3-1)} {\bf {\vartheta}}_4 \big)\,. 
 \end{align}
 Here ${\bf {\vartheta}}_{i}$, $ i=1,2,3,4$ represent for vectors of suitable dimension and $b_i=1,2$, for $i=1,2,3$.
 By  (\ref{est.DG}), (\ref{081501}) and the fact $m_1^{(a)}(z)= O(r^{(1+a)/2})$ for $a\in \mathbb{N}$, it is easy to observe that except for the first term  in (\ref{forms.1}), all the others can be bounded by $O_\prec(tp^{-1/2})$. For instance, for the factor  $\big({\bf {\vartheta}}_3^{\top} \mathcal{G}_1^{b_2}{X} (z\mathcal{G}_2)^{(b_3-1)} {\bf {\vartheta}}_4 \big)$, we can use the following estimates which are consequences of (\ref{081501}),
 \begin{align*}
&{\bf {\vartheta}}_3^{\top} \mathcal{G}_1^{b_2}{X} (z\mathcal{G}_2) {\bf {\vartheta}}_4  = {\bf {\vartheta}}_3^{\top} z\mathcal{G}_1^{b_2+1} {X} {\bf {\vartheta}}_4 = O_\prec(n^{-\frac 12}  r^{\frac{1+2b_2}{4}})\,,\notag\\
&{\bf {\vartheta}}_3^{\top} \mathcal{G}_1^{b_2}{X} (z\mathcal{G}_2)' {\bf {\vartheta}}_4 = {\bf {\vartheta}}_3^{\top} \mathcal{G}_1^{b_2}{X}{X}^{\top} \mathcal{G}_1^2 {X} {\bf {\vartheta}}_4 = {\bf {\vartheta}}_3^{\top} \mathcal{G}_1^{b_2+1} {X} {\bf {\vartheta}}_4  +  z{\bf {\vartheta}}_3^{\top} \mathcal{G}_1^{b_2+2} {X} {\bf {\vartheta}}_4 = O_\prec(n^{-\frac 12} r^{\frac{1+2b_2}{4}})\,.
 \end{align*} 
 Therefore, by the above discussion, we can further simplify (\ref{2020100701}) to get 
 {\small 
 \begin{align} \label{2020100703}
 & \frac{\mathrm{i}^2 t}{\sqrt {nr^3}}\E\sum_{i,j} \bar{u}_{1i}  ({X}^{\top}\mathcal{G}_1^2{\bf y}_1)_j \frac{\partial \mathcal{P}}{\partial x_{ij}}  h(t)\notag\\
% & =  - \mathrm{i}^2 t r^{-\frac 32}\E\sum_{i,j} \bar{u}_{1i}({X}^{\top}\mathcal{G}_1^2{\bf y}_1)_j\sum_{\begin{subarray}{c} a_1, a_2\geq 1\\ a=a_1+a_2\leq 3\end{subarray}} 
%  r^{-\frac{a-1}{2}}\Big((\mathcal{G}_1^{a_1}\bar{\bf u}_1)_{i} ({X}^{\top}\mathcal{G}_1 ^{a_2}{\bf y}_{a-2})_j +  ({X}^{\top}\mathcal{G}_1^{a_1}\bar{\bf u}_1)_{j} (\mathcal{G}_1^{a_2} {\bf y}_{a-2})_i  \Big) h(t) \notag\\
% &\qquad + O_\prec(|t|p^{-\frac 12})\notag\\
&=- \mathrm{i}^2 t \E \sum_{\begin{subarray}{c} a_1, a_2\geq 1\\ a=a_1+a_2\leq 3\end{subarray}} 
 r^{-\frac{a+2}{2}}\Big((\mathcal{G}_1^{a_1})_{\bar{\bf u}_1 \bar{\bf u}_1}(\mathcal{G}_1^2{H}\mathcal{G}_1^{a_2})_{{\bf y}_1{\bf y}_{a-2}} + (\mathcal{G}_1^{a_2})_{\bar{\bf u}_1 {\bf y}_{a-2}}(\mathcal{G}_1^2{H}\mathcal{G}_1^{a_1})_{{\bf y}_1\bar{\bf u}_1 }\Big)h(t)+ O_\prec(|t|p^{-\frac 12}) \notag\\
&= - \mathrm{i}^2 t \sum_{\begin{subarray}{c} a_1, a_2\geq 1\\ a=a_1+a_2\leq 3\end{subarray}}  
 r^{-\frac{a+2}{2}}m_1^{(a_1-1)}\frac{(zm_1)^{(a_2+1)}}{(a_2+1)!}\Big( {\bf y}_1^{\top}{\bf y}_{a-2} + \big(\bar{\bf u}_1\big)^{\top} {\bf y}_{a-2} \,{\bf y}_1^{\top}\bar{\bf u}_1 \Big)\varphi_n(t)+ O_\prec(|t|p^{-\frac 12}) \notag\\
&=  t \bigg[ \frac{m_1(zm_1)''}{2r^2} \Big( {\bf y}_1^{\top}{\bf y}_{0} + \big(\bar{\bf u}_1\big)^{\top} {\bf y}_{0} \,{\bf y}_1^{\top}\bar{\bf u}_1 \Big) + \Big( \frac{m_1'(zm_1)''}{2r^{\frac 52}} + \frac{m_1(zm_1)'''}{3!r^{\frac 52}}\Big)\Big(\Vert{\bf y}_{1} \Vert^2+ ({\bf y}_1^{\top}\bar{\bf u}_1)^2 \Big) \bigg]\varphi_n(t)\notag\\
&\qquad + O_\prec(|t|p^{-\frac 12}) \,.
 \end{align}
 }
Combining (\ref{2020100702}) and (\ref{2020100703}), by the definition of $\mathds{T}_{11}$ in (\ref{eq:Rewrite1} and the fact $m_1(z)= O(\sqrt r)$, we get 
\begin{align}\label{est:ET11P}
\mathrm{i}\E \mathds{T}_{11}h(t)&=-t m_1 \bigg[ \frac{m_1(zm_1)''}{2r^2} \Big( {\bf y}_1^{\top}{\bf y}_{0} + \big(\bar{\bf u}_1\big)^{\top}{\bf y}_{0} \,{\bf y}_1^{\top}\bar{\bf u}_1 \Big) + \Big( \frac{m_1'(zm_1)''}{2r^{\frac 52}} + \frac{m_1(zm_1)'''}{3!r^{\frac 52}}\Big)\notag\\
&\qquad \qquad \times \Big(\Vert{\bf y}_{1} \Vert^2+ ({\bf y}_1^{\top}\bar{\bf u}_1)^2\Big) \bigg]\varphi_n(t)+ O_\prec((|t|+1)n^{-\frac 12}).
\end{align}
By similar arguments, we can also derive 
\begin{align}\label{est:ET12P}
{\mathrm{i}} \E \mathds{T}_{12}h(t)
%&= \frac{m_1'}{m_1(1+r^{-\frac 12}m_2)z}  \, \frac{\mathrm{i}^2 t}{\sqrt {nr^3}}\E\sum_{i,j} \bar{u}_{1i}  ({X}^{\top}\mathcal{G}_1{\bf y}_1)_j \frac{\partial \mathcal{P}}{\partial x_{ij}}  h(t)   + O_\prec(n^{-\frac 12})\notag\\
%&=-tm_1'  \sum_{\begin{subarray}{c} a_1, a_2\geq 1\\ a=a_1+a_2\leq 3\end{subarray}}  
%r^{-\frac{a+2}{2}} m_1^{(a_1-1)}\frac{(zm_1)^{(a_2)}}{(a_2)!}\Big( {\bf y}_1^*{\bf y}_{a-2} + \big(\bar{\bf u}_1\big)^* {\bf y}_{a-2} \,{\bf y}_1^*\bar{\bf u}_1 \Big)\E h(t)
%\notag\\
%&\qquad + O_\prec((|t|+1)n^{-\frac 12}) \notag\\
&=- t m_1' \bigg[ r^{-2}{m_1(zm_1)'} \Big( {\bf y}_1^{\top}{\bf y}_{0} + \big(\bar{\bf u}_1\big)^{\top} {\bf y}_{0} \,{\bf y}_1^{\top}\bar{\bf u}_1 \Big) + r^{-\frac 52}\Big( {m_1'(zm_1)'} + \frac{m_1(zm_1)''}{2}\Big)\notag\\
&\qquad \qquad \times \Big( \Vert{\bf y}_{1} \Vert^2+ ({\bf y}_1^{\top}\bar{\bf u}_1)^2 \Big) \bigg]\varphi_n(t)+ O_\prec((|t|+1)n^{-\frac 12})\,,
\end{align}
and 
\begin{align}\label{est:ET13P}
\mathrm{i}\E \mathds{T}_{13}h(t)&=-t m_1 \bigg[ r^{-\frac 32}{m_1(zm_1)'} \Big( \Vert{\bf y}_{0}\Vert^2 +({\bf y}_0^{\top}\bar{\bf u}_1)^2 \Big) + r^{-2}\Big( {m_1'(zm_1)'} + \frac{m_1(zm_1)''}{2}\Big)\notag\\
&\qquad \qquad \times \Big( {\bf y}_0^{\top}{\bf y}_{1} + \big(\bar{\bf u}_1\big)^{\top} {\bf y}_{1} \,{\bf y}_0^{\top}\bar{\bf u}_1 \Big) \bigg]\varphi_n(t)+ O_\prec((|t|+1)n^{-\frac 12})\,.
\end{align}

Next, we turn to study $\mathrm{i}\E\mathds{T}_{2i}h(t)$, $i=1,2,3$,  as defined in (\ref{eq:Rewrite2}).
We first do the decompositions of $\mathds{T}_{2i}$'s below
\begin{align}
&\mathds{T}_{21}=\sqrt n r^{-\frac 34}\bigg( \frac{1}{1+r^{\frac 12} m_1} (\mathcal{G}_1^2{X})_{\bar{\bf u}_1\tilde{{\bf y}}_1} +  \frac{r^{\frac 12} m_1}{1+r^{\frac 12} m_1}(\mathcal{G}_1^2{X})_{\bar{\bf u}_1\tilde{{\bf y}}_1}  +  \frac{r^{\frac 12}  m_1'}{1+r^{\frac 12} m_1}(\mathcal{G}_1{X})_{\bar{\bf u}_1\tilde{{\bf y}}_1}\bigg)\,,\label{2021051601}\\
&\mathds{T}_{22}= - \frac{r^{-\frac14} m_1'}{1+r^{\frac 12}  m_1} \,\sqrt n \bigg( \frac{1}{1+r^{\frac 12}  m_1}  (\mathcal{G}_1{X})_{\bar{\bf u}_1\tilde{{\bf y}}_1} + \frac{r^{\frac 12} m_1}{1+r^{\frac 12} m_1} (\mathcal{G}_1{X})_{\bar{\bf u}_1\tilde{{\bf y}}_1}\bigg)\,,\notag\\
& \mathds{T}_{23}= \sqrt n r^{-\frac 14}\bigg( \frac{1}{1+r^{\frac 12} m_1} (\mathcal{G}_1{X})_{\bar{\bf u}_1\tilde{{\bf y}}_0} + \frac{r^{\frac 12} m_1}{1+ r^{\frac 12}  m_1} (\mathcal{G}_1{X})_{\bar{\bf u}_1\tilde{{\bf y}}_0}\bigg)\,. \nonumber
\end{align} 
And we also remark here, these seemingly artificial decompositions,  of the form $\mathcal{G}_1X=s\mathcal{G}_1X+(1-s)\mathcal{G}_1X$ for instance, in the terms $\mathds{T}_{2i}$'s, are used to facilitate our later derivations. More specifically, to prove Proposition  \ref{prop:rmeP}, we will  derive  a self-consistent equation  for the characteristic function of $\mathcal{P}$,  for which we will need to apply the basic integration by parts formula for Gaussian variables. In the sequel, very often, we will apply the integration by parts to a part such as $s\mathcal{G}_1X$ and meanwhile keep the other part $(1-s)\mathcal{G}_1X$ untouched. One will see that applying integration by parts only partially will help us gain some simple algebraic cancellations. Similar decompositions will also appear in the estimates of $\mathds{T}_{31}$ term.

 In the sequel, we only show the details of the  estimate for the $\mathds{T}_{21}$ term. The other two terms can be estimated similarly, and thus we omit the details. By Gaussian  integration by parts, we have 
\begin{align*}
&\quad \mathrm{i}E \sqrt n r^{-\frac 34}  (\mathcal{G}_1^2{X})_{\bar{\bf u}_1\tilde{{\bf y}}_1} h(t) = \frac{\mathrm{i}}{\sqrt n} r^{-\frac 54} \E\sum_{i,j} \tilde{y}_{1j}  \frac{\partial (\mathcal{G}_1^2\bar{\bf u}_1)_i h(t) }{\partial x_{ij}}\notag\\
%&=-\mathrm{i}\frac{r^{-\frac 54}}{\sqrt n} \E\sum_{i,j} \tilde{y}_{1j}   \Big( (\mathcal{G}_1{X})_{ij}(\mathcal{G}_1^2 \bar{\bf u}_1)_i + (\mathcal{G}_1)_{ii}({X}^{\top} \mathcal{G}_1^2 \bar{\bf u}_1)_j +  (\mathcal{G}_1^2{X})_{ij}(\mathcal{G}_1\bar{\bf u}_1)_i + (\mathcal{G}_1^2)_{ii}({X}^{\top} \mathcal{G}_1\bar{\bf u}_1)_j \Big)h(t)\notag\\
%&\quad + \frac{\mathrm{i}^2 t}{\sqrt n} r^{-\frac 54} \E\sum_{i,j} \tilde{y}_{1j}    (\mathcal{G}_1^2 \bar{\bf u}_1)_i
% \frac{\partial \mathcal{P}}{\partial x_{ij}}h(t) \notag\\
 &=-\mathrm{i}\E\Big(  \frac{2r^{-\frac 54}}{\sqrt n} (\mathcal{G}_1^3{X})_{ \bar{\bf u}_1 \tilde{\bf y}_1} + \sqrt n r^{-\frac 14} (\mathcal{G}_1^2{X})_{\bar{\bf u}_1 \tilde{\bf y}_1} \frac{{\rm Tr}\mathcal{G}_1}{p} + \sqrt n r^{-\frac 14} (\mathcal{G}_1{X})_{\bar{\bf u}_1 \tilde{\bf y}_1} \frac{{\rm Tr}\mathcal{G}_1^2}{p}  \Big)h(t)\notag\\
 &\quad +  \frac{\mathrm{i}^2 t}{\sqrt n} r^{-\frac 54}\E\sum_{i,j}  \tilde{y}_{1j}    (\mathcal{G}_1^2 \bar{\bf u}_1)_i
 \frac{\partial \mathcal{P}}{\partial x_{ij}}h(t) \notag\\
 &= -\mathrm{i}\E\Big(  \sqrt n r^{-\frac 14} m_1(\mathcal{G}_1^2{X})_{\bar{\bf u}_1 \tilde{\bf y}_1}  + \sqrt n r^{-\frac 14} m_1' (\mathcal{G}_1{X})_{\bar{\bf u}_1 \tilde{\bf y}_1}   \Big)h(t)\notag\\
&\quad  +  \frac{\mathrm{i}^2 t}{\sqrt n} r^{-\frac 54}\E\sum_{i,j} \tilde{y}_{1j}    (\mathcal{G}_1^2 \bar{\bf u}_1)_i
 \frac{\partial \mathcal{P}}{\partial x_{ij}}h(t) + O_\prec(n^{-\frac 12})\,,
\end{align*}
where in the last step we used (\ref{081501}), (\ref{est_m12N}) and the fact $m_1^{(a)}(z)= O(r^{(1+a)/2})$ for $a=0,1$.  Plugging the above estimate into the first term in $\mathrm{i}\E\mathds{T}_{21}h(t)$ which corresponds to the the first term inside the parenthesis in (\ref{2021051601}), we easily see that 
\begin{align*}
\mathrm{i}\E\mathds{T}_{21}h(t)= \frac{r^{-\frac 54 }}{1+\sqrt r m_1} \frac{\mathrm{i}^2 t}{\sqrt n} \E\sum_{i,j} \tilde{y}_{1j}    (\mathcal{G}_1^2 \bar{\bf u}_1)_i
 \frac{\partial \mathcal{P}}{\partial x_{ij}}h(t) + O_\prec(n^{-\frac 12})\,.
\end{align*}
Similarly to (\ref{2020100701})- (\ref{2020100703}),  we can also derive that  
\begin{align}\label{2020100801}
&\frac{r^{-\frac 54}}{\sqrt n} \, t\E\sum_{i,j} \tilde{y}_{1j}    (\mathcal{G}_1^2 \bar{\bf u}_1)_i
 \frac{\partial \mathcal{P}}{\partial x_{ij}}h(t)   \nonumber\\
 &=  -t\E\sum_{i,j} \tilde{y}_{1j}    (\mathcal{G}_1^2 \bar{\bf u}_1)_i\sum_{\begin{subarray}{c} a_1, a_2\geq 1\\ a=a_1+a_2\leq 3\end{subarray}} 
 r^{-\frac{a+1}{2}}(\mathcal{G}_1^{a_1}\bar{\bf u}_1)_{i} \big((z\mathcal{G}_2 )^{(a_2-1)}\tilde{\bf y}_{a-2}\big)_j   h(t)  + O_\prec(|t|n^{-\frac 12})\notag\\
% &= -t\E \sum_{\begin{subarray}{c} a_1, a_2\geq 1\\ a=a_1+a_2\leq 3\end{subarray}} 
% r^{-\frac{a+1}{2}}
% (\mathcal{G}_1^{a_1+2})_{ \bar{\bf u}_1\bar{\bf u}_1} \big((z\mathcal{G}_2 )^{(a_2-1)}\big)_{ \tilde{\bf y}_1\tilde{\bf y}_{a-2}} h(t) + O_\prec(|t|n^{-\frac 12})\notag\\
 &= - t\bigg[ \frac{m_1''}{2r^{\frac 32}} (zm_2) \tilde{\bf y}_1^{\top}\tilde{\bf y}_{0} + r^{-2}\Big( \frac{m_1'''}{3!} (zm_2) + \frac{m_1''}{2} (zm_2)' \Big)\Vert\tilde{\bf y}_{1}\Vert^2\bigg]\varphi_n(t) + O_\prec(|t|n^{-\frac 12})\,,
\end{align}
which leads to 
\begin{align}\label{est:ET21P}
\mathrm{i}\E\mathds{T}_{21} h(t)&=
\frac{t}{1+\sqrt r m_1} \bigg[ \frac{m_1''}{2r^{\frac 32}} (zm_2) \tilde{\bf y}_1^{\top}\tilde{\bf y}_{0} + r^{-2} \Big( \frac{m_1'''}{3!} (zm_2) + \frac{m_1''}{2} (zm_2)' \Big)\Vert\tilde{\bf y}_{1}\Vert^2\bigg]\varphi_n(t) \notag\\
 &\qquad + O_\prec((|t|+1)n^{-\frac 12})\,.
\end{align}
By analogous derivations, we can get the following estimates
\begin{align}
\mathrm{i}\E\mathds{T}_{22}h(t)&=- \frac{tm_1'}{(1+\sqrt r m_1)^2}\bigg[r^{-1} {m_1'} (zm_2) \tilde{\bf y}_1^{\top}\tilde{\bf y}_{0} + r^{-\frac 32}\Big( \frac{m_1''}{2} (zm_2) + {m_1'} (zm_2)' \Big) \Vert\tilde{\bf y}_{1}\Vert^2\bigg]\varphi_n(t) \notag\\
 &\qquad + O_\prec((|t|+1)n^{-\frac 12})\,,\label{est:ET22P}\\
\mathrm{i} \E\mathds{T}_{23}h(t)&= \frac{t}{(1+\sqrt r m_1)}\bigg[ r^{-1}{m_1'} (zm_2) \Vert\tilde{\bf y}_{0}\Vert^2 +r^{-\frac 32} \Big( \frac{m_1''}{2} (zm_2) + {m_1'} (zm_2)' \Big) \tilde{\bf y}_0^{\top}\tilde{\bf y}_{1}\bigg]\varphi_n(t)\notag\\
   &\qquad + O_\prec((|t|+1)n^{-\frac 12})\,. \label{est:ET23P}
\end{align}

 In the sequel, we focus on the derivation of the  estimate of $ \mathrm{i}\E\mathds{T}_{31}h(t)$ and directly conclude the estimates of  $ \mathrm{i}\E\mathds{T}_{32}h(t)$, $ \mathrm{i}\E\mathds{T}_{33}h(t)$ without details, since we actually only need to make some adjustments to the  estimate of $ \mathrm{i}\E\mathds{T}_{31}h(t)$.
 First, we do the following artificial decomposition for $\mathds{T}_{31}$,
 \begin{align*}
\mathds{T}_{31}= \frac{\sqrt n}{ r}\bigg( \frac{1}{1+\sqrt r m_1}({X}^\top \mathcal{G}_1^2{X})_{\bar{\bf v}_1{\bm \eta}_1} + \frac{\sqrt r m_1}{1+\sqrt r m_1}  ({X}^\top \mathcal{G}_1^2{X})_{\bar{\bf v}_1{\bm \eta}_1} 
+ \frac{\sqrt r m_1'}{1+\sqrt r m_1}  (z \mathcal{G}_2)_{\bar{\bf v}_1{\bm \eta}_1} \bigg)\,. 
 \end{align*}
 Then, by Gaussian integration by parts,  following from  (\ref{081501}) and (\ref{est_m12N}) we have
 \begin{align*}
&\quad \mathrm{i} \E\frac{\sqrt n}{ r} ({X}^{\top} \mathcal{G}^2{X})_{\bar{\bf v}_1{\bm \eta}_1}h(t)
 \notag\\
& =  \mathrm{i}\frac{r^{-\frac 32}}{\sqrt n} \E\sum_{i,j} v_{1j}^0  \frac{\partial (\mathcal{G}_1^2{X}{\bm \eta}_1)_ih(t) }{\partial x_{ij}}\notag\\
%&= \mathrm{i} \frac{r^{-\frac 32}}{\sqrt n} \E\sum_{i,j} v_{1j}^0  \Big((\mathcal{G}_1^2)_{ii} \eta_{1j} - (\mathcal{G}_1)_{ii} ({X}^{\top} \mathcal{G}_1^2 {X}{\bm \eta}_1)_j - (\mathcal{G}_1{X})_{ij} ( \mathcal{G}_1^2 {X}{\bm \eta}_1)_i - (\mathcal{G}_1^2)_{ii} ({X}^{\top} \mathcal{G}_1 {X}{\bm \eta}_1)_j \notag\\
%&\qquad - (\mathcal{G}_1^2{X})_{ij} ( \mathcal{G}_1 {X}{\bm \eta}_1)_i \Big)h(t)\notag\\
%&\quad + \frac{\mathrm{i}^2t}{\sqrt n} r^{-\frac 32} \E\sum_{i,j} v_{1j}^0  (\mathcal{G}_1^2{X}{\bm \eta}_1)_i   \frac{\partial \mathcal{P} }{\partial x_{ij}}h(t) \notag\\
&= - \mathrm{i} \E \Big( \sqrt n r^{-\frac 12} (z\mathcal{G}_2)_{\bar{\bf v}_1 {\bm \eta}_1} \frac{{\rm Tr}\, \mathcal{G}_1^2}{p} + \sqrt n r^{-\frac 12}  \big((z\mathcal{G}_2)'\big)_{\bar{\bf v}_1 {\bm \eta}_1} \frac{{\rm Tr}\, \mathcal{G}_1}{p} +\frac{2 r^{-\frac 32}}{\sqrt n} ({X}^{\top} \mathcal{G}_1^3{X})_{\bar{\bf v}_1 {\bm \eta}_1}   \Big)h(t)\notag\\
&\quad + \frac{\mathrm{i}^2 t }{\sqrt n} r^{-\frac 32} \E\sum_{i,j} v_{1j}^0  (\mathcal{G}_1^2{X}{\bm \eta}_1)_i   \frac{\partial \mathcal{P} }{\partial x_{ij}}h(t) \notag\\
&= -  \mathrm{i}\E \Big( \sqrt n r^{-\frac 12} m_1' (z\mathcal{G}_2)_{\bar{\bf v}_1 {\bm \eta}_1}  + \sqrt n r^{-\frac 12}m_1 \big((z\mathcal{G}_2)'\big)_{\bar{\bf v}_1 {\bm \eta}_1}  \Big)h(t)\notag\\
&\quad + \frac{\mathrm{i}^2 t }{\sqrt n} r^{-\frac 32} \E\sum_{i,j} v_{1j}^0  (\mathcal{G}_1^2{X}{\bm \eta}_1)_i   \frac{\partial \mathcal{P} }{\partial x_{ij}}h(t) + O_\prec(n^{-\frac 12})\,.
 \end{align*}
 This combined with definition of $\mathds{T}_{31}$, implies that 
 \begin{align*}
 \mathrm{i} \E\mathds{T}_{31}h(t)= - \frac{1}{1+ \sqrt r m_1}  \frac{ t }{\sqrt n} r^{-\frac 32} \E\sum_{i,j} v_{1j}^0  (\mathcal{G}_1^2{X}{\bm \eta}_1)_i   \frac{\partial \mathcal{P} }{\partial x_{ij}}h(t) + O_\prec(n^{-\frac 12})\,.
 \end{align*}
% We now proceed to estimate the main term above (i.e., $ \frac{t}{\sqrt n} r^{-\frac 32} \E\sum_{i,j} v_{1j}^0  (\mathcal{G}_1^2{X}{\bm \eta}_1)_i   \frac{\partial \mathcal{P} }{\partial x_{ij}}h(t)$) where we omit  the prefactor $-1/(1+ \sqrt r m_1)$. 
  Referring to (\ref{2020100801}) with slight adjustments, we can  easily obtain that 
 {\small
 \begin{align*}
& \frac{t}{\sqrt {nr^3}} \E\sum_{i,j} v_{1j}^0  (\mathcal{G}_1^2{X}{\bm \eta}_1)_i   \frac{\partial \mathcal{P} }{\partial x_{ij}}h(t)\notag\\
 &= - t\E\sum_{i,j} v_{1j}^0  (\mathcal{G}_1^2{X}{\bm \eta}_1)_i \sum_{\begin{subarray}{c} a_1, a_2\geq 1\\ a=a_1+a_2\leq3\end{subarray}}
 r^{-\frac{a+2}{2}}
 \Big( (\mathcal{G}_1^{a_1}{X}\bar{\bf v}_1)_i \big((z\mathcal{G}_2)^{(a_2-1)}\eta_{a-2}\big)_j + (\mathcal{G}_1^{a_1}{X}{\bm \eta}_{a-2})_i \big((z\mathcal{G}_2)^{(a_2-1)}\bar{\bf v}_1\big)_j\Big)h(t)   \notag\\
 &\qquad + O_\prec(|t|n^{-\frac 12})\nonumber\\
 &=- t\E  \sum_{\begin{subarray}{c} a_1, a_2\geq 1\\ a=a_1+a_2\leq3\end{subarray}}
  r^{-\frac{a+2}{2}}
\Big( ({X}^{\top} \mathcal{G}_1^{a_1+2}{X})_{\bar{\bf v}_1 {\bm \eta}_1} \big( (z\mathcal{G}_2)^{(a_2-1)}\big)_{\bar{\bf v}_1 {\bm \eta}_{a-2}} + 
({X}^{\top} \mathcal{G}_1^{a_1+2}{X})_{ {\bm \eta}_{a-2}\, {\bm \eta}_1} \big( (z\mathcal{G}_2)^{(a_2-1)}\big)_{\bar{\bf v}_1 \bar{\bf v}_1 } \Big)h(t) \notag\\
&\qquad + O_\prec(|t|n^{-\frac 12})\nonumber\\
&= - t\Big[ \frac{(zm_2)''}{2r^2} (zm_2) \Big( {\bm \eta}_{0}^{\top} {\bm \eta}_1+ \bar{\bf v}_1^{\top} {\bm \eta}_1\, \bar{\bf v}_1^{\top} {\bm \eta}_{0}\Big) \notag\\
&\qquad + r^{-\frac 52}\Big( \frac{(zm_2)'''}{3!} (zm_2)  + \frac{(zm_2)''}{2} (zm_2)'  \Big) \Big( \Vert {\bm \eta}_{1}\Vert^2+\big( \bar{\bf v}_1^{\top} {\bm \eta}_1\big)^2 \Big)  \Big]\varphi_n(t) + O_\prec(|t|n^{-\frac 12})\,.
 \end{align*}
 }
 Therefore, 
 \begin{align}\label{est:ET31P}
  \mathrm{i}\E\mathds{T}_{31}h(t) & =  \frac{t}{(1+\sqrt r m_1)}\Big[ \frac{(zm_2)''}{2r^2} (zm_2) \Big( {\bm \eta}_{0}^{\top} {\bm \eta}_1+ \bar{\bf v}_1^{\top} {\bm \eta}_1\,\bar{\bf v}_1^{\top} {\bm \eta}_{0}\Big) \notag\\
&+ r^{-\frac 52} \Big( \frac{(zm_2)'''}{3!} (zm_2)  + \frac{(zm_2)''}{2} (zm_2)'  \Big) \Big( \Vert {\bm \eta}_1\Vert^2+\big( \bar{\bf v}_1^{\top} {\bm \eta}_1\big)^2\Big)  \Big]\varphi_n(t)+ O_\prec((|t|+1)n^{-\frac 12})\,.
 \end{align}
 Similarly, we also get 
 \begin{align}\label{est:ET32P}
 \mathrm{i} \E\mathds{T}_{32}h(t) &=  -\frac{tm_1'}{(1+\sqrt r m_1)^2}\Big[ r^{-\frac 32} {(zm_2)'} (zm_2) \Big( {\bm \eta}_{0}^{\top} {\bm \eta}_1+ \bar{\bf v}_1^{\top} {\bm \eta}_1\, \bar{\bf v}_1^{\top} {\bm \eta}_{0}\Big) \notag\\
& + r^{-2}\Big( \frac{(zm_2)''}{2} (zm_2)  + {(zm_2)'}(zm_2)'  \Big) \Big( \Vert{\bm \eta}_1\Vert^2+\big( \bar{\bf v}_1^{\top} {\bm \eta}_1\big)^2\Big)  \Big]\varphi_n(t) + O_\prec((|t|+1)n^{-\frac 12})\,.
 \end{align}
 and 
  \begin{align}\label{est:ET33P}
\mathrm{i}  \E\mathds{T}_{33} h(t) & =   \frac{t}{(1+\sqrt r m_1)}\Big[r^{-\frac 32}{(zm_2)'} (zm_2) \Big(\Vert {\bm \eta}_{0}\Vert^2+\big(  \bar{\bf v}_1^{\top} {\bm \eta}_{0}\big)^2\Big) \notag\\
&+ r^{-2} \Big( \frac{(zm_2)''}{2} (zm_2)  + {(zm_2)'}(zm_2)'  \Big)\Big( {\bm \eta}_{1}^{\top} {\bm \eta}_0+ \bar{\bf v}_1^{\top} {\bm \eta}_0\, \bar{\bf v}_1^{\top} {\bm \eta}_{1}\Big)  \Big]\varphi_n(t) + O_\prec((|t|+1)n^{-\frac 12})\,.
 \end{align}
 
Combining (\ref{est:ET11P})- (\ref{est:ET13P}), (\ref{est:ET21P})- (\ref{est:ET23P}) and (\ref{est:ET31P})- (\ref{est:ET33P}), together with the definition of ${\bf y}_{0,1}, \tilde{\bf y}_{0,1}, {\bm \eta}_{0,1} $ in (\ref{def:y_ty_eta}), after elementary computations, we can then conclude  that 
\begin{align*}
  \varphi'(t)= \mathrm{i}\E  \sum_{i,j=1}^3 \mathds{T}_{ij} h(t)=
 - \big({\bf c}^{\top} \mathcal{M} {\bf c}\big) t \varphi_n(t) +  O_\prec((|t|+1) n^{-\frac 12})\,.
\end{align*}
 Hence, we finish the proof of Proposition  \ref{prop:rmeP}.
\end{myPro}

\section{Additional numerical results} \label{sec:appen_simu}

\subsection{Additional figures and tables for simulation settings in Section \ref{sec: numerical}}

\begin{figure}[h]
\caption{Examples 1a and 1b,  type I and type II errors for competing methods with increasing balanced sample sizes (1a) and increasing $n_0$ only (1b) .  \label{fig::ex 1a 1b}}

 \begin{subfigure}[t]{0.5\textwidth}
        \centering
        \includegraphics[scale=0.24]{ex1a_type_I_error.png}
        \caption{Example 1a, type I error}
    \end{subfigure}%
      \hspace{+0.1cm}
    \begin{subfigure}[t]{0.5\textwidth}
        \centering
        \includegraphics[scale=0.24]{ex1a_type_II_error.png}
        \caption{Example 1a, type II error}
    \end{subfigure}%    \begin{subfigure}[t]{0.5\textwidth}
    
    \begin{subfigure}[t]{0.5\textwidth}
        \centering
        \includegraphics[scale=0.24]{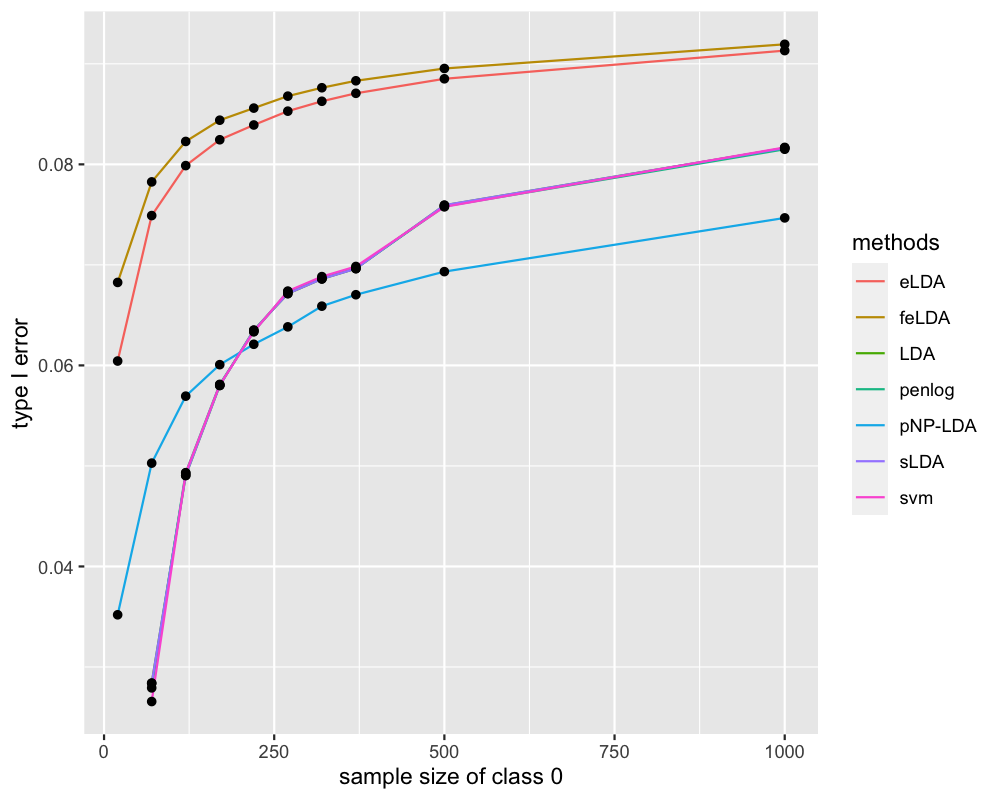}
        \caption{Example 1b, type I error}
    \end{subfigure}%    \begin{subfigure}[t]{0.5\textwidth}
      \hspace{+0.1cm}
    \begin{subfigure}[t]{0.5\textwidth}
        \centering
        \includegraphics[scale=0.24]{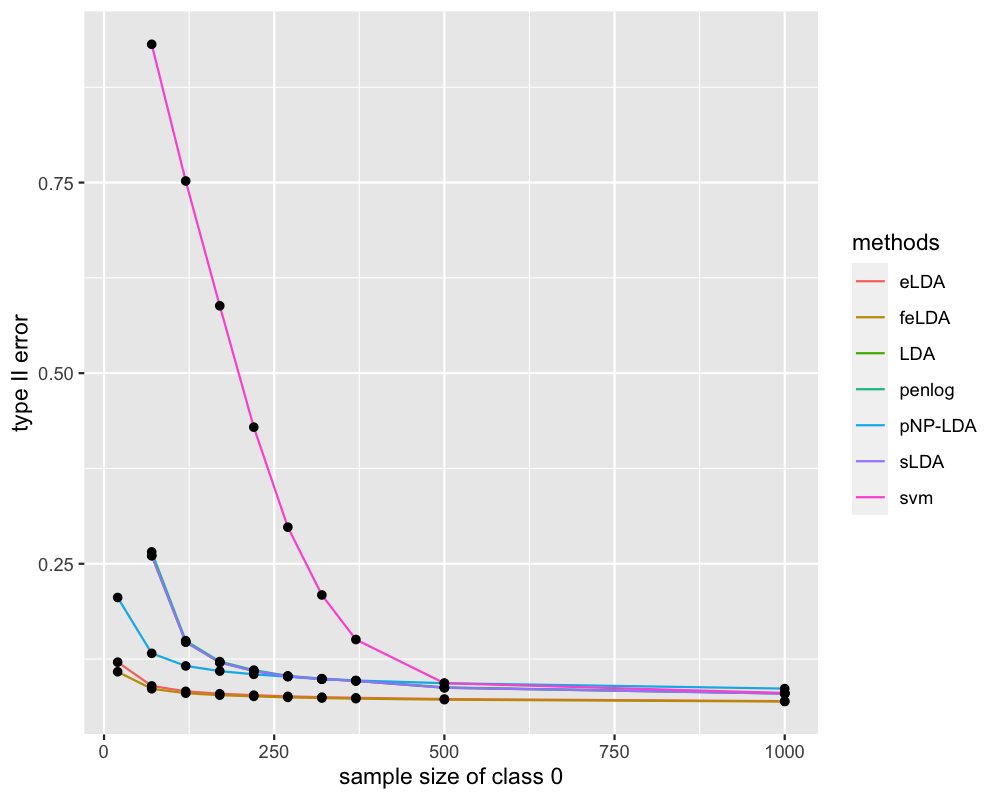}
        \caption{Example 1b, type II error}
    \end{subfigure}%
\end{figure}

\begin{figure}[h]
\caption{Examples 1c, 1c' and 1c*, type I and type II error for competing methods with increasing dimension $p$ and different $\delta$'s.  $\delta=0.1$ in Example 1c, $\delta = 0.05$ in Example 1c', and $\delta = 0.01$ in Example 1c*.  \label{fig::ex 1c 1c'}}

 \begin{subfigure}[t]{0.5\textwidth}
        \centering
        \includegraphics[scale=0.22]{ex1c_type_I_error.png}
        \caption{Example 1c, type I error}
    \end{subfigure}%
    \begin{subfigure}[t]{0.5\textwidth}
        \centering
        \includegraphics[scale=0.22]{ex1c_type_II_error.png}
        \caption{Example 1c, type II error}
    \end{subfigure}%    \begin{subfigure}[t]{0.5\textwidth}
    \vspace{-0.1cm}
    \begin{subfigure}[t]{0.5\textwidth}
        \centering
        \includegraphics[scale=0.22]{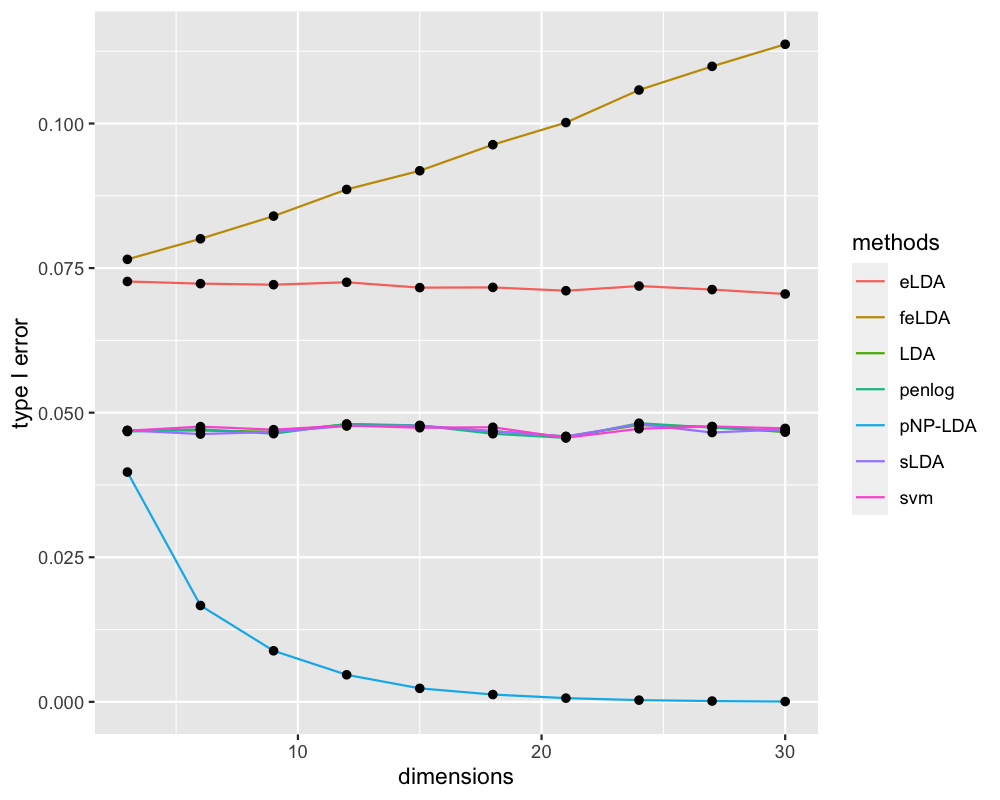}
        \caption{Example 1c', type I error}
    \end{subfigure}%    \begin{subfigure}[t]{0.5\textwidth}
    \begin{subfigure}[t]{0.5\textwidth}
        \centering
        \includegraphics[scale=0.22]{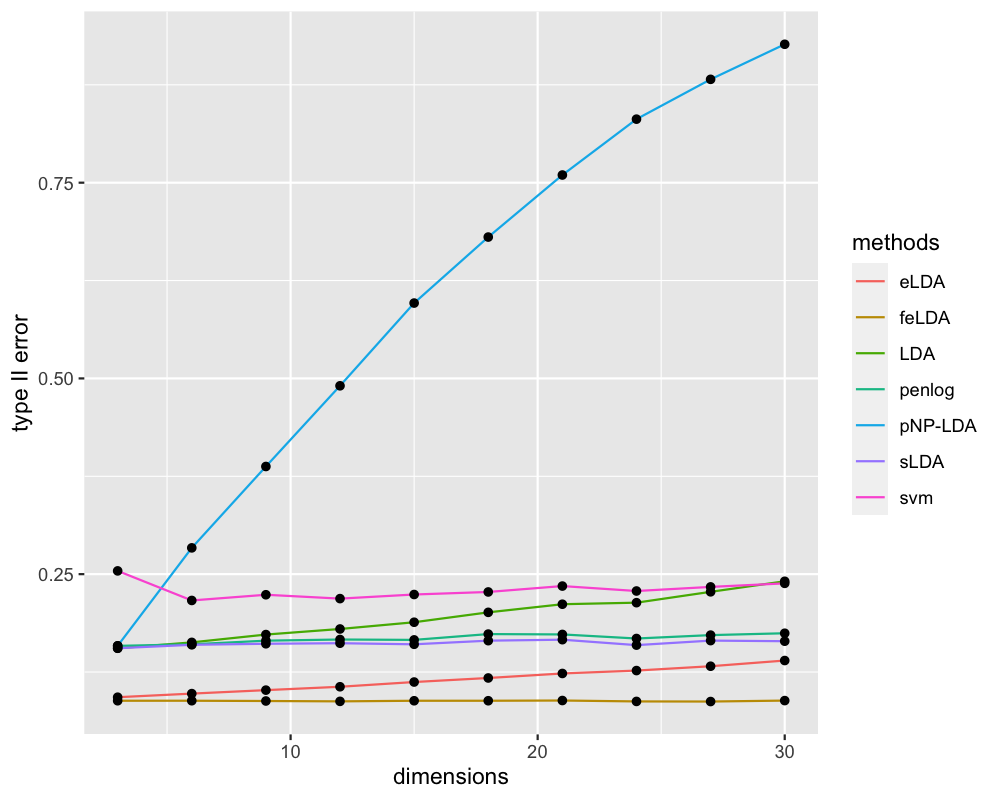}
        \caption{Example 1c', type II error}
    \end{subfigure}%
   \vspace{-0.1cm} 
\begin{subfigure}[t]{0.5\textwidth}
        \centering
        \includegraphics[scale=0.22]{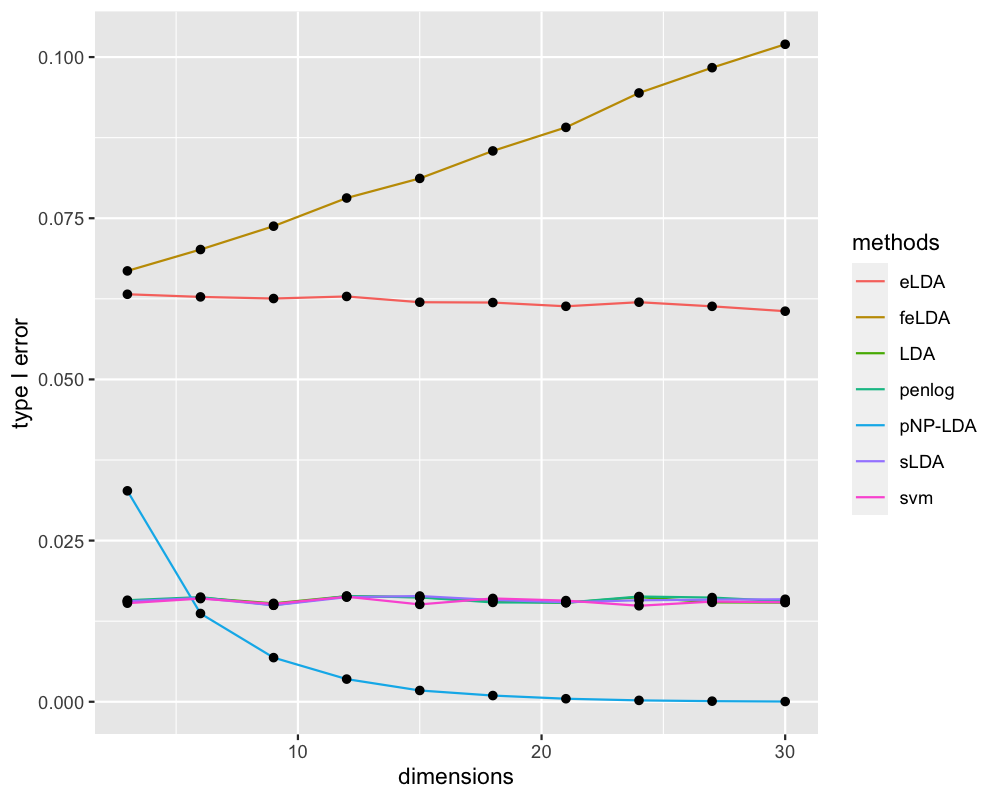}
        \caption{Example 1c* type I error}
    \end{subfigure}%    \begin{subfigure}[t]{0.5\textwidth}
    \begin{subfigure}[t]{0.5\textwidth}
        \centering
        \includegraphics[scale=0.22]{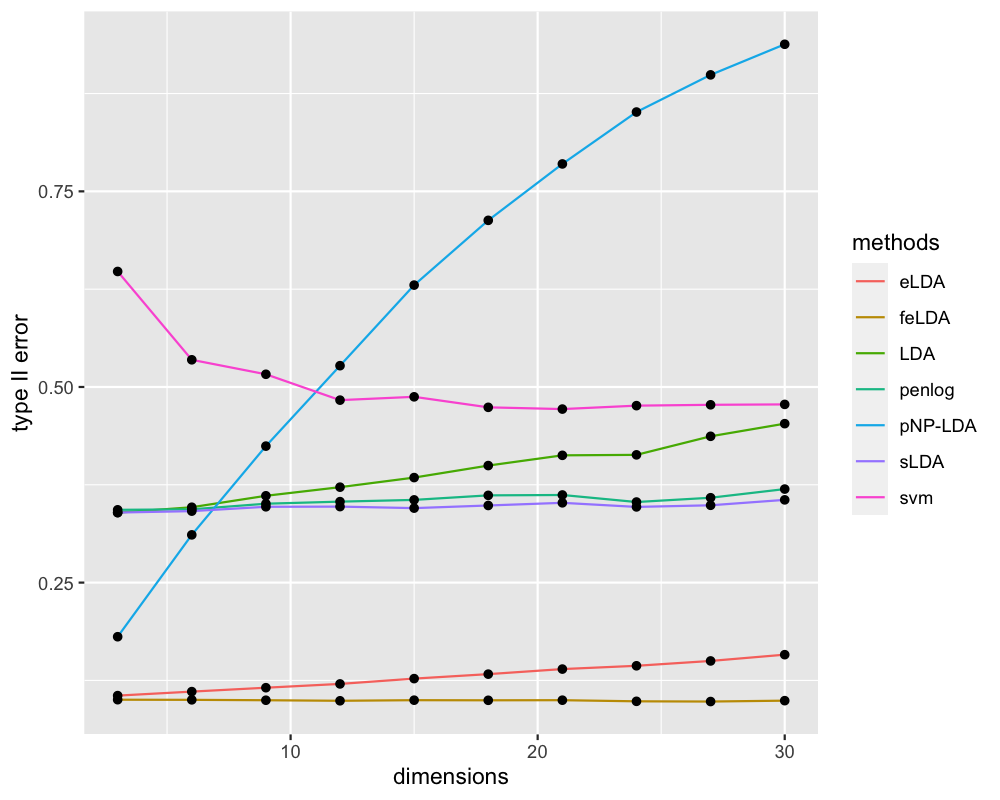}
        \caption{Example 1c*, type II error}
    \end{subfigure}%
\end{figure}

\begin{figure}[h]
\caption{Examples 1d and 1d',  imbalanced sample sizes with larger $n_1$.  Type I and type II error for competing methods with increasing dimension $p$, but different $\delta$'s:  $\delta=0.1$ in Example 1d and $\delta = 0.05$ in Example 1d'.  \label{fig::ex 1d 1d'}}

 \begin{subfigure}[t]{0.5\textwidth}
        \centering
        \includegraphics[scale=0.24]{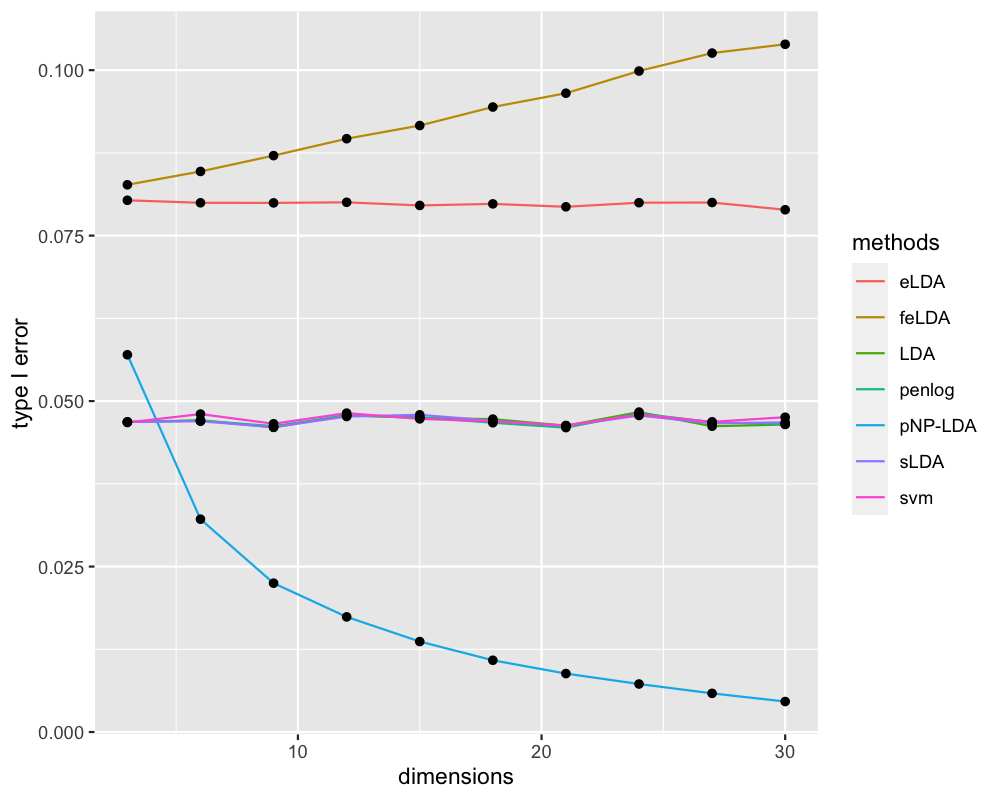}
        \caption{Example 1d, type I error}
    \end{subfigure}%
      \hspace{+0.1cm}
    \begin{subfigure}[t]{0.5\textwidth}
        \centering
        \includegraphics[scale=0.24]{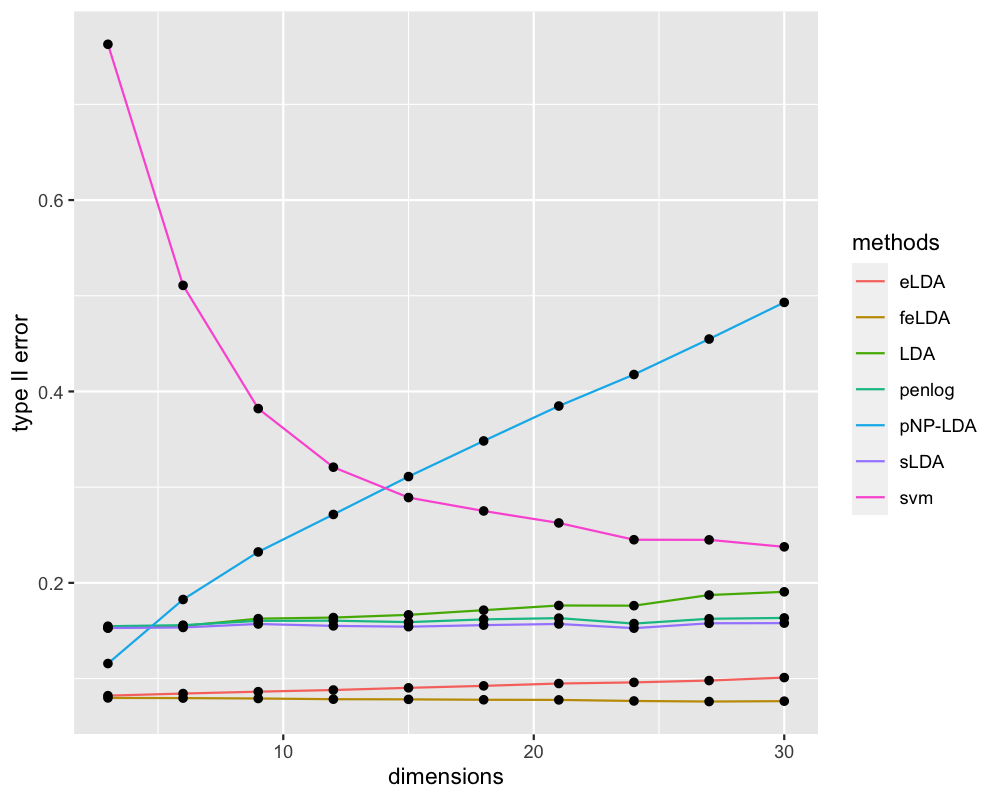}
        \caption{Example 1d, type II error}
    \end{subfigure}%    \begin{subfigure}[t]{0.5\textwidth}
    
    \begin{subfigure}[t]{0.5\textwidth}
        \centering
        \includegraphics[scale=0.24]{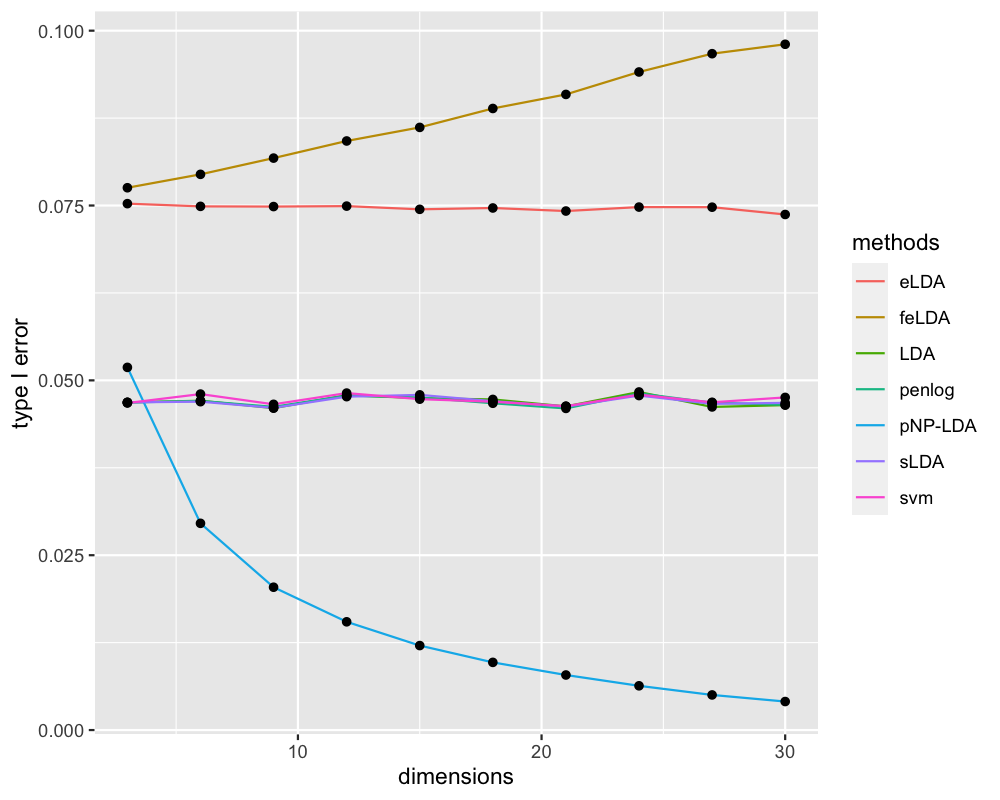}
        \caption{Example 1d', type I error}
    \end{subfigure}%    \begin{subfigure}[t]{0.5\textwidth}
      \hspace{+0.1cm}
    \begin{subfigure}[t]{0.5\textwidth}
        \centering
        \includegraphics[scale=0.24]{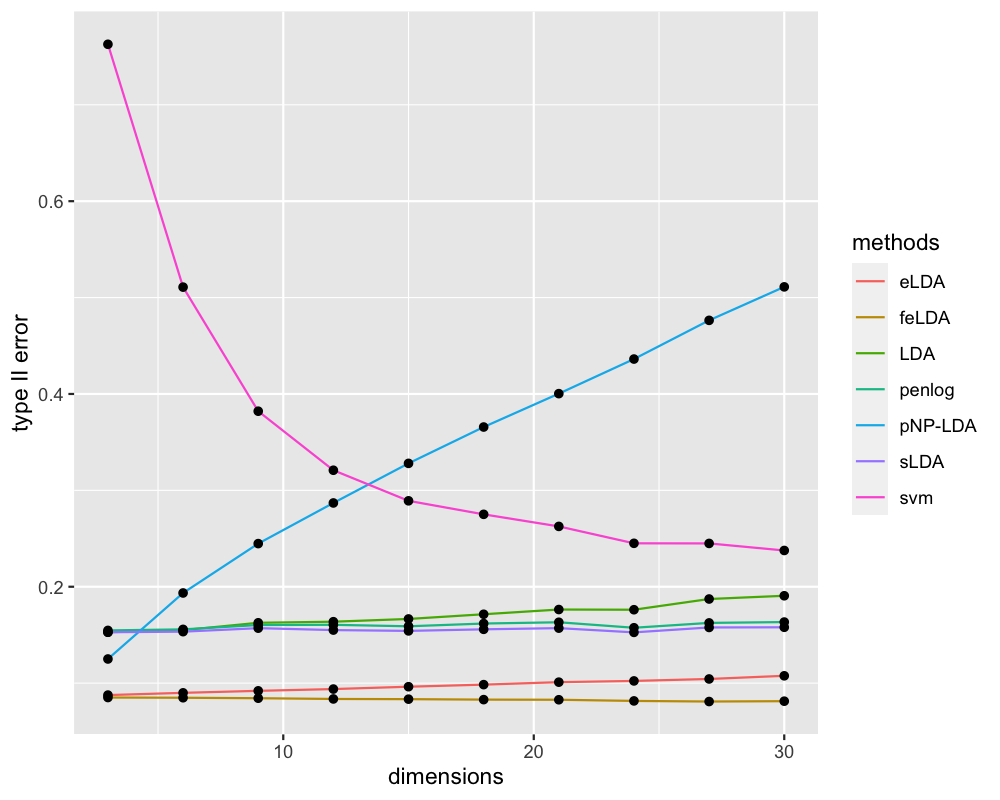}
        \caption{Example 1d', type II error}
    \end{subfigure}%

\end{figure}

\begin{figure}[h]
\caption{Examples 2a and 2b,   type I and type II error for competing methods with increasing dimension $p$. Example 2a has balanced sample sizes and Example 2b has imbalanced sample sizes. \label{fig::ex 2a 2b}}

 \begin{subfigure}[t]{0.5\textwidth}
        \centering
        \includegraphics[scale=0.24]{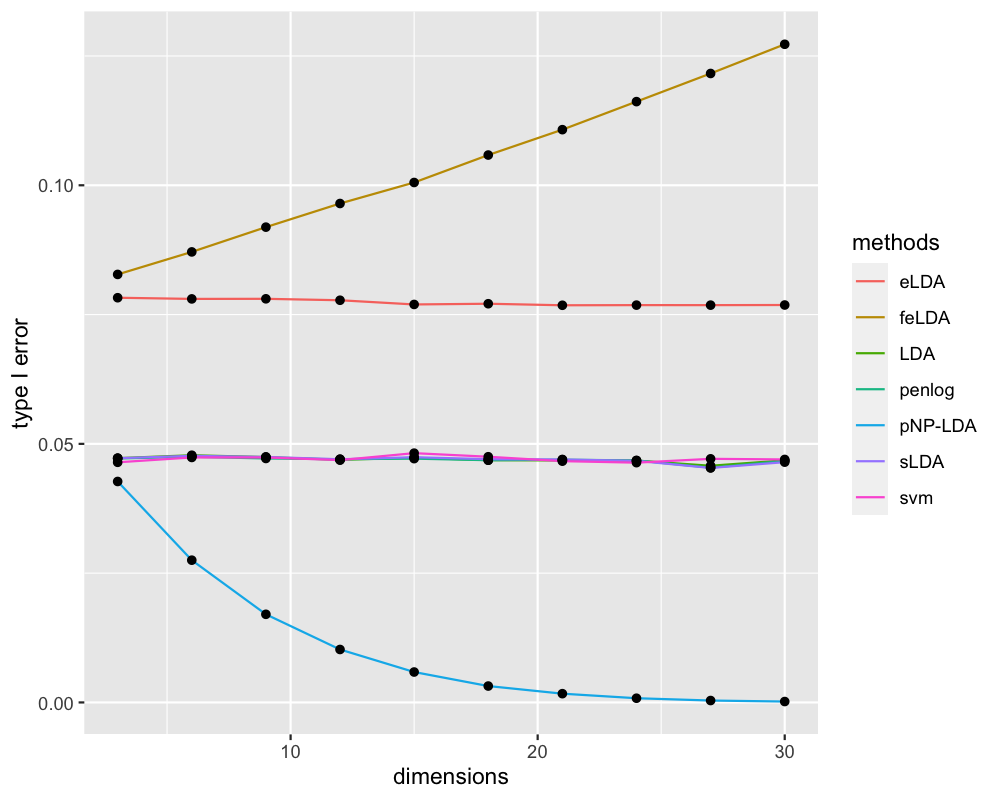}
        \caption{Example 2a, type I error}
    \end{subfigure}%
      \hspace{+0.1cm}
    \begin{subfigure}[t]{0.5\textwidth}
        \centering
        \includegraphics[scale=0.24]{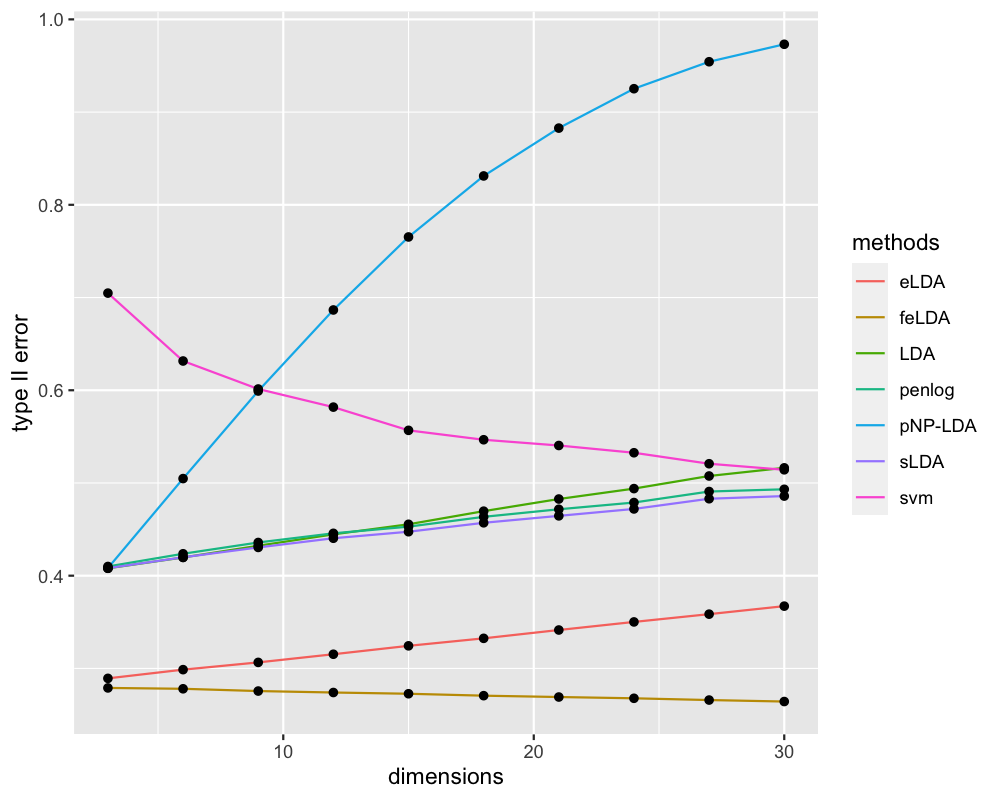}
        \caption{Example 2a type II error}
    \end{subfigure}%    \begin{subfigure}[t]{0.5\textwidth}
    
    \begin{subfigure}[t]{0.5\textwidth}
        \centering
        \includegraphics[scale=0.24]{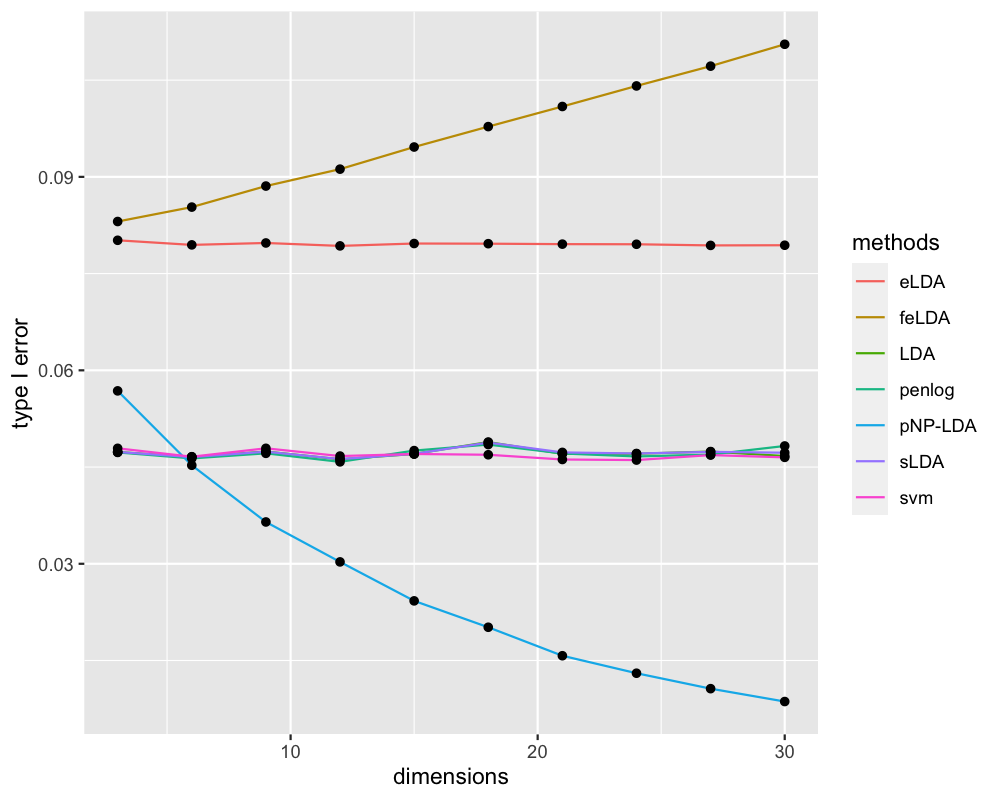}
        \caption{Example 2b, type I error}
    \end{subfigure}%    \begin{subfigure}[t]{0.5\textwidth}
      \hspace{+0.1cm}
    \begin{subfigure}[t]{0.5\textwidth}
        \centering
        \includegraphics[scale=0.24]{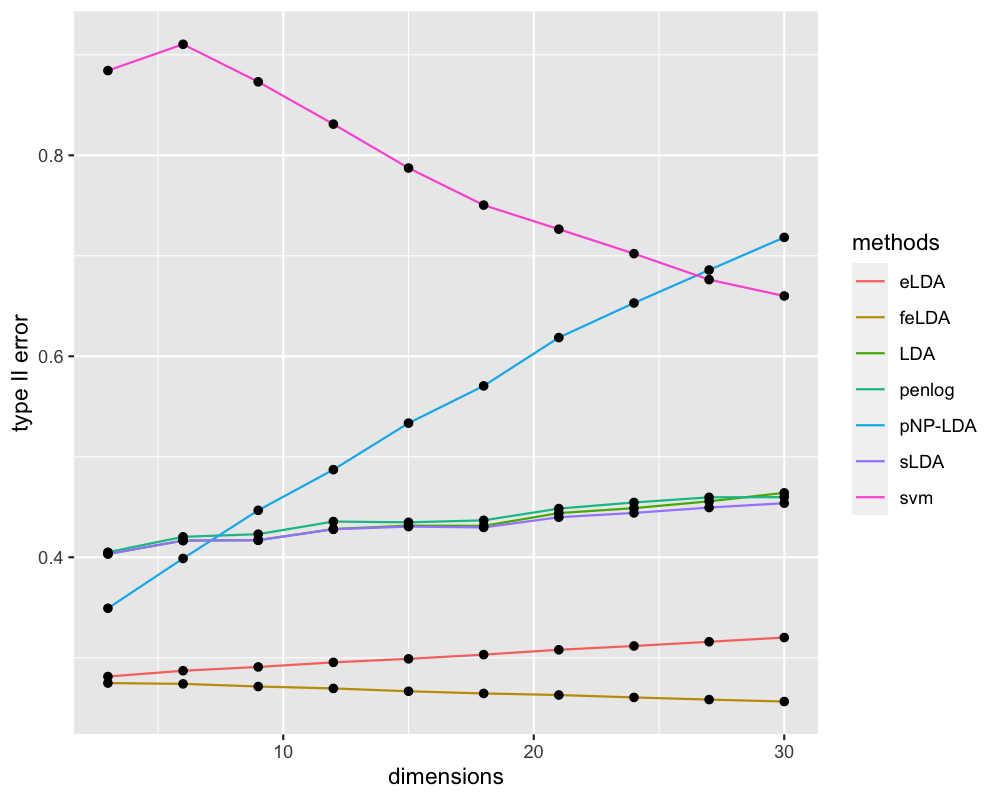}
        \caption{Example 2b, type II error}
    \end{subfigure}%

\end{figure}

\beginsupplement 
Figures \ref{fig::ex 1a 1b},  \ref{fig::ex 1c 1c'} and \ref{fig::ex 1d 1d'},  Tables \ref{tb::simu1ab}, \ref{tb::simu1cc'} correspond to Examples 1 in Section \ref{sec: numerical}.  Figure \ref{fig::ex 2a 2b} corresponds to Examples 2.  

\begin{table}[t]
\caption{Examples 1a and 1b,  violation rates over different $n_0$ and methods.\label{tb::simu1ab}}
\centering
\renewcommand{\arraystretch}{0.6}
\begin{tabular}{r|rrrrrrrrrrr}
\hline
&Methods&$n_0=20$&$70$&$120$&$170$&$220$&$270$&$320$&$370$&$500$&$1000$\\
\hline
\multirow{7}{2cm}{Example 1a}
&\texttt{NP-lda}&NA&.016&.047&.062&.071&.087&.074&.074&.078&.080\\
 &\texttt{NP-slda}&NA&.016&.046&.062&.071&.086&.074&.074&.077&.079\\
&\texttt{NP-penlog}&NA&.018&.050&.064&.075&.096&.071&.071&.084&.078\\
&\texttt{NP-svm}&NA&.020&.045&.064&.077&.084&.068&.064&.082&.084\\
 &\texttt{pNP-lda}&.000&.000&.004&.004&.002&.001&.001&.002&.002&.007\\
 &\texttt{elda}&.091&.084&.108&.105&.103&.104&.104&.100&.101&.081\\
&\texttt{felda}&.220&.144&.145&.138&.134&.141&.141&.126&.121&.100\\
\hline
\multirow{7}{2cm}{Example 1b}
&\texttt{NP-lda}&NA&.017&.043&.055&.072&.090&.078&.069&.078&.078\\
 &\texttt{NP-slda}&NA&.017&.043&.056&.072&.090&.075&.069&.077&.078\\
&\texttt{NP-penlog}&NA&.016&.047&.063&.076&.091&.075&.072&.084&.074\\
&\texttt{NP-svm}&NA&.022&.058&.066&.072&.089&.070&.065&.082&.075\\
 &\texttt{pNP-lda}&.028&.015&.012&.010&.005&.005&.003&.005&.002&.000\\
 &\texttt{elda}&.083&.087&.090&.095&.095&.090&.099&.102&.101&.091\\
&\texttt{felda}&.138&.122&.112&.118&.122&.121&.121&.122&.121&.112\\
\hline
\end{tabular}
\end{table}

\begin{table}[t]
\caption{Examples 1c, 1c' and 1c*,  violation rates over different $p$ and methods. \label{tb::simu1cc'}}
\centering
\renewcommand{\arraystretch}{0.6}
\begin{tabular}{r|rrrrrrrrrrr}
\hline
&Methods&$p=3$&$6$&$9$&$12$&$15$&$18$&$21$&$24$&$27$&$30$\\
\hline
\multirow{7}{2cm}{Example 1c   $\delta=0.1$}
&\texttt{NP-lda}&.044&.039&.039&.045&.058&.046&.035&.049&.044&.048\\
 &\texttt{NP-slda}&.045&.033&.037&.050&.047&.043&.034&.045&.038&.041\\
&\texttt{NP-penlog}&.037&.042&.035&.056&.050&.044&.031&.049&.043&.041\\
&\texttt{NP-svm}&.041&.040&.041&.044&.041&.042&.043&.039&.035&.048\\
 &\texttt{pNP-lda}&.001&.000&.000&.000&.000&.000&.000&.000&.000&.000\\
 &\texttt{elda}&.105&.091&.084&.107&.104&.079&.105&.099&.082&.082\\
&\texttt{felda}&.147&.206&.274&.362&.435&.548&.597&.712&.790&.817\\
\hline
\multirow{7}{2cm}{Example 1c' $\delta=0.05$}
&\texttt{NP-lda}&.044&.039&.039&.045&.058&.046&.035&.049&.044&.048\\
 &\texttt{NP-slda}&.045&.033&.037&.050&.047&.043&.034&.045&.038&.041\\
&\texttt{NP-penlog}&.037&.042&.035&.056&.050&.044&.031&.049&.043&.041\\
&\texttt{NP-svm}&.041&.040&.041&.044&.041&.042&.043&.039&.035&.048\\
 &\texttt{pNP-lda}&.001&.000&.000&.000&.000&.000&.000&.000&.000&.000\\
 &\texttt{elda}&.061&.049&.043&.052&.046&.042&.057&.054&.044&.044\\
&\texttt{felda}&.087&.115&.161&.260&.431&.410&.472&.599&.679&.732\\
\hline
\multirow{7}{2cm}{Example 1c* $\delta=0.01$}
&\texttt{NP-lda}&.001&.001&.000&.004&.002&.000&.000&.000&.002&.001\\
 &\texttt{NP-slda}&.001&.001&.001&.003&.004&.000&.002&.000&.000&.001\\
&\texttt{NP-penlog}&.001&.001&.000&.003&.004&.000&.001&.000&.001&.001\\
&\texttt{NP-svm}&.000&.002&.001&.002&.000&.002&.000&.000&.001&.001\\
 &\texttt{pNP-lda}&.000&.000&.000&.000&.000&.000&.000&.000&.000&.000\\
 &\texttt{elda}&.014&.007&.008&.009&.003&.011&.011&.016&.010&.010\\
&\texttt{felda}&.025&.032&.053&.100&.146&.188&.259&.361&.436&.530\\
\hline
\end{tabular}
\end{table}

\subsection{An extra example on t-distributions}\label{sec:heavy-tail}

{\bf Example 3.} The setting is the same as Example 1a in Section \ref{sec: numerical} in the main text,  except that instead of from multivariate Gaussian distributions,  data are generated from multivariate t-distributions with degrees of freedom 4. 

\begin{figure}[t]
\caption{Example 3,  type I and type II error for competing methods with increasing and balanced sample sizes \label{fig::ex 3}}

 \begin{subfigure}[t]{0.5\textwidth}
        \centering
        \includegraphics[scale=0.24]{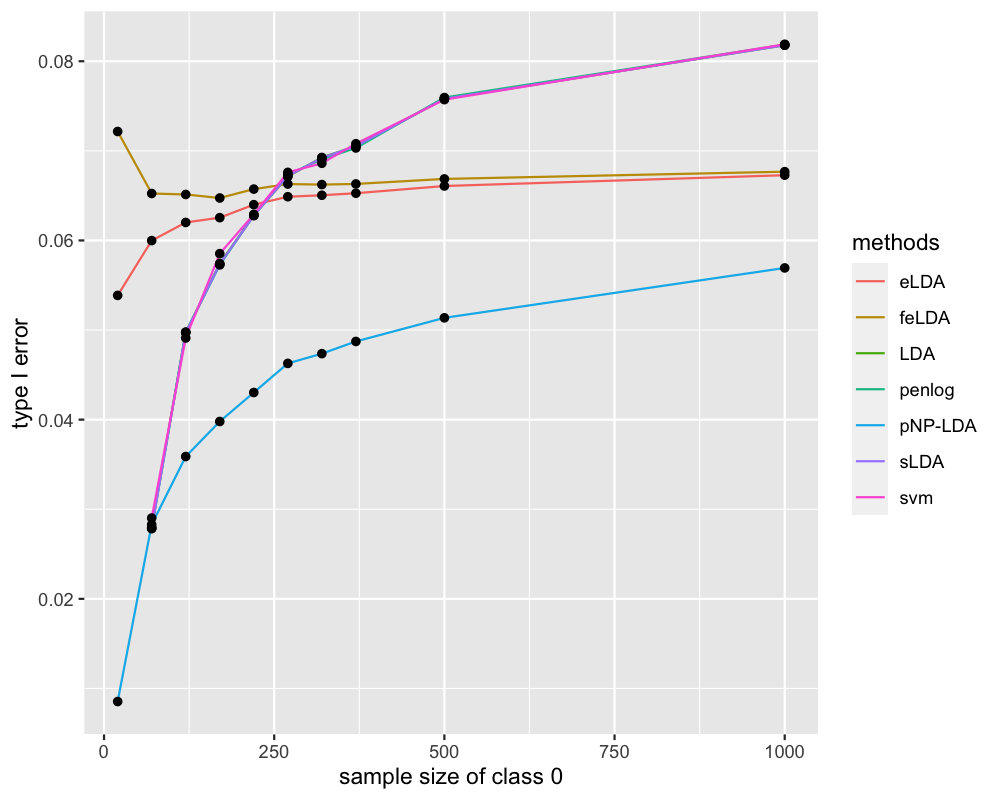}
        \caption{Example 3, Type I error}
    \end{subfigure}%
      \hspace{+0.1cm}
    \begin{subfigure}[t]{0.5\textwidth}
        \centering
        \includegraphics[scale=0.24]{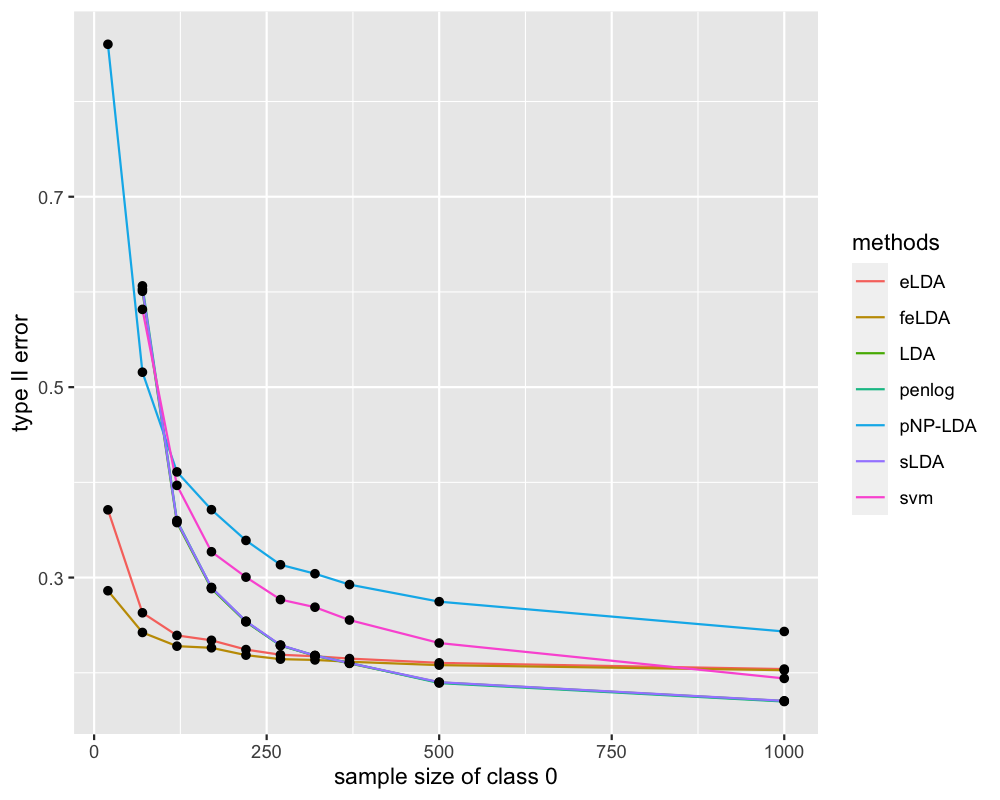}
        \caption{Example 3 Type II error}
    \end{subfigure}
    \end{figure}

Example 3 helps provide a broader understanding of the newly proposed classifiers under non-Gaussian distributions.  Figure \ref{fig::ex 3} depicts type I and type II errors, and Table \ref{tb::simu3} summarizes the observed violation rates.  We have two observations as follows: 1) among \verb+pNP-lda+, \verb+elda+ and \verb+felda+,  which are implementable for all sample sizes,  \verb+elda+ and \verb+felda+ clearly dominate \verb+pNP-lda+.  \verb+elda+ and \verb+felda+ have the type I error bounded under $\alpha$ and enjoy much smaller type II errors comparing to \verb+pNP-lda+;  2) comparing \verb+elda+ and \verb+felda+ with other umbrella algorithm based NP classifiers,  we observe that when sample size of class 0 is very small (in the current setting, less than 220),  the umbrella algorithm based classifiers either cannot be implemented ($n_0=20$) or have much worse type II errors than \verb+elda+ and \verb+felda+.  As the sample size further increases,  the performances of most umbrella algorithm based classifiers begin to catch up and eventually outperform  \verb+elda+ and \verb+felda+.  We believe this phenomenon is due to the fine calibration of the LDA model in the development of \verb+elda+ and \verb+felda+, which leads to conservative results in heavy-tail distribution settings.  On the other hand,  the nonparametric NP umbrella algorithm does not rely on any distributional assumptions and benefit from larger sample sizes.  

\begin{table}[t]
\caption{Example 3,  violation rates over different $n_0$ and methods.\label{tb::simu3}}
\centering
\renewcommand{\arraystretch}{0.6}
\begin{tabular}{r|rrrrrrrrrrr}
\hline
&Methods&$n_0=20$&$70$&$120$&$170$&$220$&$270$&$320$&$370$&$500$&$1000$\\
\hline
\multirow{7}{2cm}{Example 3}
&\texttt{NP-lda}&NA&.024&.067&.064&.068&.074&.073&.078&.073&.057\\
 &\texttt{NP-slda}&NA&.024&.071&.062&.069&.078&.069&.076&.074&.056\\
&\texttt{NP-penlog}&NA&.021&.061&.059&.069&.077&.073&.074&.075&.058\\
&\texttt{NP-svm}&NA&.026&.063&.066&.066&.084&.065&.080&.081&.068\\
 &\texttt{pNP-lda}&.000&.001&.000&.000&.000&.000&.000&.000&.000&.000\\
 &\texttt{elda}&.075&.026&.009&.006&.003&.001&.003&.001&.000&.000\\
&\texttt{felda}&.191&.043&.017&.011&.004&.001&.003&.001&.001&.000\\
\hline
\end{tabular}
\end{table}

\subsection{Lung cancer dataset continued}
For the lung cancer dataset we explored in the real data section,  we selected another set of parameters $\alpha = 0.1$,  and $\delta=0.4$ for a comparison among all five methods, including the umbrella algorithm based NP methods.  We present the results in Table \ref{tb::realdata1-2}. We observe that,  \verb+elda+ dominates \verb+NP-slda+,   \verb+NP-penlog+,  and \verb+NP-svm+ in both the type I and the type II errors.  \verb+pNP-lda+ again produces a type I error of 0 and a type II error of 1: not informative at all.  \verb+elda+ outperforms all other competing methods.
 \begin{table}[t]
\caption{Lung cancer dataset  \label{tb::realdata1-2}}
\centering
\begin{tabular}{r|rrrrrrrrrr}
\hline
&&\texttt{NP-slda}&\texttt{NP-penlog}&\texttt{NP-svm}&\texttt{pNP-lda}&\texttt{elda}\\
\hline
\multirow{2}{2cm}{$\alpha=0.1$ $\delta=0.4$}
&\texttt{type I error}&.083&.078&.081&.000&.031\\
&\texttt{type II error}&.015&.026&.022&1.000&.013\\
&\texttt{observed violation rate}&.49&.45&.46&.00&.28\\
\hline
\end{tabular}
\end{table}

\end{appendices}

\bibliographystyle{chicago}

\bibliography{npnonsplit}

%\textcolor{red}{Important lemmas, the stetch of the proofs, but not the proofs themselves will be in this section.}

\end{document}